\documentclass[a4paper,11pt]{article} \pdfoutput=1
\usepackage[normalem]{ulem}
\usepackage[utf8]{inputenc}
\usepackage{arydshln} % Dashed lines in array
\usepackage{xcolor}
\usepackage{float}
\usepackage{epsf}

\usepackage{graphicx}

\usepackage{verbatim} % allows block comments

\usepackage{amsmath}
\usepackage{amssymb}
\usepackage{mathrsfs}
\usepackage{slashed}

\usepackage{amsfonts}
\usepackage{xfrac} %for nice small fractions: sfrac
\usepackage{booktabs} % for nice tables
\usepackage{soul} % for striking out \st{text}
\usepackage{pifont} %for the symbols in the table, second try
\graphicspath{{figs/}}

\usepackage{cite}
\usepackage[
colorlinks=true
,urlcolor=black
,anchorcolor=black
,citecolor=black
,filecolor=black
,linkcolor=black
,menucolor=black
 ,pagecolor=black
,linktocpage=true
,pdfproducer=medialab
,pdfa=true
]{hyperref}

\usepackage[left=2cm,right=2cm,top=2cm,bottom=3cm]{geometry}

\usepackage{color}
\definecolor{cred}{RGB}{180,50,40}
\definecolor{purple}{RGB}{180,90,180}
\definecolor{darkgreen}{RGB}{0, 100, 0}

\def\be{\begin{equation}}
\def\ee{\end{equation}}
\newcommand{\bea}{\begin{eqnarray}}
\newcommand{\eea}{\end{eqnarray}}

\allowdisplaybreaks
\begin{document}

\setcounter{footnote}{0}
%\vspace*{-1.5cm}
%\begin{flushright}
%LPT Orsay-19-17 \\

%\vspace*{2mm}
%\today
%\end{flushright}
\begin{center}
\vspace*{1mm}

\vspace{1cm}
{\Large\bf
  A nonunitary interpretation for a single vector leptoquark   \\}

\vspace*{0.2cm}{\Large\bf combined explanation to the $B$-decay anomalies}

\vspace*{0.8cm}
\renewcommand*{\thefootnote}{\fnsymbol{footnote}}

{\bf C.~Hati\,\footnote{chandan.hati@clermont.in2p3.fr},
J.~Kriewald\,\footnote{jonathan.kriewald@clermont.in2p3.fr},
J. Orloff\,\footnote{jean.orloff@clermont.in2p3.fr} and
A.~M.~Teixeira\,\footnote{ana.teixeira@clermont.in2p3.fr}}

\renewcommand*{\thefootnote}{\arabic{footnote}}
\setcounter{footnote}{0}

\vspace*{.5cm}
Laboratoire de Physique de Clermont (UMR 6533), CNRS/IN2P3,\\
Univ. Clermont Auvergne, 4 Av. Blaise Pascal, F-63178 Aubi\`ere Cedex,
France
\end{center}

\vspace*{2mm}

\begin{abstract}
  In order to simultaneously account for both $R_{D^{(\ast)}}$ and
  $R_{K^{(\ast)}}$ anomalies in $B$-decays, we consider an extension
  of the Standard Model by a single vector leptoquark field, and study
  how one can achieve the required lepton flavour non-universality,
  starting from a priori universal gauge couplings.  While the unitary
  quark-lepton mixing induced by $SU(2)_L$ breaking is insufficient,
  we find that effectively nonunitary mixings hold the key to
  simultaneously address the $R_{K^{(\ast)}}$ and $R_{D^{(\ast)}}$
  anomalies.  As an intermediate step towards various UV-complete
  models, we show that the mixings of charged leptons with additional
  vector-like heavy leptons successfully provide a nonunitary
  framework to explain $R_{K^{(\ast)}}$ and $R_{D^{(\ast)}}$.
  These realisations have a strong impact for electroweak
  precision observables and for flavour violating ones: isosinglet
  heavy lepton realisations are already excluded due to excessive
  contributions to lepton flavour violating $Z$-decays.
  Furthermore, in the near future, the expected progress in the sensitivity
  of charged lepton flavour violation experiments should allow to fully probe
  this class of vector leptoquark models.
\end{abstract}
\vspace*{1mm}

\section{Introduction}
In the Standard Model (SM), gauge interactions are strictly
flavour universal, as confirmed by precision measurements of
several electroweak observables, such as $Z\to \ell \ell$
decays~\cite{Tanabashi:2018oca, ALEPH:2005ab}.
Recently, a number of observables related to $B$-meson semileptonic
decays has started exhibiting slight deviations,
from their SM predictions, a.k.a.~{\it anomalies}, suggesting the possibility of lepton
flavour universality violation (LFUV).
The most robust LFU-sensitive measurements arise from
ratios of individual decay modes, where the theoretical hadronic
uncertainties
(e.g. from form factors) cancel out, such as the ratio $R_{D^{(*)}}$ between
charged current decays, or the ratio $R_{K^{(\ast)}}$ between neutral
current decays, respectively defined as %\vspace{-2mm}
\begin{equation}
  R_{D^{(*)}} \,= \,\frac{\text{BR}(B \to D^{(*)} \,\tau^- \,\bar\nu)}{
    \text{BR}(B \to  D^{(*)}\, \ell^- \,\bar\nu)}\, , \quad R_{K^{(\ast)}} \,= \,
  \frac{\text{BR}(B \to K^{(*)}\, \mu^+\,\mu^-)}{\text{BR}(B \to
  K^{(*)}\, e^+\,e^-)}\,,
\end{equation}
where $\ell=e,\,\mu$.
Several
experiments have reported deviations from the theoretical LFU SM expectations~\cite{Lees:2012xj,Lees:2013uzd, Huschle:2015rga,
  Adachi:2009qg, Bozek:2010xy, Aaij:2015yra,Hirose:2016wfn, Abdesselam:2019dgh, Aaij:2019wad, Aaij:2017vbb, Abdesselam:2019wac, Aaij:2015esa, Wehle:2016yoi}.
Quantitatively, the current measured values of $R_D$~\cite{Amhis:2016xyh, Abdesselam:2019dgh} and
$R_{D^\ast}$~\cite{Aaij:2015yra,Hirose:2016wfn,Amhis:2016xyh,Abdesselam:2019dgh}
exceed the SM predictions by about
$1.4\,\sigma$ and $2.5\,\sigma$ respectively~\cite{Bigi:2016mdz,Bigi:2017jbd}, and their combination
leads to a deviation of $3.1\,\sigma$ from the SM
prediction~\cite{Ligeti:2016npd,Crivellin:2016ejn,
  Amhis:2016xyh}.
On the other hand, and independently of the charged current modes,
the measurement of $R_K$ for the dilepton invariant mass squared bin
$[1.1,6]~\text{GeV}^{2}$~\cite{Aaij:2019wad} displays a $2.5\,\sigma$
deviation below the SM
prediction~\cite{Bordone:2016gaq,Capdevila:2017bsm}. Likewise, the measurement
of $R_{K^*}$~\cite{Aaij:2017vbb} translates into $2.3\,\sigma$ and
$2.6\,\sigma$ deviations, also below the expected SM
values for the dilepton
invariant mass squared bins $[0.045, 1.1]~\text{GeV}^{2}$ and
$[1.1, 6]~\text{GeV}^{2}$, respectively~\cite{Bordone:2016gaq,Capdevila:2017bsm}.

\noindent Further neutral current anomalies have emerged, for instance in the
observable $\Phi \equiv d \text{BR}(B_s\to\phi\mu\mu)/ dm_{\mu\mu}^2$,
in a similar kinematic regime
($m_{\mu\mu}^2\in[1,6]\,{\rm
  GeV}^2$)~\cite{Aaij:2015esa,Altmannshofer:2014rta,Straub:2015ica},
also with a deviation of about $3\,\sigma$. Deviations from
the SM expectations have also been found in the angular observable
$P_5^{\prime}$ of the $B \to K^\ast \ell^+ \ell^-$ decay.

Although these anomalies by no means invalidate the SM at this stage, their
persistence and relatively coherent pattern
inevitably raise the question of which
(minimal) new ingredients beyond the SM (BSM) would be required to
explain them.
A first model-independent
approach~\cite{Alguero:2019ptt,Aebischer:2019mlg,Ciuchini:2019usw,Datta:2019zca,Arbey:2019duh,Shi:2019gxi,Bardhan:2019ljo,Alok:2019ufo,Alok:2017qsi,Ghosh:2014awa,Glashow:2014iga,
  Bhattacharya:2014wla,Freytsis:2015qca,Ligeti:2016npd,Ciuchini:2017mik,Bigi:2017jbd}
is to introduce higher dimensional effective operators, coupling two
quarks with two leptons. Despite the large number of possibilities, it
is nevertheless remarkable that only a reduced number
of such non-standard couplings
significantly eases the tensions with the SM predictions.

It is thus desirable to consider which BSM constructions could be at
the origin of these effective operators. Among the most minimal scenarios
studied, one has flavour-sensitive $Z^\prime$
exchanges~\cite{Altmannshofer:2014cfa,Crivellin:2015mga,
  Crivellin:2015lwa,Sierra:2015fma,
  Crivellin:2015era,Celis:2015ara,Bhatia:2017tgo,Kamenik:2017tnu,Chen:2017usq,
  Camargo-Molina:2018cwu,Darme:2018hqg,Baek:2018aru,Biswas:2019twf,
  Allanach:2019iiy},
leptoquark exchanges~\cite{Hiller:2014yaa,
  Gripaios:2014tna,Sahoo:2015wya,Varzielas:2015iva,Alonso:2015sja,Bauer:2015knc,
  Hati:2015awg,Fajfer:2015ycq,Das:2016vkr,Becirevic:2016yqi,Sahoo:2016pet,Cox:2016epl,
  Crivellin:2017zlb,Becirevic:2017jtw,Cai:2017wry,Dorsner:2017ufx,Greljo:2018tuh,
  Sahoo:2018ffv,Becirevic:2018afm,Hati:2018fzc,deMedeirosVarzielas:2018bcy,
  Aebischer:2018acj,deMedeirosVarzielas:2019okf,Yan:2019hpm,Bigaran:2019bqv,Popov:2019tyc},
$R-$parity violating supersymmetric
models~\cite{Deshpand:2016cpw,Altmannshofer:2017poe,Das:2017kfo,Earl:2018snx,
  Trifinopoulos:2018rna,Trifinopoulos:2019lyo}, and various
other constructions~\cite{Greljo:2015mma,Arnan:2017lxi,Geng:2017svp,Choudhury:2017qyt,
  Choudhury:2017ijp,Grinstein:2018fgb,Cerdeno:2019vpd,Bhattacharya:2019eji,
  Crivellin:2019dun,Arnan:2019uhr}.

In this work, we focus on the exchange of a vector leptoquark $V_1$ transforming
as $(\mathbf{3},\mathbf{1},2/3)$ under the SM gauge group,
which has been
shown to be particularly
attractive for its ability to provide a single particle solution
simultaneously to
both charged and neutral current anomalies~\cite{Assad:2017iib,Buttazzo:2017ixm,Calibbi:2017qbu,Bordone:2017bld,Blanke:2018sro,Bordone:2018nbg,Kumar:2018kmr,Angelescu:2018tyl,Balaji:2018zna,
  Fornal:2018dqn,Baker:2019sli,Cornella:2019hct,DaRold:2019fiw}. Complying with
the experimental measurements suggests that $V_1$ should have
non-universal couplings to quarks and leptons.
We assume $V_1$ to be an elementary spin-1 gauge boson; since it carries charges (as a
leptoquark must), the underlying gauge symmetry is necessarily
non-abelian, with universal (gauge) couplings as long as it remains
unbroken\footnote{In this work we are interested in the minimal (gauge extension) scenario where the vector leptoquark is an elementary gauge boson corresponding to a gauge group under which the SM fermion generations are universally charged and no additional protection or symmetry is introduced to induce non-universality. For models in which the vector leptoquark appears as a composite field, see for instance \cite{Barbieri:2016las}; for other models where the gauge group is non-minimal and/or the gauge charges of the SM fermion generations are non-universal, see e.g. \cite{Greljo:2018tuh}.}.
As an example, such a field $V_1$ is naturally
contained within the theoretically well-motivated Pati-Salam model
(PS) as an $SU(4)$ gauge boson. However, the current bounds on the
charged lepton
flavour violating (cLFV) decays $K_L\rightarrow \mu e$ and
$K\rightarrow \pi \mu e$ lead to dramatic (lower) bounds
on the mass of such a vector
leptoquark ($m_V$), typically above the 100~TeV scale for $\mathcal O(1)$
couplings~\cite{Hung:1981pd,Valencia:1994cj,Smirnov:2007hv,Carpentier:2010ue,Kuznetsov:2012ai,Smirnov:2018ske}. In turn, this renders the new state
excessively heavy to account for the $B$-meson decay
anomalies.

The cLFV bound on $m_V$ turns out to effectively preclude a viable
solution to both charged and neutral current anomalies: in the
unbroken phase, $V_1$ has a single universal coupling to matter;
$SU(2)$-breaking introduces a possible misalignment of the quark and
lepton mass eigenbases, thus resulting in LFU-violating $V_1$
couplings, proportional to a $3\times3$ unitary matrix.  In order to
explain the $R_{D^{(*)}}$ anomalies, the $b\tau$ and $s\tau$ couplings
are required to be large\footnote{To satisfy the constraints from the $\tau$ decays, the $c\nu$ coupling induced by $b\tau$ via CKM mixing is in general not sufficient to comfortably explain $R_{D^{(\ast)}}$ \cite{Feruglio:2017rjo}; on the other hand the maximum $c\nu$ coupling induced by $d_i\mu$ and $d_ie$ (for neutrino flavour in $c\nu$  different from $\nu_\tau$) are fixed by $R_K^{(*)}$ data (for $i=2,3$) and kaon decays (for $i=1$), while the $c\nu$ coupling induced by $d\tau$ is highly CKM suppressed. In view of this and working in a unitary parametrisation of the leptoquark couplings, the only viable possibility therefore is to maximise the $b\tau$ and $s\tau$ entries.}, which in turn leads to large couplings
between the first two generations of quarks and leptons (as a
consequence of the unitarity of the mixing matrix), leading to
excessive contributions to cLFV.  The question that naturally emerges
is whether one can find a minimal embedding of $V_1$
that successfully allows to overcome the cLFV constraints and address
both $R_{K^{(*)}}$ and $R_{D^{(*)}}$ anomalies. In other words, can
one go beyond the tight constraints arising from a ($3\times 3$)
unitary mixing of quarks and leptons?
This necessarily requires the addition of new fields, beyond $V_1$,
and along these lines,
one possibility is to add other vector
leptoquark fields (thus implying a larger gauge group), whose mixing
would allow to overcome the above mentioned constraints, as
explored in~\cite{Bordone:2017bld}.

In the present study,
we avoid this further enlarging of the gauge group,
adhering to the single vector leptoquark
hypothesis, and pursue a distinct avenue.
In particular, and motivated
by the phenomenological impact of having nonunitary left-handed
leptonic mixings
in the presence of (heavy) sterile neutral leptons~\cite{Xing:2007zj, Blennow:2016jkn,
  Fernandez-Martinez:2016lgt, Escrihuela:2015wra}, we consider the
possibility of nonunitary $V_1$ couplings, as arising from
the presence of $n$ additional vector-like heavy leptons $L$
(also present in the construction of~\cite{Bordone:2017bld}).
In the broken phase, the
$V_1$ couplings are then given by a $(3+n)\times (3+n)$
mixing matrix, so that the couplings to SM fermions now correspond to
a $3\times 3$ sub-block, which is no longer
unitary. We argue that
this departure from unitary mixings might indeed hold the key to
simultaneously address $R_{K^{(\ast)}}$ and $R_{D^{(\ast)}}$ data,
while satisfying existing cLFV constraints.

The addition of vector-like heavy charged leptons\footnote{Heavy
  vector-like quarks will not be considered, as they are not required
  for a minimal working model.} can be seen as an intermediary step
towards a full ultraviolet-complete model, providing a better
framework for the peculiar structure of leptoquark couplings required
by the anomalies. In this framework, the nonunitary mixings will also
lead to the modification of SM-like charged and neutral lepton
currents, establishing an inevitable link to electroweak precision
(EWP) observables, such as lepton flavour violating and/or LFUV $Z$-decays.  The latter
observables will prove to be extremely constraining, ultimately
leading to the exclusion of isosinglet vector-like heavy leptons as a
source of non-universality in $B$-meson decays.

These constraints are much milder for isodoublet heavy leptons:
after arguing that for a single additional heavy charged lepton, cLFV
constraints exclude an explanation of even $R_{D^{(*)}}$
alone, we show that the addition of $n=3$ vector-like
isodoublet leptons allows a simultaneous explanation of both
$R_{K^{(*)}}$ and $R_{D^{(*)}}$ anomalies, while respecting all
available constraints.

This work is organised as follows: in Section~\ref{sec:framework} we
describe the underlying framework; Section~\ref{sec:pheno:constraints}
is devoted to a comprehensive analysis of the phenomenological
implications of the nonunitary framework, regarding the $B$-meson
anomalies, and several flavour and EWP observables. A summary and
concluding remarks can be found in Section~\ref{sec:concs}.

\section{Towards a nonunitarity interpretation of vector leptoquark
  couplings}\label{sec:framework}

As mentioned in the Introduction, we consider here a SM extension by a single
vector leptoquark $V_1$, which transforms under the SM
gauge group
${SU}(3)_c\times {SU}(2)_L \times U(1)_Y$ as
$(\mathbf{3},\mathbf{1},2/3)$.
Without loss of generality, we assume that $V_1$ is a gauge boson
of an unspecified gauge extension of $SU(3)_c$ with a
universal (i.e. flavour blind) gauge coupling; without relying
on a specific gauge embedding and/or
Higgs sector, our only working assumption is that
all fermions acquire a mass after electroweak symmetry breaking (EWSB), and that the physical
eigenstates are obtained from the diagonalisation of the corresponding
(generic) mass matrices.
In the weak basis, the interaction of $V_1$ with the SM
matter fields can be written as
\begin{eqnarray}
\label{eq:lagrangian:Vql0}
\mathcal{L} \supset \sum_{i=1}^{3}
V_{1}^\mu \left [
  \frac{\kappa_{L}}{\sqrt{2}}
  \left(\bar{{d}}_{L}^{0,i} \gamma_\mu {\ell}_{L}^{0,i} +
  \bar{{u}}_{L}^{0,i} \gamma^\mu  {\nu}_{L}^{0,i} \right)
  + \frac{\kappa_{R}}{\sqrt{2}}
  \bar{{d}}_{R}^{0,i} \gamma_\mu {\ell}_{R}^{0,i}
  + \frac{\bar{\kappa}_{R}}{\sqrt{2}}
  \bar{{u}}_{R}^{0,i} \gamma_\mu {\nu}_{R}^{0,i}
\right]+\text{H.c.}\, ,
\end{eqnarray}
in which the ``0'' superscript denotes interaction states, and $i=1 - 3$ are
family indices. The couplings $\kappa_{L,R}$ are flavour diagonal, and
universal.
Since left-handed couplings are the minimal essential ingredient
frequently called upon to
simultaneously explain the neutral and charged current
anomalies~\cite{Buttazzo:2017ixm}, for simplicity we will henceforth only consider the
latter (i.e., taking $\kappa_{L} \neq 0$ and
$\kappa_{R}=\bar{\kappa}_R=0$). Furthermore, notice that this can be
easily realised in chiral PS
models~\cite{Buttazzo:2017ixm,Balaji:2018zna,Fornal:2018dqn},
and is moreover  phenomenologically well-motivated\footnote{
In the context of PS unification
it has been noted in the literature
that if the vector leptoquark couples to
both left- and right-handed fermion fields with similar gauge
strength, then in the absence of some helicity suppression,
bounds from various searches for lepton flavour
violating mesonic decay modes put a lower limit on the vector
leptoquark mass around 100~TeV~\cite{Valencia:1994cj,Smirnov:2007hv,Carpentier:2010ue,Kuznetsov:2012ai,Smirnov:2018ske}.}.

\noindent
In terms of physical fields, the Lagrangian can be written as
\begin{eqnarray}
\label{eq:lagrangian:Vql_phys3}
\mathcal{L} \supset \sum_{i,j,k=1}^{3}
V_{1}^\mu \left( \bar{d}_{L}^{i}\, \gamma_\mu\,
K_1^{ik}\,\ell_{L}^{k} +\bar{u}_{L}^{j}\, V^{\dagger}_{ji}\,\gamma_\mu\,
K_1^{ik}\, U^{\text P}_{kj}\, \nu_{L}^{j}\right)
+\text{H.c.}\, ,
\end{eqnarray}
where $V$ is the Cabibbo-Kobayashi-Maskawa (CKM) mixing matrix
and $U^{\text P}\equiv U^{\ell\dagger}_L U^{\nu}_L$ the
Pontecorvo-Maki-Nakagawa-Sakata (PMNS) leptonic mixing matrix; we have
also introduced $K_1\equiv \frac{\kappa_{L}}{\sqrt{2}} U^{\ell}_L$ to
denote the ``effective'' leptoquark couplings in the physical fermion
basis. Being proportional to an
  arbitrary unitary matrix (which we hereby denote $V_0$),
  $K_1$ can be further cast as
\begin{equation}\label{eq:V0:3x3}
  K_1=\frac{\kappa_L}{\sqrt{2}}\, V_0=\frac{\kappa_L}{\sqrt{2}}
\begin{pmatrix}
c_{12}c_{13} & s_{12}c_{13} & s_{13}%\mathrm e^{-i\delta}
\\
-s_{12}c_{23}- c_{12}s_{23}s_{13}%\mathrm e^{-i\delta}
&
c_{12}c_{23} - s_{12}s_{23} s_{13}%\mathrm e^{-i\delta}
 & s_{23}c_{13} \\
s_{12} s_{23} - c_{12} c_{23} s_{13} %\mathrm e^{-i\delta}
&
- c_{12}s_{23} - s_{12} c_{23} s_{13}%\mathrm e^{-i\delta}
 & c_{23} c_{13}
\end{pmatrix}\,,
\end{equation}
in which we used the standard parametrisation of a real $3\times 3$ unitary
matrix in terms of three angles $\theta_{12,23,13}$ (with $c_{ij}$ and
$s_{ij}$ respectively denoting $\cos \theta_{ij}$ and $\sin
\theta_{ij}$), and we restrict ourselves to real parameter space in all our analysis. (We emphasise here that $V_0$ is not the PMNS matrix, and that the
above angles are not those associated with neutrino
oscillation data.)

\subsection{Accounting for $\pmb{R_{K^{(\ast)}}}$ and $\pmb{R_{D^{(\ast)}}}$
  in a minimal $\pmb{V_1}$ leptoquark
  framework}\label{sec:unitarycase}

The presence of a vector leptoquark, whose interactions with
quarks and leptons are defined in Eqs.~(\ref{eq:lagrangian:Vql0},
\ref{eq:lagrangian:Vql_phys3}), can induce new operators, contributing
to $b$-decays (both neutral and charged currents). In the SM,
$b \to s \ell\ell$ and $b \to c \ell \nu$ decays respectively occur at
one-loop and at tree-level; on the other hand, the new $V_1$-mediated
contributions to both decays arise at the tree-level.  Thus, the new
contributions required to explain $R_{K^{(\ast)}}$ data are
comparatively smaller than those needed to account for the discrepancy
in $R_{D^{(*)}}$ data: in particular, $R_{D^{(*)}}$ requires the mass
scale of $V_1$ to be quite low $\sim \mathcal{O}(1\,\mathrm{TeV})$,
while it is possible to explain $R_{K^{(\ast)}}$ for leptoquark masses
$m_V\sim \mathcal{O}(10\,\mathrm{TeV})$ (taking into account all the
constraints from rare transitions and decays).  The low mass scale
required to explain the $R_{D^{(\ast)}}$ anomaly effectively precludes
a simultaneous (combined) explanation for both anomalies, due to the
excessive associated contributions to cLFV kaon decays, in particular
to $K_L \rightarrow e^\pm \mu^\mp$ (which occurs at the tree level).
Consequently, both modes ($K_L \to e^{+} \mu^{-} $ and
$K_L \to e^{-} \mu^{+} $) have to be suppressed separately.  In terms
of the parametrisation of Eq.~\eqref{eq:V0:3x3}, saturating
$R_{D^{(\ast)}}$ requires maximising the $23$ and $33$ entries of
$V_0$ (thus leading to $\theta_{13} \sim 0$ and
$\theta_{23} \sim \frac{\pi}{4}$).  This implies that the branching
fractions of the tree-level kaon decay modes are proportional to
$\sin^2\theta_{12}$ and $\cos^2\theta_{12}$, respectively.  A
sufficient and simultaneous suppression of contributions to these
modes is then clearly not possible.

\begin{figure}
\centering
\includegraphics[width = 0.5\textwidth]{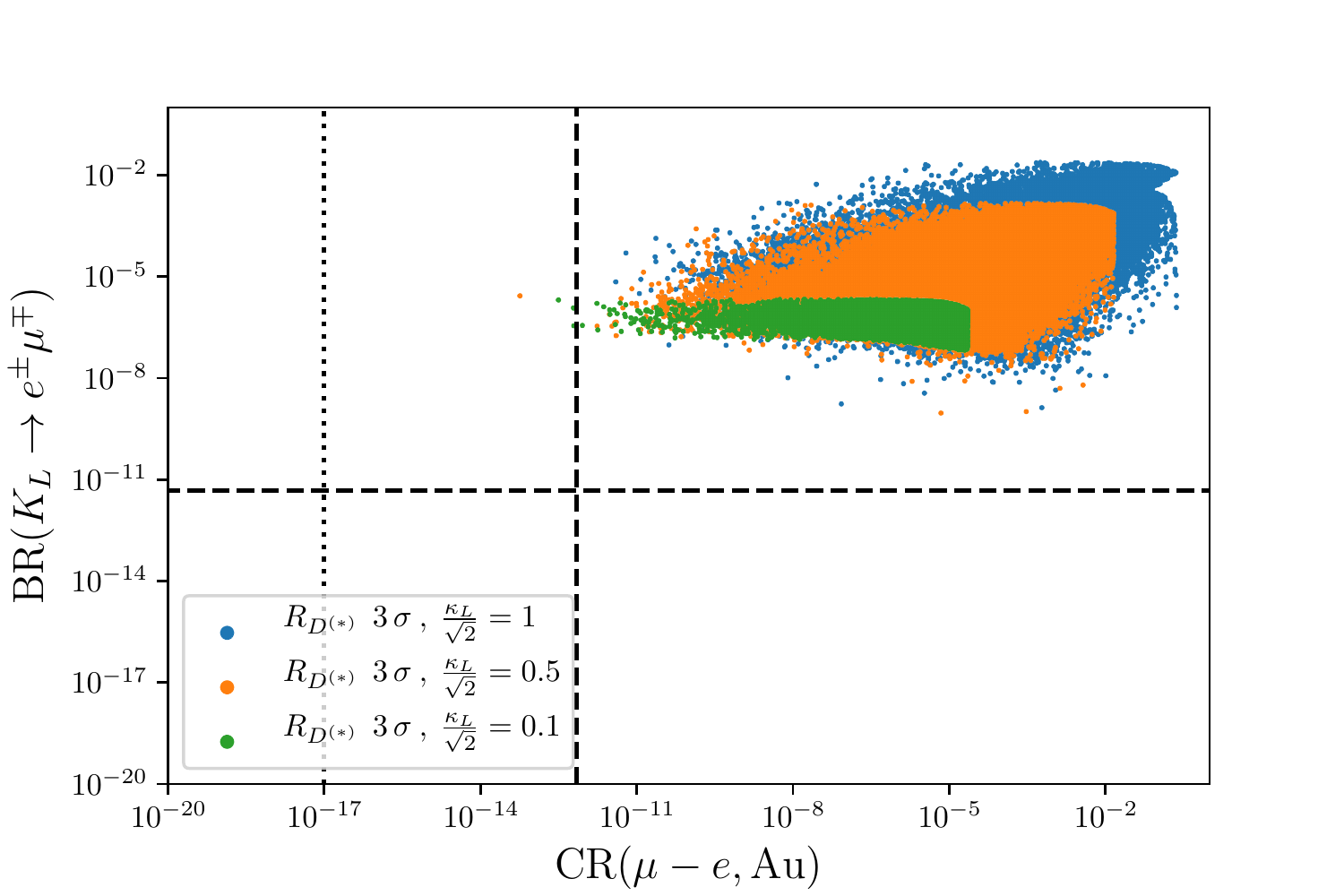}
\caption{Associated predictions for CR($\mu - e$, Au) and
  BR($K_L \to e^\pm \mu^\mp$) for sample points satisfying
  $R_{D^{(\ast)}}$ at the $3\:\sigma$ level for the (unitary)
  parametrisation of Eq.~(\ref{eq:V0:3x3}). The dashed lines represent
  the current experimental upper bounds (see Tables~\ref{tab:flavour}
  and~\ref{tab:cLFV} in Section~\ref{sec:constraints}), and the dotted
  line a benchmark future sensitivity to CR($\mu - e$, Al).
  All mixing angles have been varied randomly between $-\pi$ and $\pi$ and the leptoquark mass is set to $m_V \sim 1.5 \mathrm{TeV}$. The blue, orange and green points respectively correspond to three benchmark choices, $\frac{\kappa_L}{\sqrt{2}}=1,\, 0.5,\, 0.1$.}
\label{fig:RDKaon}
\end{figure}

Likewise, excessively large contributions to $\mu-e$ conversion (also
occurring at tree-level) further exclude a low scale realisation, with
$m_V \sim \mathcal{O}(1\,\mathrm{TeV})$.  The above arguments are
illustrated by Fig.~\ref{fig:RDKaon}, in which we display the
predictions for neutrinoless $\mu - e$ conversion and
$K_L \to e^\pm \mu^\mp$ associated with having contributions to
$R_{D^{(\ast)}}$ within $3 \sigma$ of the current best fit (for
$m_V \sim \mathcal{O}(1\,\mathrm{TeV})$ and three different values of $\sfrac{\kappa_L}{\sqrt{2}}$).

\subsection{Vector-like fermions and ``effective'' nonunitary mixings
  in the light sector}\label{sec:nonunitarycase}
The above discussion suggests that the minimal flavour structure
encoded in the (unitary) parametrisation of the
leptoquark-quark-lepton currents (Eq.~(\ref{eq:lagrangian:Vql_phys3}))
is insufficient to
account for both anomalies.
A stronger enhancement of LFUV in the leptoquark couplings can be
achieved if one hypothesises that the ``effective'' leptoquark
mixings - i.e. the $3\times 3$ matrix $V_0$ is nonunitary.
As we proceed to discuss, in order to explain
$R_{K^{(*)}}$ and $R_{D^{(*)}}$ data simultaneously, and for
universal gauge couplings, a highly nonunitary flavour misalignment
between quarks and leptons is in fact required.

Such a nonunitary flavour misalignment can be understood in the
presence of heavy vector-like fermions,
${SU}(2)_L$ singlets or doublets, which have non-negligible
mixings with the SM fermions. This can be encoded by generalising the
charged lepton mixing matrix to a $3\times (3+n)$ semi-unitary matrix, so
that SM interaction fields and physical states are related as
$\ell^0_L=U^{\ell}_L \ell_L$ (for $n$ additional heavy states).

\noindent
The Lagrangian of Eq.~(\ref{eq:lagrangian:Vql_phys3}) can thus be
recast as
\begin{eqnarray}
\label{eq:lagrangian:Vql_phys3N}
\mathcal{L} \supset \sum_{i,j=1}^{3} \sum_{k=1}^{3+n}
V_{1}^{\mu} \left(
\bar{d}_{L}^{i} \gamma_\mu  K_L^{ik} \ell_{L}^{k} +
\bar{u}_{L}^{j} V^{\dagger}_{ji} \gamma_\mu K_L^{ik} U^{\text P}_{kj}
\nu_{L}^{j}
\right)
+\text{H.c.}\, .
\end{eqnarray}
Notice that in the above equation, the effective leptoquark coupling
$K_L$ generalises $K_1$ of Eq.~\eqref{eq:V0:3x3}, and now corresponds to a rectangular
$3\times (3+n)$ matrix, which can be written in terms
of $U_L^\ell$ as $K_L\equiv \frac{\kappa_{L}}{\sqrt{2}} U^{\ell}_L$.

Finally, $K_L$ can be further decomposed as $K_L=\left( K_1 , K_2\right)$,
so that $K_1$ can be now identified with the nonunitary mixings in
the light sectors (contrary to the simple limit of Eq.~(\ref{eq:V0:3x3})). $K_2$ is a $3\times n$ matrix which corresponds to the
$n$ heavy degrees of freedom describing the coupling parameters of the
heavy (vector-like) states.
Inspired by the approach frequently adopted in the context of neutrino
physics, the deviation from unitarity in the $K_1$ block can now be
parametrised as~\cite{Xing:2007zj,Blennow:2016jkn,
Fernandez-Martinez:2016lgt,Escrihuela:2015wra}
\begin{equation}\label{eq:Ndescopm_C1}
  K_1=\frac{\kappa_L}{\sqrt{2}} A\,
  V_0=\frac{\kappa_L}{\sqrt{2}}\begin{pmatrix}\alpha_{11} & 0 & 0\\
\alpha_{21} & \alpha_{22} & 0\\
\alpha_{31} & \alpha_{32} & \alpha_{33}
\end{pmatrix}V_0\,,
\end{equation}
with $V_0$ given in Eq.~(\ref{eq:V0:3x3}).
The left-triangle matrix $A$, characterises the deviation from
unitarity and encodes the effects of the mixings with the heavy states.

As already mentioned in the Introduction, we assume
that the vector leptoquark $V_1$ appears as a gauge boson in an
unspecified $SU(3)_c$ extension. Since neither the gauge embedding nor the
Higgs sector is explicitly specified, our only assumption is that
after EWSB all fermions (SM and vector-like)
are massive, and that the physical eigenstates are obtained from
the diagonalisation of an (effective) generic $(3+n)\times (3+n)$
lepton mass matrix.
For simplicity  (see Section~\ref{sec:results}), we take $n=3$ generations of
  heavy leptons in what follows; the $6\times6$ charged lepton mass matrix $\mathcal M_\ell$ can be
diagonalised by a bi-unitary transformation
\begin{equation}
  \mathcal{M_\ell}^{\text{diag}} \,= \,
 U^{\ell\dagger}_L \, \mathcal{M_\ell} \, U^\ell_R\,.
\end{equation}
Being a unitary $6\times6$ matrix, $U^\ell_L$ can be parametrised by
15 real angles and 10 phases, and cast as a the
product of 15 unitary rotations, $\mathcal R_{ij}$. By choosing a
convenient ordering for the products of the complex rotation matrices,
one can establish a parametrisation that allows isolating
the information relative to the heavy leptons in a simple and compact form.
Schematically, this can be described by the following
($2\times2$ block matrix) decomposition~\cite{Xing:2007zj}, to which we adhere for the remainder of our discussion,
\begin{equation}\label{eq:UL:ARBS}
U^\ell_L = \begin{pmatrix}
A & R\\
B & S
\end{pmatrix}
\begin{pmatrix}
V_0 & \mathbf 0\\
\mathbf 0 & \mathbf 1
\end{pmatrix}
\end{equation}
further defining
\begin{equation}
\begin{split}
\begin{pmatrix}
A & R\\
B & S
\end{pmatrix}
& =
\mathcal R_{56} \mathcal R_{46} \mathcal R_{36} \mathcal R_{26}
\mathcal R_{16} \mathcal R_{45} \mathcal R_{35}
\mathcal R_{25} \mathcal R_{15} \mathcal R_{34}
\mathcal R_{24} \mathcal R_{14}\:\text,\\
\begin{pmatrix}
V_0 & \mathbf 0\\
\mathbf 0 & \mathbf 1
\end{pmatrix}
& = \mathcal R_{23} \mathcal R_{13} \mathcal R_{12}\,.
\end{split}
\end{equation}
Under the above decomposition, one can still identify the SM-like
mixings, given by $V_0$ (cf. Eq.~(\ref{eq:V0:3x3})); the leptoquark
couplings\footnote{Note that this is an identification of the mixing
  elements with the effective leptoquark couplings by choosing the basis
  in which the down-type quarks are diagonal.
} are now parametrised by the $3\times 6$ (rectangular) matrix,
\begin{equation}\label{eq:KL:K1K2}
	K_L\, =\,(K_1, K_2)\, = \,\frac{\kappa_L}{\sqrt{2}}(A \,V_0, R)\,.
\end{equation}
The diagonal elements of the triangular matrix $A$, $\alpha_{ii}$, can be expressed as
\begin{eqnarray}\label{eq:Aalphaii}
\alpha_{ii} \,= \,c_{i6}\,c_{i5}\,c_{i4}\,,
\end{eqnarray}
in terms of the cosines of the mixing angles,
${c}_{ij}=\cos\theta_{ij}$. (The SM-like limit can be recovered for
$A \to \mathbf 1$.)
The off-diagonal elements can be cast
as~\cite{Xing:2007zj}
\begin{eqnarray}\label{eq:Aalphaij}
\alpha_{21} &=& -c^{}_{14} \,c^{}_{15} \,\hat{s}^{}_{16} \,\hat{s}^\ast_{26} -
c^{}_{14} \,\hat{s}^{}_{15} \,\hat{s}^\ast_{25} \,c^{}_{26}
-\hat{s}^{}_{14}\, \hat{s}^\ast_{24} \,c^{}_{25}\, c^{}_{26}\,, \nonumber\\
\alpha_{32} &=& -c^{}_{24}\, c^{}_{25}\, \hat{s}^{}_{26} \,\hat{s}^\ast_{36} -
c^{}_{24}\, \hat{s}^{}_{25} \,\hat{s}^\ast_{35} \,c^{}_{36} -\hat{s}^{}_{24}
\,\hat{s}^\ast_{34}\, c^{}_{35}\, c^{}_{36}\,, \nonumber\\
\alpha_{31} &=&-c^{}_{14} \,c^{}_{15}\, \hat{s}^{}_{16}\, c^{}_{26}\,
\hat{s}^\ast_{36}
+ c^{}_{14} \,\hat{s}^{}_{15} \,\hat{s}^\ast_{25} \,\hat{s}^{}_{26}
\,\hat{s}^\ast_{36}
- c^{}_{14} \,\hat{s}^{}_{15}\, c^{}_{25} \,\hat{s}^\ast_{35}\,
c^{}_{36}\nonumber\\
&+&\hat{s}^{}_{14} \,\hat{s}^\ast_{24}\, c^{}_{25}\, \hat{s}^{}_{26}\,
\hat{s}^\ast_{36}
+ \hat{s}^{}_{14} \,\hat{s}^\ast_{24} \,\hat{s}^{}_{25} \,
\hat{s}^\ast_{35} \,c^{}_{36}
- \hat{s}^{}_{14} \,c^{}_{24} \,\hat{s}^\ast_{34} \,c^{}_{35}c^{}_{36}\,,
\end{eqnarray}
where
$\hat{s}_{ij} \equiv e^{i\delta^{}_{ij}} \sin\theta^{}_{ij}$,
with $\theta^{}_{ij}$ and
$\delta^{}_{ij}$ respectively being the angles and CP phases associated
with the $\mathcal R_{ij}$ rotation. Finally, it is worth emphasising
that not only the full $6\times6$ matrix $U_L^\ell$ is
unitary,
but its upper $3\times6$ block $(A V_0, R)$ is also semi-unitary on its
own, with $\frac{2}{\kappa_L^2}K_L K_L^\dagger = 1$.

This formalism, which can be easily generalised to $n$ extra
generations, offers the possibility of successfully
separating the information relative to the heavy leptons
in a simple and compact form. Although the couplings (in particular the
$\alpha_{ij}$ entries) can be in general complex, in what follows we
consider a minimal scenario where all couplings are taken to be real.

\section{Explaining LFUV data with nonunitary couplings:
  phenomenological viability}\label{sec:pheno:constraints}

We recall that, as mentioned in
the Introduction, we work under the minimal assumptions that
the singlet vector leptoquark $V_1$ (colour triplet) should correspond
to a gauge extension of $SU(3)_c$ unifying quarks and leptons
with a universal (i.e., flavour independent) gauge coupling. We first
describe the effects of the vector leptoquark on the
neutral and charged current $b$ decays, and then summarise the most
stringent constraints arising from numerous flavour violating and
flavour conserving observables (meson oscillations and decays, as well
as charged lepton flavour violation processes). We then present our main numerical results.

\subsection{New contributions to $\pmb{R_{K^{(\ast)}}}$ and $\pmb{R_{D^{(\ast)}}}$}
In what follows, we proceed to explore whether the relaxation of the
unitarity requirement on the ${SU}(2)_L$-singlet vector leptoquark $V_1$ couplings
to SM matter does allow addressing $R_{K^{(\ast)}}$ and
$R_{D^{(\ast)}}$ data simultaneously.

\paragraph{Anomalies in neutral current $b$ decays: $\pmb{R_{K^{(\ast)}}}$}
As mentioned in the Introduction, several measurements of the ratio of
branching ratios of $B \to K^{(*)} \ell \ell$ ($\ell = e, \mu$)
exhibit tensions when compared to the SM predictions.
The most recent averages
(and SM estimations) are associated with the following
deviations~\cite{Aaij:2019wad,Aaij:2017vbb,Abdesselam:2019wac}, in which
the dilepton invariant mass squared bin (in $\text{GeV}^{2}$)
is identified by the subscript:
\begin{eqnarray}\label{eq:RK*:expSMsigma}
  R_{K [1.1,6]}^{\text{LHCb}}\, &=&\, 0.846\,\pm_{0.054}^{0.060}\,
  \pm_{0.014}^{0.016}\,, \quad
 R_{K}^{\text{SM}}\, =\, 1.0003\, \pm\, 0.0001
  \,,
 \nonumber\\
 R_{K^*[0.045, 1.1]}^{\text{LHCb}}\, &=&\, 0.66 ^{+0.11}_{-0.07} \,\pm\,
 0.03\,, \quad
 R_{K^*[0.045, 1.1]}^{\text{Belle}}\, =\, 0.52 ^{+0.36}_{-0.26} \,\pm\,
 0.05\,, \quad
R_{K^*[0.045, 1.1]}^{\text{SM}}\,  \sim\,  0.93 \,,
 \nonumber\\
 R_{K^* [1.1, 6]}^{\text{LHCb}} \,&=& \,0.69 ^{+0.11}_{-0.07}\, \pm
0.05\,, \quad
 R_{K^* [1.1, 6]}^{\text{Belle}} \,= \,0.96 ^{+0.45}_{-0.29}\, \pm
0.11\,, \quad
 R_{K^* [1.1, 6]}^{\text{SM}} \,\sim \, 0.99\, .
\end{eqnarray}
Other anomalies in the neutral current mode of $B$ meson decays have
also emerged concerning the observable $\Phi \equiv d {\rm
  BR}(B_s\to\phi\mu\mu)/ dm_{\mu\mu}^2$ in the analogous bin
($m_{\mu\mu}^2\in[1,6]\,{\rm GeV}^2$)~\cite{Aaij:2015esa}, with a
similar deviation (at a level of approximately $3\sigma$), as well as in
the angular observable $P_5^{\prime}$ in $B \to K^\ast \ell^+ \ell^-$
processes. While LHCb's results for $P_5^{\prime}$ in $B \to K^\ast
\mu^+ \mu^-$ decays manifest a slight discrepancy with respect to the
SM, the Belle Collaboration~\cite{Wehle:2016yoi} reported that, when
compared to the muon case, $P_5^{\prime}$ results for electrons show a
better agreement with theoretical SM expectations. Nevertheless,
there has been an ongoing discussion about the possibility that
incorrectly estimated hadronic uncertainties might be at the origin of the
observed anomalies in the mode $B\to K^{*}\mu^+\mu^-$, due to power
corrections to the form factors, or charm-loop
contributions~\cite{Jager:2012uw,Jager:2014rwa,Ciuchini:2015qxb,Ciuchini:2016weo}.

\begin{figure}[t!]
\centering
\includegraphics[width = 0.6\textwidth]{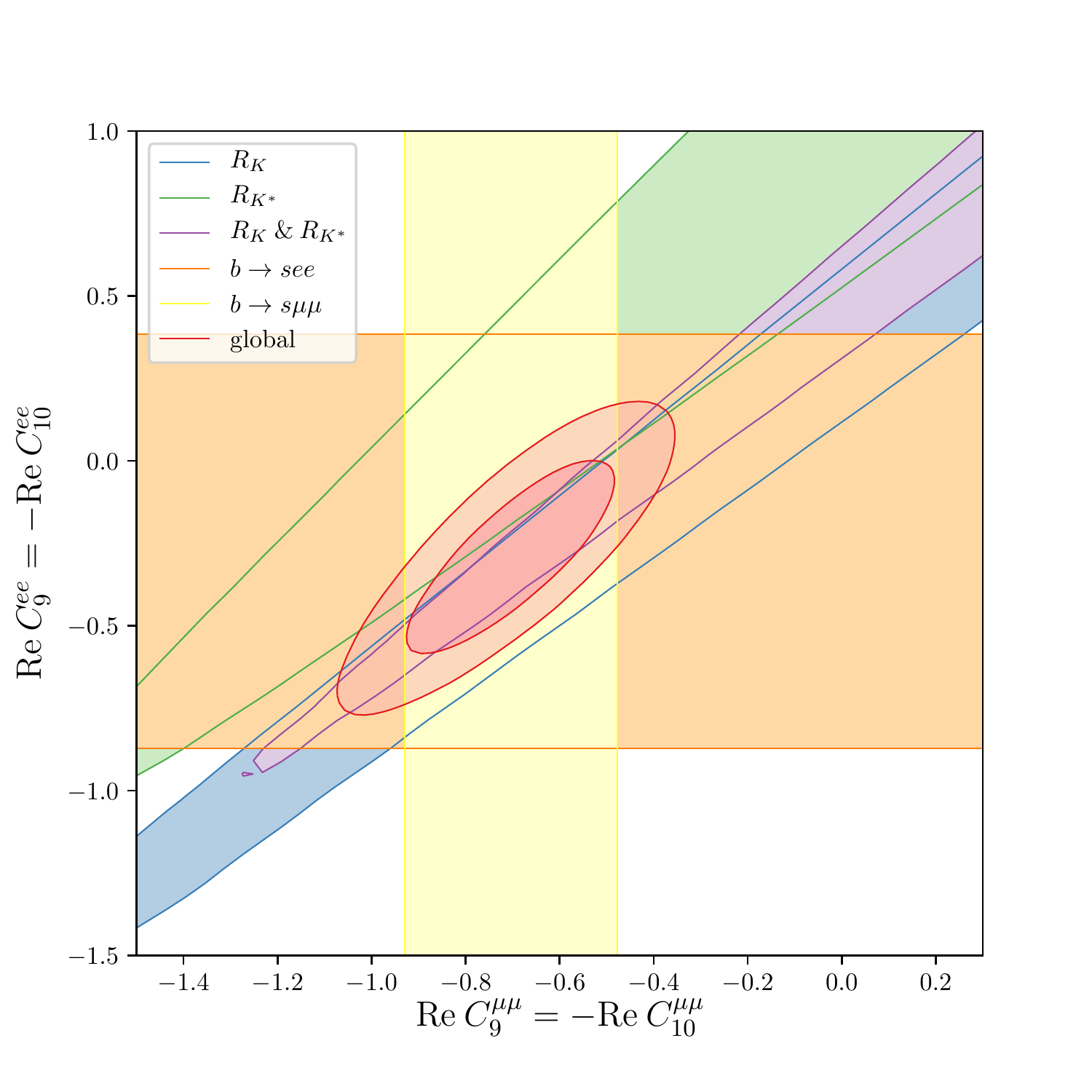}
\caption{Likelihood $1\sigma$ contours from  the latest
  measurements of $R_K^{(*)}$,
  $b\rightarrow s \mu \mu$, and $b\rightarrow s e e$ observables, as
  well as
  the combined global fit ($1\sigma$ and $2\sigma$) in the  $C_{9,sb}^{\mu\mu}$ vs.
  $C_{9,sb}^{ee}$ plane, assuming the 2D hypothesis
  $C_{9,sb}^{\mu\mu}=-C_{10,sb}^{\mu\mu}$ and
  $C_{9,sb}^{ee}=-C_{10,sb}^{ee}$. The global fit is obtained at the $b$-quark mass scale using the package \texttt{flavio}  \cite{Straub:2018kue}.}
\label{fig:RK_global}
\end{figure}

The effective Hamiltonian describing the neutral current effects at
the level of quark transitions $d_j \to d_i \ell^-\ell'^+$ is given in
Appendix~\ref{app:meson}. A model-independent analysis of the current data at the $b$-quark mass scale can be made using the
2D hypothesis $C_{9, \text{NP}}^{\mu \mu}\,=-\,C_{10, \text{NP}}^{\mu
  \mu}$ and $C_{9, \text{NP}}^{ee}\,=-\,C_{10, \text{NP}}^{ee}$,
allowing for the possibility of $(V-A)$ new physics (NP) effects in both electron
and muon channels. Using the available experimental measurements of
$R_{K^{(*)}}$ in different high and low $q^2$ bins, including the
latest updates from LHCb and Belle collaborations, the available
experimental measurements on the angular
observables for $b\rightarrow s \mu \mu$ and $b\rightarrow s ee$, and
the latest average for BR$(B_s\rightarrow \mu\mu)$ -- which already includes the
latest measurement from the ATLAS collaboration -- we find the global fit
ranges
\begin{eqnarray}
-0.50 \,(-0.58) &\geq  C_{9,sb}^{\mu\mu}=-C_{10,sb}^{\mu\mu} &\geq (-0.83)\, -0.91\,,\nonumber\\
0.00 \,(-0.12) &\geq  C_{9,sb}^{ee}=-C_{10,sb}^{ee} &\geq (-0.45)\, -0.55\,,
\end{eqnarray}
at the $2\,\sigma$ ($1\,\sigma$) level. In Fig.~\ref{fig:RK_global} we
show the $1\sigma$ likelihood contours from the latest experimental
measurements of $R_K^{(*)}$, $b\rightarrow s \mu \mu$ and
$b\rightarrow s e e$ observables, as well as their combined global fit
($1\sigma$ and $2\sigma$) in the $C_{9,sb}^{\mu\mu}$ vs.
$C_{9,sb}^{ee}$ plane, assuming the 2D hypothesis
$C_{9,sb}^{\mu\mu}=-C_{10,sb}^{\mu\mu}$ and
$C_{9,sb}^{ee}=-C_{10,sb}^{ee}$.

The vector leptoquark $V_1$
contributes at tree level, yielding the following
Wilson coefficients at the leptoquark mass scale for the $\mathcal{O}_{9,10}^{ij;\ell\ell'}$
operators~\cite{Dorsner:2016wpm}
\begin{equation}\label{eq:effHam:C9C10}
  C_{9}^{ij;\ell\ell'}\, =\, -C_{10}^{ij;\ell\ell'}\,= \,
  -\frac{\pi}{{\sqrt{2}{G_F} \,\alpha_\text{em}\, {V_{3j}}\,V_{3i}^*}}\,
  \frac{1}{{{m^2_V}}}K^{i\ell'}_{L}\,K^{j\ell\ast}_{L}\,,
\end{equation}
in which $G_F$ denotes the Fermi constant, $\alpha_\text{em}$ is the fine-structure constant, $V$ the CKM matrix, $K_L$
the ``effective'' leptoquark couplings (cf. Eq.~\eqref{eq:KL:K1K2})
and $m_V$ the vector leptoquark mass. The matching of the model parameters with the Wilson coefficients is performed at the scale of leptoquark mass and the Wilson coefficients are subsequently run down to the $b$-quark mass scale~\cite{Feruglio:2017rjo}. In particular, it is interesting to note that due to RG-running a large $K^{i\tau}_{L}$ coupling can potentially induce a non-negligible lepton-universal contribution to $b \to s \ell \ell$ transitions via a $\log$-enhanced anapole photon penguin contribution, as noted in Ref.~\cite{Crivellin:2018yvo}. The dominant $\log$-enhanced contribution is given by
\begin{align}\label{eq:anapole}
\begin{aligned}
\Delta C_{9}^{ij;\text{univ.}} &\approx
-\sum_{\ell=e,\mu,\tau}\frac{\sqrt{2}}{G_F V_{tb}V_{ts}^* m^2_V}\,\frac{1}{6}\,
K^{i\ell}_{L}\,K^{j\ell\ast}_{L}\,\log(m^2_b/m_V^2)~.
\end{aligned}
\end{align}
In our analysis, the running from the scale of leptoquark mass to the $b$-quark mass scale, and to any other relevant process (observable) scale, is taken into account using the \texttt{wilson} package~\cite{Aebischer:2018bkb} in association with the \texttt{flavio} package~\cite{Straub:2018kue}.

\paragraph{Anomalies in charged current $b \to c \ell \bar \nu$
  transitions: $\pmb{R_{D^{(\ast)}}}$}

Important deviations from the SM prediction of lepton flavour
universality in $B \to D^{(*)} \ell \nu$ decays have also been
reported by several experimental collaborations. The most recent
averages for the $R_{D}$ and $R_{D^{(*)}}$ ratios by the HFLAV
Collaboration~\cite{Amhis:2016xyh} are

\begin{eqnarray}\label{eq:RD*:expSMsigma}
 R_{D}\, =\, 0.340\,\pm\, 0.027\,\pm\,0.013\,,\quad
& R_{D}^{\text{SM}}\, =\, 0.299\, \pm\, 0.003 \, \quad
& (1.4 \sigma) \,;
 \nonumber\\
R_{D^*}\, =\, 0.295\,\pm\, 0.011\,\pm\,0.008\,,\quad
& R_{D^*}^{\text{SM}}\, =\, 0.258\, \pm\, 0.005 \, \quad
& (2.5 \sigma) \,.
 \end{eqnarray}

\begin{figure}[t!]
\centering
\includegraphics[width = 0.5\textwidth]{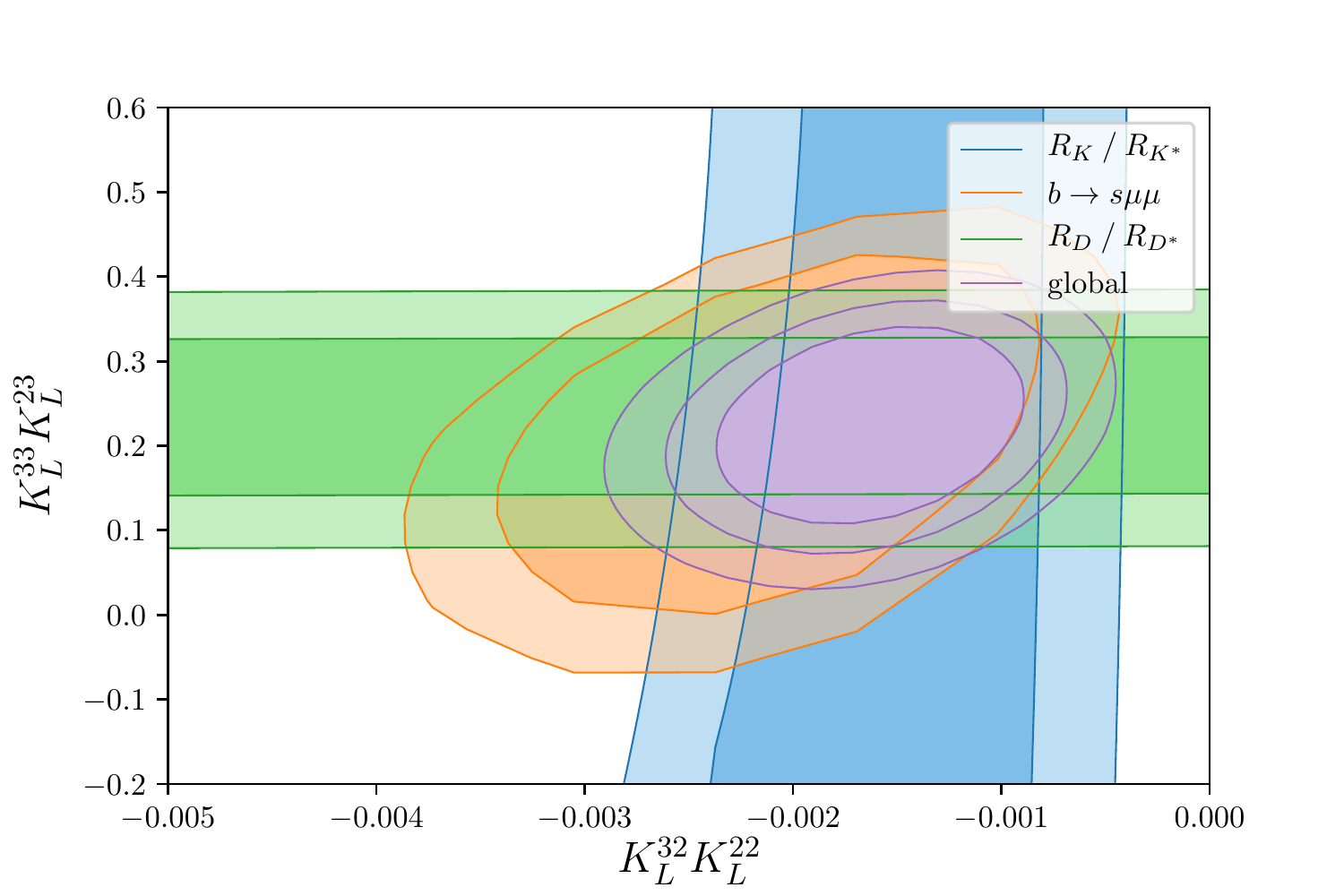}
\caption{Likelihood $1\sigma$ and $2\sigma$ contours from $R_{K^{(*)}}$, $R_{D^{(*)}}$,
  $b\rightarrow s \mu \mu$ observables and the combined global fit
  ($1\sigma$, $2\sigma$ and $3\sigma$) in the plane
  $K_L^{33}K_L^{23}-K_L^{32}K_L^{22}$, defined at the leptoquark mass scale cf.~Eq.~(\ref{eq:lagrangian:Vql_phys3}).  The first generation lepton
  and quark couplings are set to zero.}
\label{fig:RK_RD_global}
\end{figure}

\noindent
We define the effective Hamiltonian for the charged current
transitions $d_k\to u_j\bar{\nu}\ell^{-}$ as
\begin{equation}\label{eq:Heff:RD}
\mathcal{H}_\text{eff}^{\ell_f\nu_i} \,=\,
  \frac{4\,G_F}{\sqrt 2}\, V_{jk}\, C_{jk}^{fi}\, \left(
  \bar u_{j} \, \gamma ^\mu  \,P_L \,d_k \right) \,
  \left( {\bar \ell_f}  \,\gamma _\mu  \,P_L  \,\nu _i \right)\,,
\end{equation}
where, in the SM, $C_{jk,{\rm SM}}^{fi}=\delta_{fi}$.
The contribution from the vector leptoquark $V_1$ is given by
\begin{equation}\label{eq:RD:Cij}
  C_{jk,V_1}^{fi}\,=\,\frac{\sqrt{2}}{4\,G_{F}\,m_V^2}\,
  \frac{1}{V_{jk}}\,
  (V\,K_{L}\, U^P)_{ji}\, K_{L}^{kf\ast}\,.
\end{equation}
One can further construct the double ratios
\begin{equation}\label{eq:RDoverRDSM-th}
R_{D}/R_{D}^\text{SM} \,= \,
R_{D^\ast}/R_{D^\ast}^\text{SM}\,=\,
\sum_{i = 1}^3
\left| \delta _{3i} \,+\, C_{cb,V_1}^{\tau i} \right|^2\,,
\end{equation}
(equal to unity when NP decouples, i.e., $\kappa_L \to 0$ and $m_V \to
\infty$).
After combining current experimental world averages with the SM
predictions, the current anomalous data can be summarised
as $R_{D}/R_D^\text{SM} \,= \,1.14 \pm 0.10\, , \;
R_{D^{\ast}}/R_{D^{\ast}}^\text{SM} \,= \,1.14 \pm 0.06\, ,$
in which the statistical and systematical errors have been added in
quadrature.
In Fig.~\ref{fig:RK_RD_global}, we display
the $1\sigma$ and $2\sigma$ likelihood contours from $R_{K^{(*)}}$, $R_{D^{(*)}}$,
$b\rightarrow s \mu \mu$ observables and the combined global fit
($1\sigma$, $2\sigma$ and $3\sigma$) in the plane
$K_L^{33}K_L^{23}-K_L^{32}K_L^{22}$, where the couplings (see Eq.~(\ref{eq:lagrangian:Vql_phys3})) are varied independently and the
others are set to zero.

As can be seen in Fig. \ref{fig:RK_RD_global}, both
$b\to c\ell\nu$ and $b\to s\ell\ell$ anomalies can be indeed accommodated
simultaneously in a minimal $V_1$ model.  However, we will subsequently show
how this is realised in the nonunitary framework taking all couplings
into account, as suggested by the $b\to s \ell\ell$ data.

\subsection{Constraints from (rare) flavour processes, EW
  precision observables and direct searches}\label{sec:constraints}

The extended framework called upon to address the $B$ meson decay
anomalies - not only the additional vector leptoquark, but also the
presence of extra vector-like fermions, which are the origin of the
nonunitarity of the $V_1$ effective couplings - opens the door to
extensive contributions to numerous observables.

While most of the NP contributions occur via higher order (loop)
exchanges, it is important to notice that $V_1$ can also mediate very
rare (or even SM forbidden) processes already at the tree level. As we
proceed to discuss, the
latter observables prove to be particularly constraining, and put
stringent bounds on the degrees of freedom of these leptoquark
realisations.

\bigskip
Leptoquark SM extensions aiming at addressing the anomalies in $R_{K^{(\ast)}}$
and $R_{D^{(\ast)}}$ data receive strong constraints from
$d_j \to d_i \bar \nu \nu$ transitions (in particular
$s\to d \nu\nu$ and $b\to s \nu\nu$). However, the vector leptoquark
$V_1$ does not generate contributions at tree level,
and the first non-vanishing contribution appears at one loop. Consequently, we find that even with significant uncertainties,
the semileptonic decays into charged dileptons
$d_j \to d_i \ell^-\ell^{\prime +}$ often lead to tighter
constraints (both the lepton flavour conserving and the lepton flavour violating modes).
In the present analysis, we therefore include bounds from
$K \to \pi \ell \ell^\prime$ and $B \to K \ell \ell^\prime$, as well
as the stringent limits on NP contributions arising from the observed
decay mode $B_s \to \mu^+ \mu^-$.
Furthermore, we also take into account the lepton flavour violating
leptonic decays
$B \to e^{\pm} \mu^{\mp}$, $B_s \to e^{\pm} \mu^{\mp}$ and
$K_L \to e^{\pm} \mu^{\mp}$.
Leptoquark contributions to
neutral meson oscillations and mixings, such as
$K^0-\bar K^0$ and
$B_s^0 - \bar B_s^0$, are also evaluated.
Table~\ref{tab:flavour} provides a brief summary
of the current experimental status for these mesonic
observables (current bounds and future
sensitivities). A detailed discussion of the formalism used to
evaluate the vector leptoquark contributions
is provided in Appendix~\ref{app:meson}.

{\small
\renewcommand{\arraystretch}{1.}
\begin{table}[h!]
\begin{center}
\begin{tabular}{|c|c|c|}
\hline
Observables & SM prediction & Experimental data  \\
\hline
$\text{BR}(K^+ \to \pi^+ \nu\bar \nu)$ &
$(8.4 \pm 1.0) \times 10^{-11}
\phantom{|}^{\phantom{|}}_{\phantom{|}}$\cite{Buras:2015qea} &
\begin{tabular}{l}
$17.3^{+11.5}_{-10.5} \times 10^{-11}
\phantom{|}^{\phantom{|}}_{\phantom{|}}$\cite{Artamonov:2008qb}
\\
$< 11 \times 10^{-10}
\phantom{|}^{\phantom{|}}_{\phantom{|}}$\cite{Na62:2018}
\end{tabular}
\\
\hline
$\text{BR}(K_L \to \pi^0 \nu\bar \nu)$ &
$(3.4 \pm 0.6) \times 10^{-11}
\phantom{|}^{\phantom{|}}_{\phantom{|}}$\cite{Buras:2015qea} &
$ \leq  2.6 \times 10^{-8}
\phantom{|}^{\phantom{|}}_{\phantom{|}}$\cite{Ahn:2009gb}\\
\hline
$R_{K^{(\ast)}}^{\nu\nu}$  ($B\to  K^{(\ast)} \nu\bar \nu$)
&
$R_{K^{(\ast)}}^{\nu\nu}=1$
&
\begin{tabular}{l}
$R_K^{\nu\nu} < 3.9$
$\phantom{a}^{\phantom{|}}_{\phantom{|}}$\cite{Grygier:2017tzo}
\\
$R_{K^\ast}^{\nu\nu} < 2.7
\phantom{|}^{\phantom{|}}_{\phantom{|}}$\cite{Grygier:2017tzo}
\end{tabular}
\\
\hline
$B_s^0 - \bar B_s^0$ {(mixing parameters)}
&
\begin{tabular}{l}
$\Delta_{s} = |\Delta_{s}|e^{i\phi_s}=1$ $\phantom{a}^{\phantom{|}}_{\phantom{|}}$
\\
$\phi_{s}=0$
\end{tabular}
&
\begin{tabular}{l}
$|\Delta_s|=1.01^{+0.17}_{-0.10}
\phantom{|}^{\phantom{|}}_{\phantom{|}}$\cite{Charles:2015gya},\\
$\phi_s [^{\circ}]=1.3^{+2.3}_{-2.3}
\phantom{|}^{\phantom{|}}_{\phantom{|}}$\cite{Charles:2015gya}
\end{tabular}
\\
\hline
$K^0 - \bar K^0$:
$\Delta m_K/(10^{-15}\text{GeV})$
&
\begin{tabular}{c}
$3.1\pm1.2
\phantom{|}^{\phantom{|}}_{\phantom{|}}$\cite{Brod:2011ty}
\end{tabular}
&
\begin{tabular}{c}
$3.484\pm0.006
\phantom{|}^{\phantom{|}}_{\phantom{|}}$\cite{Tanabashi:2018oca}
\end{tabular}
\\
\hline
$\text{BR}(B_s \to \mu\mu)$ &
$(3.23 \pm 0.27) \times 10^{-9}
\phantom{|}^{\phantom{|}}_{\phantom{|}}$\cite{Buras:2012ru} &
$  2.7_{-0.5}^{+0.6} \times 10^{-9}
\phantom{|}^{\phantom{|}}_{\phantom{|}}$\cite{Tanabashi:2018oca}\\
\hline
$\text{BR}(K_L \to \mu^\pm e^\mp)$ &
--- &
$< 4.7\times 10^{-12}
\phantom{|}^{\phantom{|}}_{\phantom{|}}$\cite{Tanabashi:2018oca}\\
\hline
$\text{BR}(B_s \to \mu^\pm e^\mp)$ &
--- &
$< 1.1 \times 10^{-8}
\phantom{|}^{\phantom{|}}_{\phantom{|}}$\cite{Tanabashi:2018oca}\\
\hline
$\text{BR}(B^{0} \to \mu^\pm e^\mp)$ &
--- &
$< 2.8\times 10^{-9}
\phantom{|}^{\phantom{|}}_{\phantom{|}}$\cite{Tanabashi:2018oca}\\
\hline
$\text{BR}(B^{0} \to \tau^\pm e^\mp)$ &
--- &
$< 2.8 \times 10^{-5}
\phantom{|}^{\phantom{|}}_{\phantom{|}}$\cite{Tanabashi:2018oca}\\
\hline
$\text{BR}(B^{0} \to \tau^\pm \mu^\mp)$ &
--- &
$< 2.2\times 10^{-5}
\phantom{|}^{\phantom{|}}_{\phantom{|}}$\cite{Tanabashi:2018oca}\\
\hline
$\text{BR}(B^{0} \to K^0 \mu^\pm e^\mp)$ &
--- &
$< 2.7\times 10^{-8}
\phantom{|}^{\phantom{|}}_{\phantom{|}}$\cite{Tanabashi:2018oca}\\
\hline
$\text{BR}(B^{0} \to \pi^0 \mu^\pm e^\mp)$ &
--- &
$< 1.4\times 10^{-7}
\phantom{|}^{\phantom{|}}_{\phantom{|}}$\cite{Tanabashi:2018oca}\\
\hline
$\text{BR}(B^\pm \to K^+ \mu^\pm e^\mp)$ &
--- &
$< 9.1 \times 10^{-8}
\phantom{|}^{\phantom{|}}_{\phantom{|}}$\cite{Tanabashi:2018oca}\\
\hline
$\text{BR}(B^\pm \to K^+ \tau^\pm e^\mp)$ &
--- &
$< 3.0 \times 10^{-5}
\phantom{|}^{\phantom{|}}_{\phantom{|}}$\cite{Tanabashi:2018oca}\\
\hline
$\text{BR}(B^\pm \to K^+ \tau^\pm \mu^\mp)$ &
--- &
$< 4.8 \times 10^{-5}
\phantom{|}^{\phantom{|}}_{\phantom{|}}$\cite{Tanabashi:2018oca}\\
\hline
$\text{BR}(B^\pm \to \pi^+ \mu^\pm e^\mp)$ &
--- &
$< 1.7 \times 10^{-7}
\phantom{|}^{\phantom{|}}_{\phantom{|}}$\cite{Tanabashi:2018oca}\\
\hline
$\text{BR}(B^\pm \to \pi^+ \tau^\pm e^\mp)$ &
--- &
$< 7.5 \times 10^{-5}
\phantom{|}^{\phantom{|}}_{\phantom{|}}$\cite{Tanabashi:2018oca}\\
\hline
$\text{BR}(B^\pm \to \pi^+ \tau^\pm \mu^\mp)$ &
--- &
$< 7.2 \times 10^{-5}
\phantom{|}^{\phantom{|}}_{\phantom{|}}$\cite{Tanabashi:2018oca}\\
\hline
\end{tabular}
\caption{Relevant observables and current
  experimental status for leptonic and semi-leptonic meson decays;
  when appropriate, the associated SM prediction is also included.
}\label{tab:flavour}
\end{center}
\end{table}
\renewcommand{\arraystretch}{1.}
}

\bigskip
The lepton flavour non-universal couplings of vector
leptoquarks (in general nonunitary in the present framework) induce new contributions to cLFV
observables: radiative decays $\ell_i \rightarrow \ell_j \gamma$ and 3-body decays
$\ell_i \rightarrow 3 \ell_j$ at loop level, and neutrinoless
$\mu - e$ conversion in nuclei both at tree and loop level. Further
taking into account the impressive associated experimental
sensitivity, it is clear that these observables lead to important
constraints on the vector leptoquark couplings to SM fermions. It is
important to stress that although the radiative decays are generated
at higher order, relevant anapole contributions can add to the
Wilson coefficients accounting for the tree-level contributions to
neutrinoless $\mu - e$ conversion and $\mu\rightarrow 3e$. The higher
order anapole contributions can have a magnitude comparable to the
tree level ones (or even account for the dominant contribution).  In
addition, dipole operators also contribute significantly to radiative
decays and to neutrinoless $\mu -e$ conversion. Although we do take tauonic modes into
account, we notice here that due to the associated current
experimental sensitivity, (semi)leptonic tau decays in general lead to
comparatively looser constraints; likewise, semileptonic meson decays
into final states including tau leptons are typically less
constraining. However, the expected improvements in sensitivity from
dedicated experiments might render the tau modes important probes of
SM extensions via vector leptoquarks\footnote{For a detailed
  discussion regarding semileptonic meson decays into final states
  with tau leptons see, for example,~\cite{Capdevila:2017iqn}.}. As done for the first time in this work, all these contributions must be systematically included to thoroughly constrain the vector leptoquark couplings.

A summary of the current experimental status (current bounds and
future sensitivities) is given in Table~\ref{tab:cLFV}; for simplicity, in the numerical analysis we consider a benchmark future sensitivity to $\mu - e$ conversion in Aluminium of $\mathcal O (10^{-17})$.
The relevant details of the computation of the cLFV
observables considered in this work can be found in
Appendix~\ref{app:cLFV}.

\begin{table}[h!]
\begin{center}
\begin{tabular}{|c|l|l|}
\hline
cLFV process & Current experimental bound & Future sensitivity   \\
\hline
$\text{BR}(\mu\to e \gamma)$	&
 \quad $ 4.2\times 10^{-13}$ \quad (MEG~\cite{TheMEG:2016wtm})	&
 \quad $6\times 10^{-14}$ \quad (MEG II~\cite{Baldini:2018nnn}) \\
$\text{BR}(\tau \to e \gamma)$	&
 \quad $3.3\times 10^{-8}$ \quad (BaBar~\cite{Aubert:2009ag})	 &
 \quad $10^{-9}$ \quad (Belle II~\cite{Kou:2018nap}) 	 	\\
$\text{BR}(\tau \to \mu \gamma)$	&
 \quad $ 4.4\times 10^{-8}$ \quad (BaBar~\cite{Aubert:2009ag})	 &
 \quad $10^{-9}$ \quad (Belle II~\cite{Kou:2018nap})		\\
\hline
$\text{BR}(\mu \to 3 e)$	&
 \quad $1.0\times 10^{-12}$ \quad (SINDRUM~\cite{Bellgardt:1987du}) 	&
 \quad $10^{-15(-16)}$ \quad (Mu3e~\cite{Blondel:2013ia})  	\\
$\text{BR}(\tau \to 3 e)$	&
 \quad $2.7\times 10^{-8}$ \quad (Belle~\cite{Hayasaka:2010np})&
 \quad $10^{-9}$ \quad (Belle II~\cite{Kou:2018nap})  	\\
$\text{BR}(\tau \to 3 \mu )$	&
 \quad $3.3\times 10^{-8}$ \quad (Belle~\cite{Hayasaka:2010np})	 &
 \quad $10^{-9}$ \quad (Belle II~\cite{Kou:2018nap})		\\
\hline
$\text{CR}(\mu- e, \text{N})$ &
 \quad $7 \times 10^{-13}$ \quad  (Au, SINDRUM~\cite{Bertl:2006up}) &
 \quad $10^{-14}$  \quad (SiC, DeeMe~\cite{Nguyen:2015vkk})    \\
& &  \quad $7 \times 10^{-15} (3\times10^{-18})$  \quad (Al, COMET~\cite{Krikler:2015msn,KunoESPP19})  \\
& &  \quad $8 \times 10^{-17}$  \quad (Al, Mu2e~\cite{Bartoszek:2014mya})\footnotemark \\
% & &  \quad $10^{-18}$  \quad (Ti, PRISM/PRIME~\cite{Kuno:2005mm})  \\
\hline
\end{tabular}
\caption{
Current experimental bounds and
future sensitivities of various cLFV processes considered in the
analysis.
}
\label{tab:cLFV}
\end{center}
\end{table}

\bigskip
\footnotetext{On the long term, and for Titanium targets, Mu2e-II~\cite{Abusalma:2018xem} is expected to improve the sensitivity by a factor of ten or more.}

The presence of heavy vector-like fermions (at the source
of the nonunitary couplings of the vector leptoquark to the light
fermions) can have a non-negligible impact on the couplings of SM
fermions to gauge bosons. In turn, this can be manifest in new
contributions to several EW precision observables - potentially in
conflict with  SM expectations and precision data -, which will prove to
play a key role in constraining the mixings of the SM charged leptons
with the heavy states.

For the $Z$-couplings, which are modified at the tree level, the most stringent constraints are expected to
arise from leptonic $Z$ decays; in our analysis, we take into account
the LFU ratios and cLFV decay modes of the $Z$ boson. We summarise in
Table~\ref{tab:EWPO} the EWP observables which are of relevance for
our study (experimental measurements and SM predictions).  A
discussion on how the $\bar f f Z$ couplings are modified as a
consequence of nonunitary effective leptoquark couplings, as well as
their impact for several observables, is detailed in
Appendix~\ref{app:ew}.

{\small
\renewcommand{\arraystretch}{1.}
\begin{table}[h!]
\begin{center}
\begin{tabular}{|c|c|c|}
\hline
Observables  & Experimental data & SM Prediction\\
\hline
$\Gamma_Z$ & $2.4952\pm0.0023\:\mathrm{GeV}$ &
$2.4942\pm0.0008\:\mathrm{GeV}$\\
$\Gamma(Z\to\ell^+\ell^-)$ & $83.984\pm0.086\:\mathrm{MeV} $ &
$83.959\pm0.008 \:\mathrm{MeV}$\\
\hline
$R_e$ & $20.804\pm0.050$ & $ 20.737\pm0.010$\\
$R_\mu$ & $20.785\pm0.033$ & $ 20.737\pm0.010$\\
$R_\tau$ & $20.764\pm0.045$ & $ 20.782\pm0.010$\\
\hline
$\mathrm{BR}(Z\to e^\pm\mu^\mp)$ & $ < 7.5 \times 10^{-7} $ & --\\
$\mathrm{BR}(Z\to e^\pm\tau^\mp)$ & $< 9.8\times 10^{-6} $ & --\\
$\mathrm{BR}(Z\to \mu^\pm\tau^\mp)$ & $ < 1.2\times10^{-5}$ & --\\
\hline
\end{tabular}
\caption{Subset of EWP observables affected by the modified
  $Z$ couplings, with the corresponding experimental measurements and
  SM predictions~\cite{Tanabashi:2018oca}. The ratios $R_\ell$ are
  defined as
  $R_\ell = {\Gamma_\text{had}}/{\Gamma(Z\to\ell^+\ell^-)}$, and
\mbox{$\Gamma(Z\to\ell^+\ell^-)$} denotes an average over $\ell =
  e,\,\mu,\,\tau$.
}\label{tab:EWPO}
\end{center}
\end{table}
\renewcommand{\arraystretch}{1.}
}

\bigskip
Finally, it is clear that the (negative)
results of direct searches for the exotic states must be taken into account.
At the LHC, pairs of vector leptoquarks can be abundantly produced
in various processes (via $t$-channel lepton exchange and direct
couplings to one or two gluons).
Due to the underlying gauge structure of possible
ultraviolet (UV) completions, the production cross section
strongly depends on the coupling to gluons which makes the
theoretical predictions for the production of vector leptoquarks
less robust than those for scalar leptoquarks. On the other hand, if the vector leptoquark corresponds to a spontaneously broken non-abelian gauge symmetry, gauge invariance completely fixes the couplings between the vector leptoquark and the gluons, implying lower limits on the vector leptoquark mass (albeit still depending on the branching fractions of the leptoquark) from the negative results in the direct searches for pair production at the LHC, see for instance \cite{Angelescu:2018tyl}.
As a natural consequence of the favoured structure called upon to
maximise the effects on $B$-meson decay anomalies, $V_1$ is expected to
dominantly decay into either $t\bar \nu_\tau$ or $b\bar \tau$. The ATLAS and CMS collaborations have
conducted extensive searches, assuming that
leptoquarks couple exclusively to third generation quarks and
leptons~\cite{Khachatryan:2014ura,Aad:2015caa,Sirunyan:2017yrk,Sirunyan:2018vhk,Sirunyan:2018kzh},
which have led to lower limits on the $V_1$ mass $\sim 1.5$~TeV.
Further important collider signatures are
$pp \to \tau \bar\tau + X$, arising from $t$-channel
leptoquark exchange or from
single leptoquark production in association with a charged lepton.
As argued in~\cite{Cerri:2018ypt,CidVidal:2018eel}, the projected vector
leptoquark reach of HL-LHC with $3$~ab$^{-1}$ is close to
$1.8$~TeV.
Much higher luminosities and/or more sophisticated search strategies
are required to probe the preferred mass scale and couplings of states
at the origin of a combined explanation
of the anomalies (for instance, as suggested by the study
of~\cite{Buttazzo:2017ixm}).
Other potentially interesting search modes could include $b\mu b\tau$ and $b \mu b \mu$ final
states.

In the present analysis, we select (working) benchmark values for the
mass and gauge coupling of the vector leptoquark
allowing to comply with the current available limits. In particular, for the following numerical analysis we set $\frac{\kappa_L}{\sqrt{2}} = 1$ as a benchmark choice. Nevertheless, for any other choice consistent with the constraints from direct searches (as discussed above) the qualitative behaviour and the conclusions drawn remain the same. However, for very small values $\kappa_L$ ($\kappa_L\lesssim 0.1$ for $m_V\gtrsim 1.5$~TeV) the number of points in the best fit region for $R_{K^{(*)}}$ and $R_{D^{(*)}}$ anomalies starts to decrease drastically, so that the NP effects become negligible with respect to the SM.

\subsection{Results and discussion}{\label{sec:results}}
We now finally address the question of whether a SM extension via vector
leptoquarks can simultaneously provide an explanation to both the
$R_{K^{(\ast)}}$ and $R_{D^{(\ast)}}$ data, working under the
hypothesis of universal gauge couplings for the vector leptoquark
$V_1$. In this framework, the required flavour non-universality arises
from nonunitary mixings among the SM fermions - a consequence of the
existence of heavy vector-like fermions which have non-negligible
mixings with the light fields.

We first begin by considering the most minimal scenario where the
nonunitary flavour misalignment is due to the presence of a
{\it single generation of heavy vector-like charged leptons}
(i.e., $n=1$ in Eq.~(\ref{eq:lagrangian:Vql_phys3N})).
Such a minimal field content already leads to a sufficient amount of
LFUV to account for both $R_{K^{(*)}}$ and $R_{D^{(*)}}$ anomalies.
Although new contributions to rare meson decays and transitions are
still in good agreement with current experimental bounds, this
scenario is ruled out due to the stringent constraints on cLFV\footnote{
Despite being loop-suppressed in the present $V_1$
  leptoquark  SM extension, $B\to K^{(*)}\nu\bar{\nu}$ decays can in
  general lead to significant constraints; nonetheless, in
  the scenarios here discussed, we find that constraints from LFV meson
  decays, and most importantly cLFV observables,
  provide tighter constraints.} modes.
Excessive contributions to (tree-level) muon-electron conversion
in nuclei play a crucial role in ruling out this realisation, as well
as the closely related radiative decays.
In order to reconcile the model's prediction with
the current bounds on CR($\mu-e$, Au), the photon-penguin contributions
must (at least) partially cancel the sizeable tree-level ones;
however, such large photon-penguin contributions then translate into
unacceptably large $\mu \to e
\gamma$ decay rates, already in conflict with current bounds.

\medskip The required flavour non-universality can be recovered for a
less dramatic unitarity violation; this can be achieved by extending
the particle content by {\it two or more additional heavy charged
  lepton states}, or formally for $n\geq 2$ in
Eq.~(\ref{eq:lagrangian:Vql_phys3N}).  Although $n=2$ provides more
freedom to evade the constraints found in the case $n=1$, no generic
solution was found in this case, which might then require an extreme fine
tuning to become viable. In the subsequent discussion, we therefore take $n=3$ which
is the minimal case and conveniently replicates the number
of generations in the SM.

\begin{figure}[h!]
   \mbox{\hspace{-8mm}\includegraphics[width=.40\linewidth]{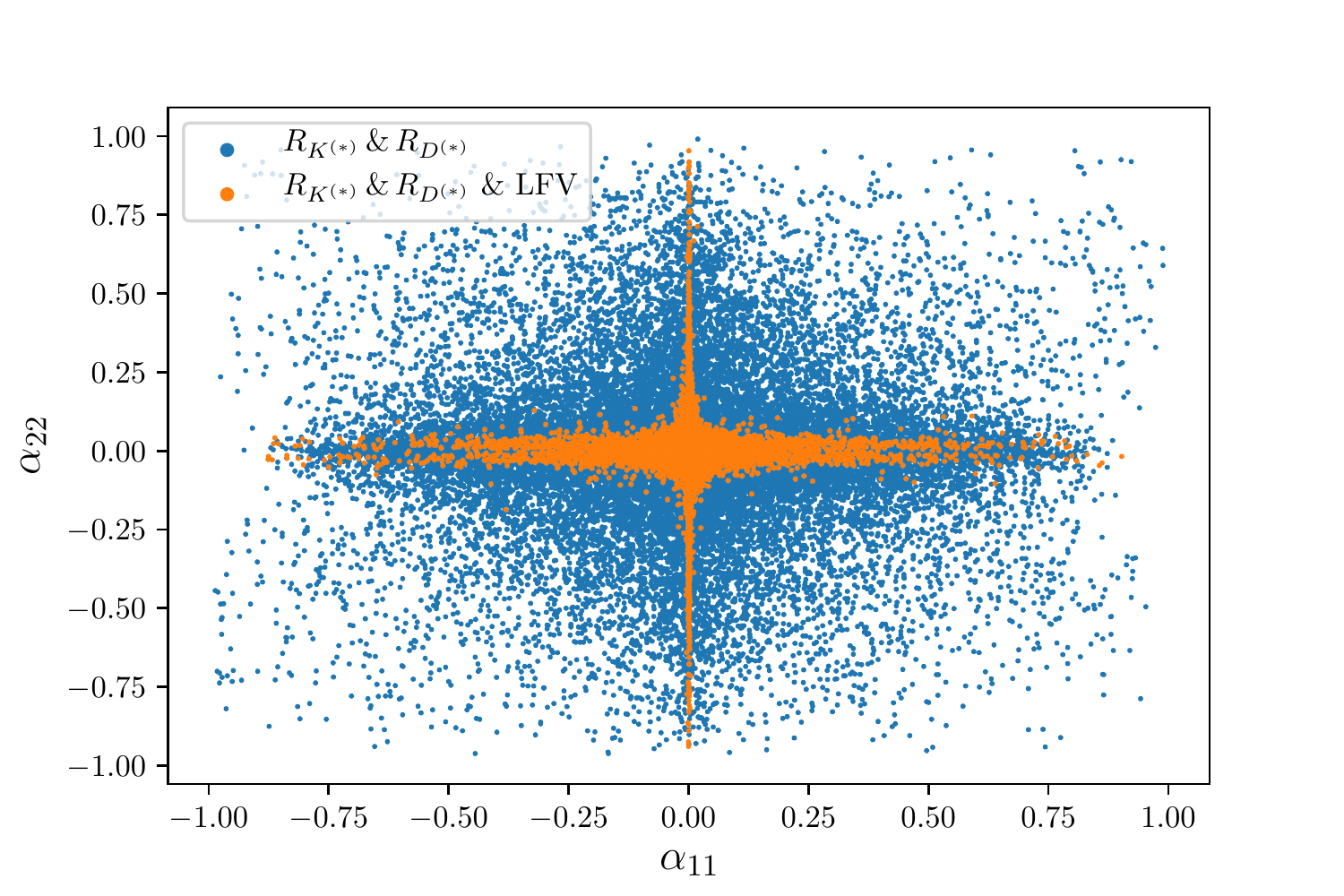}
         \hspace{-6mm}\includegraphics[width=.40\linewidth]{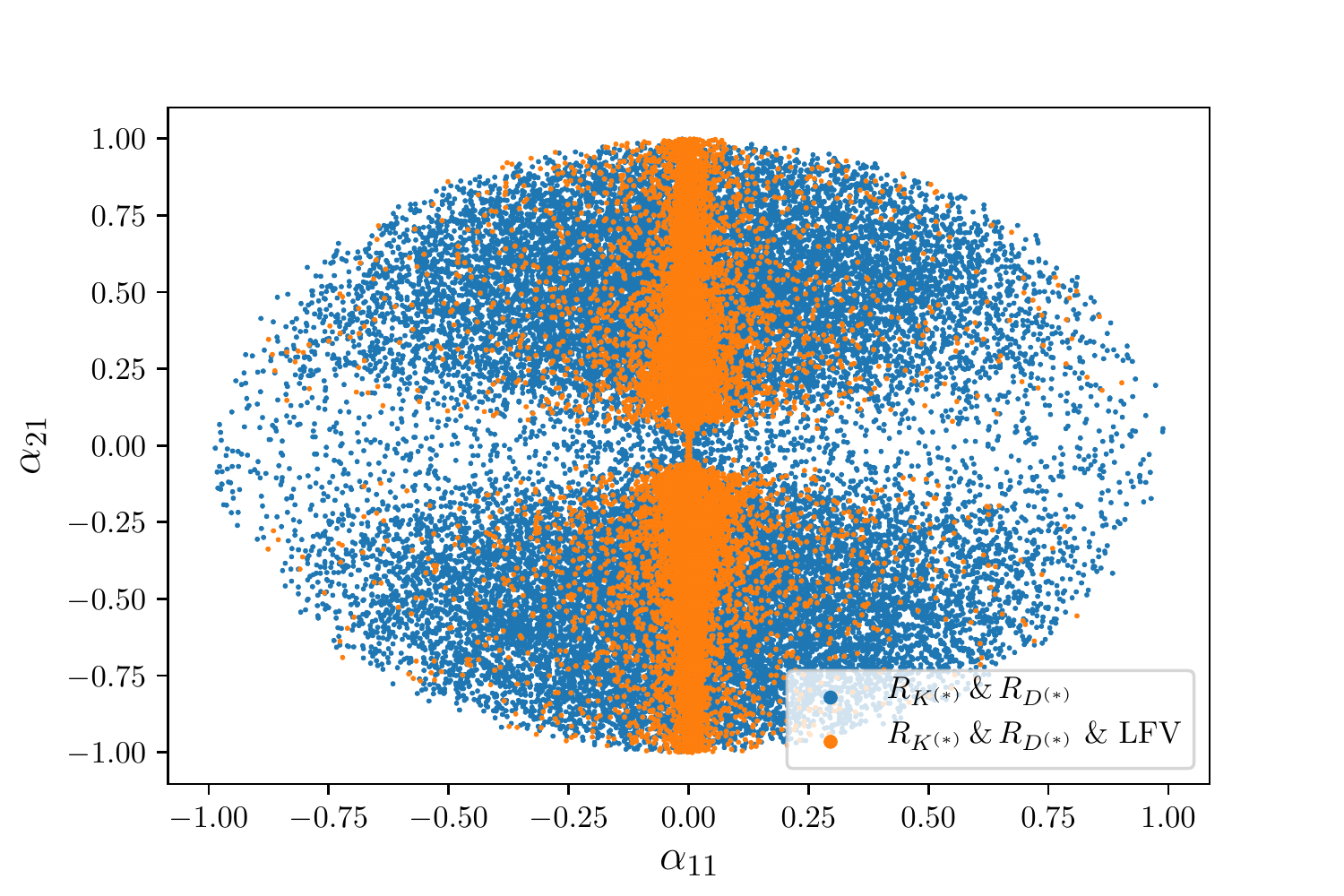}
         \hspace{-6mm}\includegraphics[width=.40\linewidth]{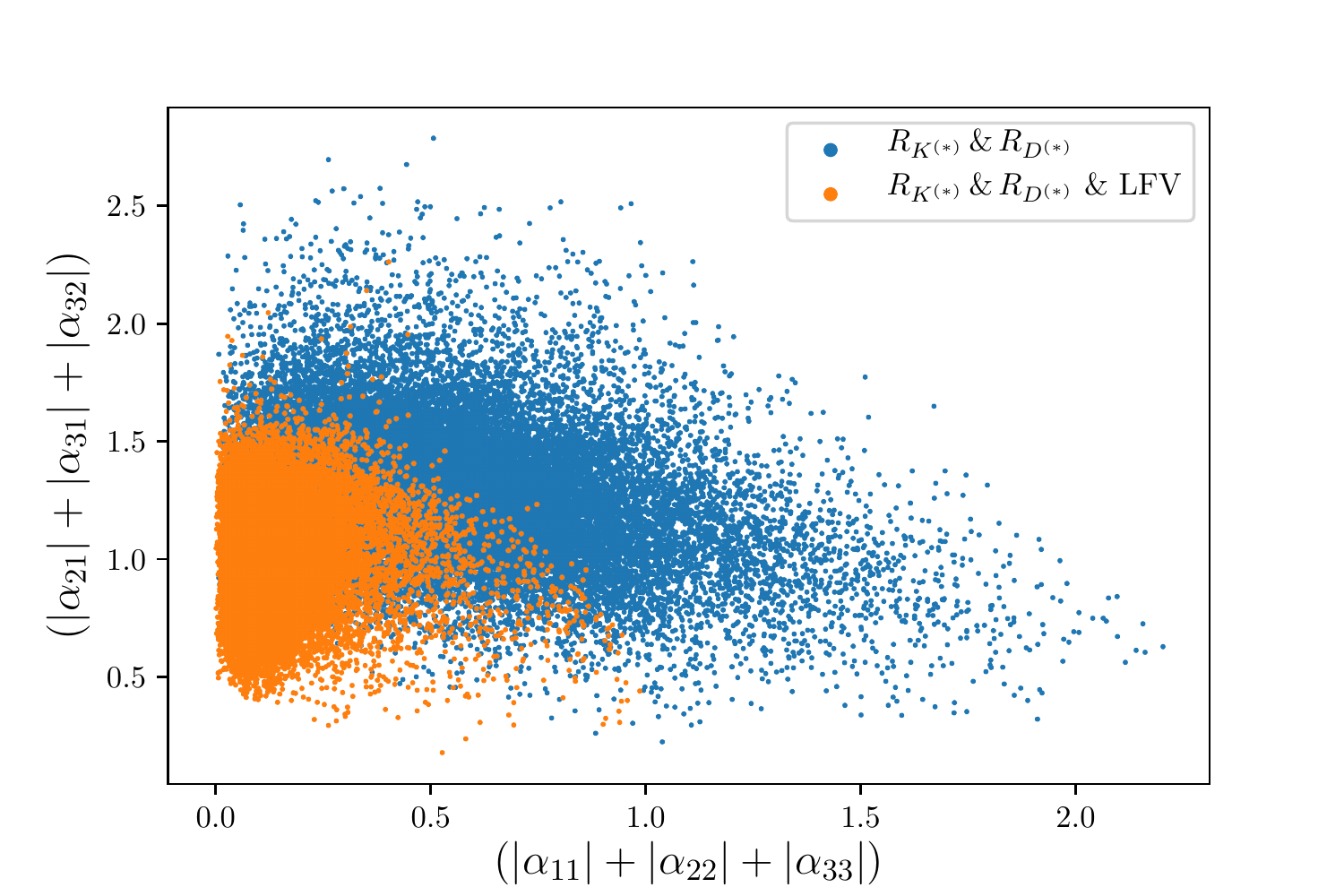}}  \\
    \includegraphics[width=.30\linewidth]{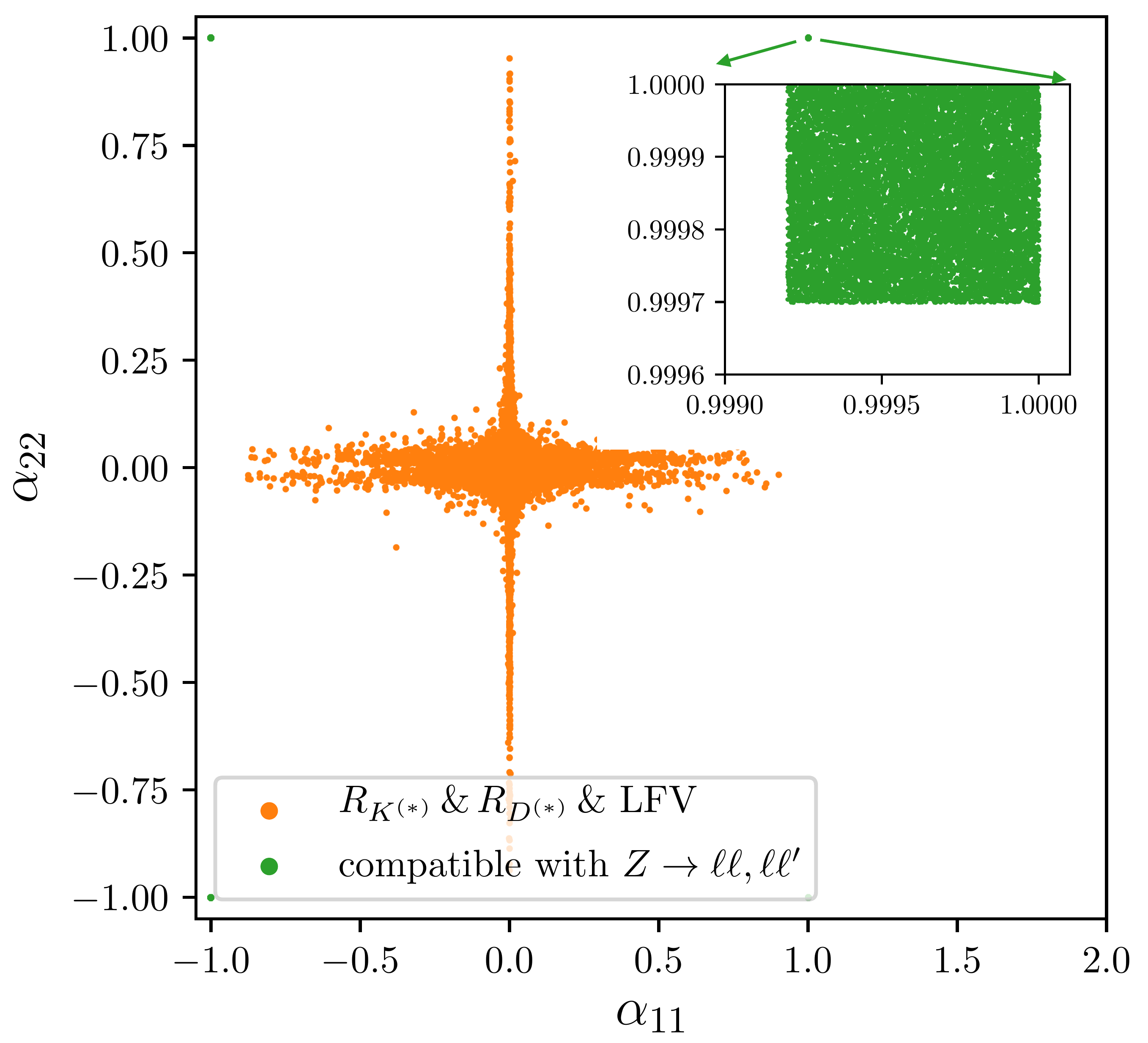}
    \hspace{.5 cm}
    \includegraphics[width=.302\linewidth ]{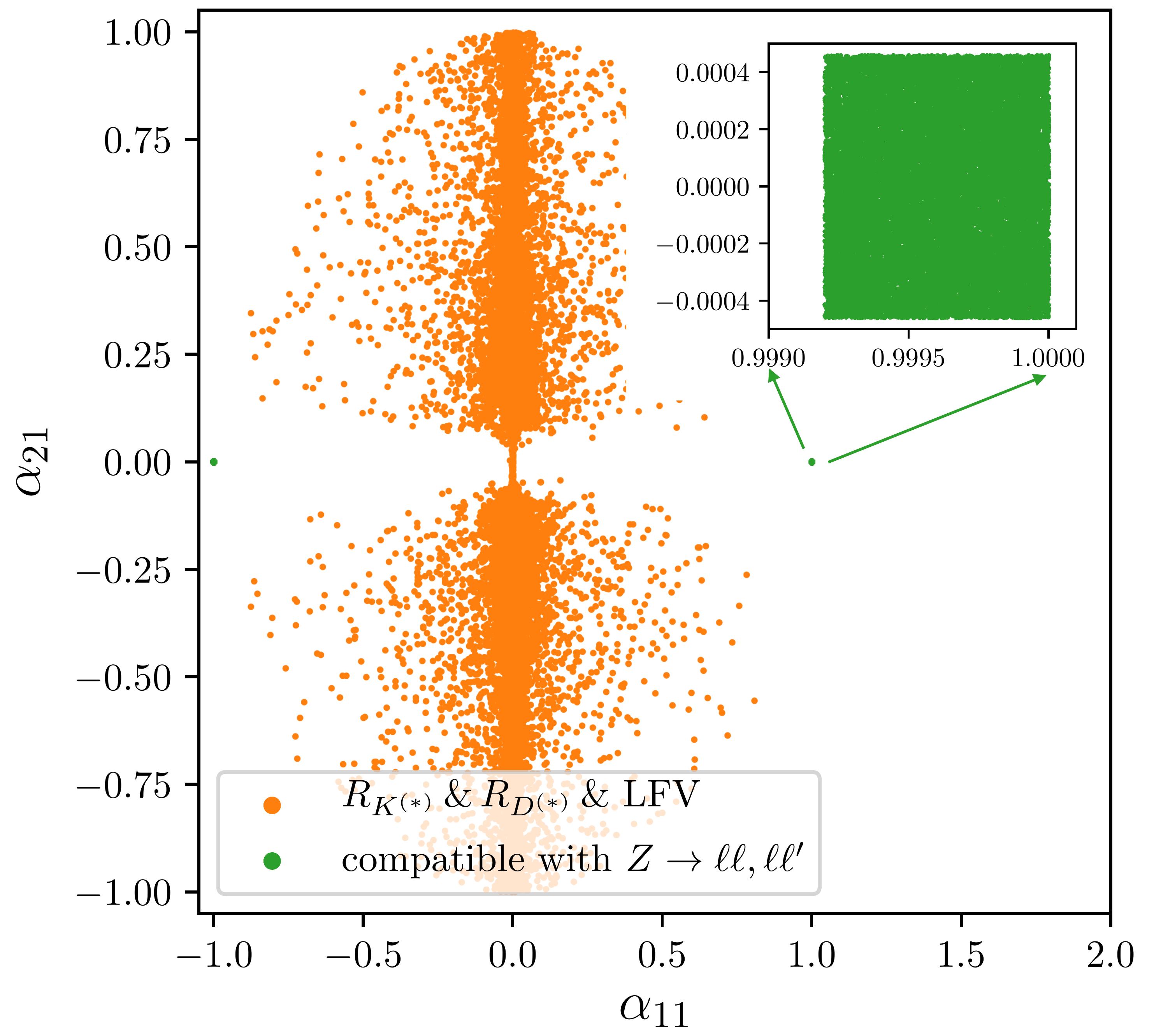}
	 \hspace{.3 cm}
    \includegraphics[width=.305\linewidth]{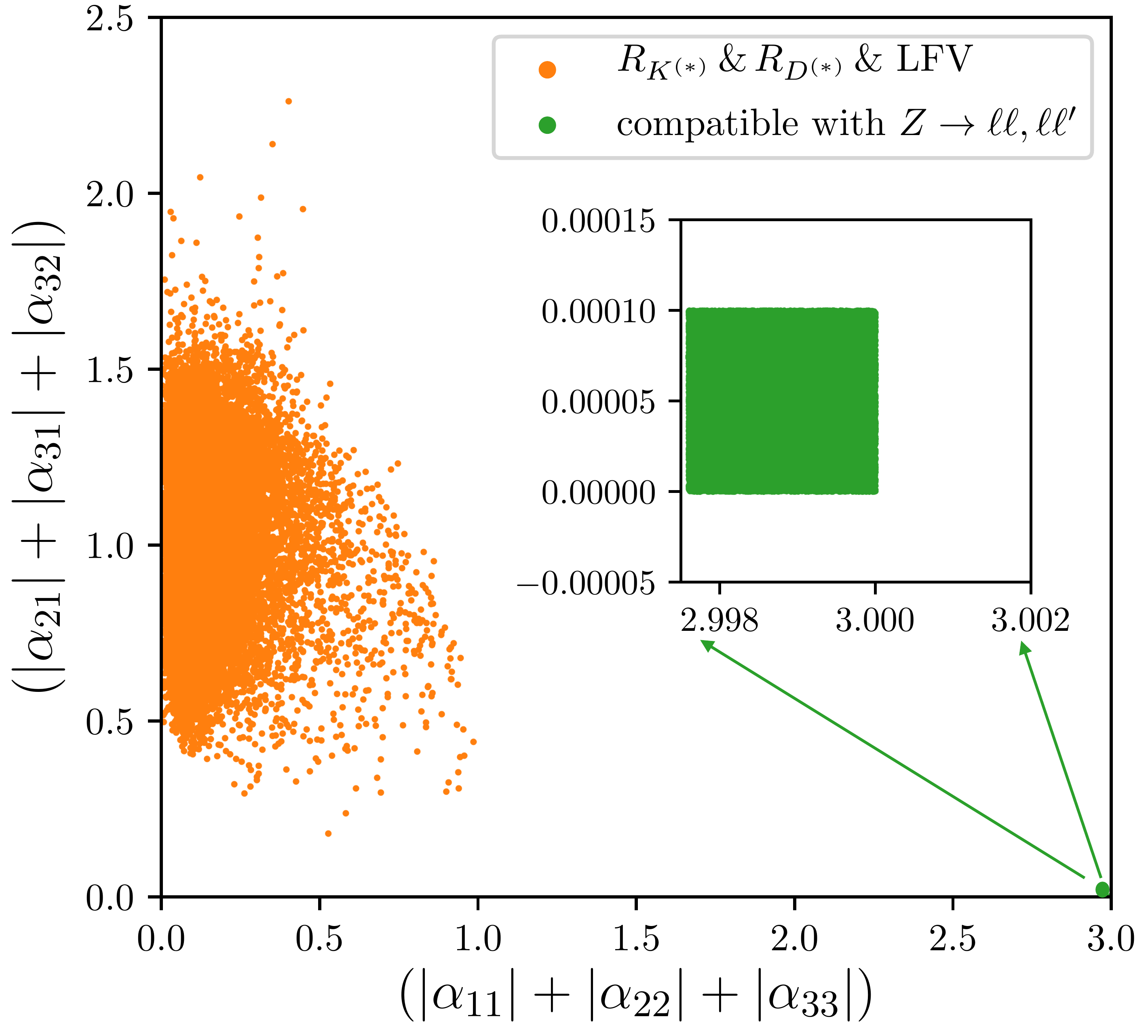}
    \caption{On the
    upper row, $A$-matrix entries complying with
    $R_{K^{(*)}}$ and $R_{D^{(*)}}$ data (blue) and those which in
    addition
    respect all imposed flavour constraints (yellow).
    Lower row: for the case of {\it isosinglet
    heavy leptons},
    $A$-matrix entries complying with
    $R_{K^{(*)}}$ and $R_{D^{(*)}}$ data as well as LFV bounds (yellow),
    and those which now further comply with bounds from
    $Z$ decays (green). (Notice that the green inset area
    corresponds to a zoom-out of what would otherwise be a tiny region close to
    the border of the parameter space.)
    In all panels we have taken $m_V=1.5$~TeV and all mixing angles have been varied randomly between $-\pi$ and $\pi$.}
    \label{fig:alphaii_ij:RDRKcLFV}
\end{figure}

For $n=3$, our study suggests that it is in general possible to
find regions in the parameter space in which the required non-universal
flavour structure to explain both $R_{K^{(*)}}$ and $R_{D^{(*)}}$
arises in a natural way, while still complying with all constraints
from flavour violating processes (meson and lepton sectors).
This can be seen from the upper row of
Fig.~\ref{fig:alphaii_ij:RDRKcLFV}, in which we display the regimes
for the entries of the matrix $A$ (cf.
Eqs.~(\ref{eq:UL:ARBS}) - (\ref{eq:Aalphaij})) which account for both $R_{K^{(*)}}$ and
$R_{D^{(*)}}$ data, as well as regions respecting the constraints
arising from the several flavour violating modes considered in our
analysis. Concerning the latter, we find it worth mentioning that
the most stringent constraints arise from $K_L \to \mu^\pm e^\mp$,
$\mu \to e \gamma$, and $\mu-e$ conversion in nuclei;
$B$-meson cLFV decays, or  (semi)leptonic $B$ and
$K$ decays lead to comparatively milder constraints
(or are systematically satisfied).

\begin{figure}[t!]
\hspace*{-4mm}
\includegraphics[width=0.52\textwidth,]{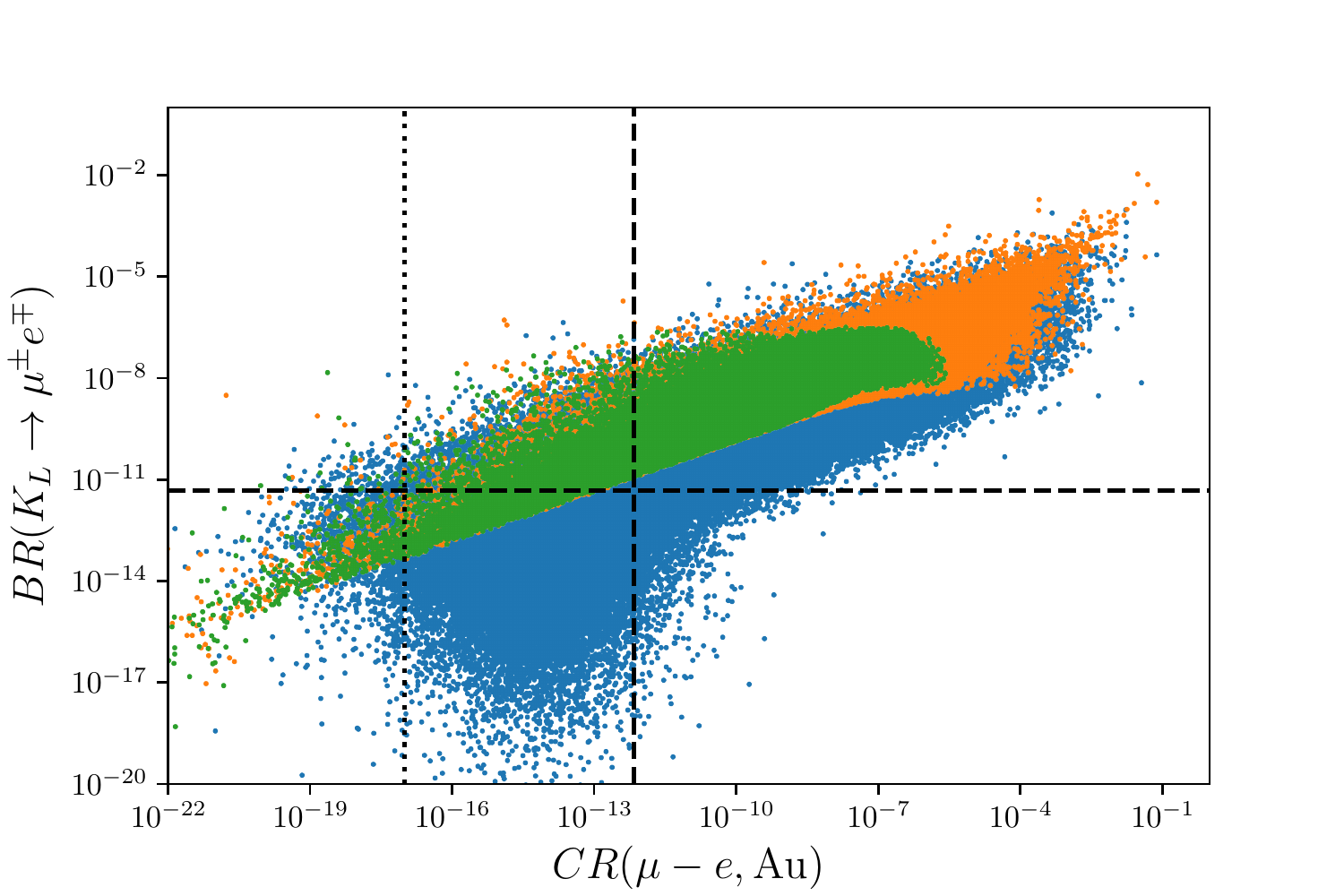}
\hspace*{2mm}
\includegraphics[width=0.52\textwidth,]{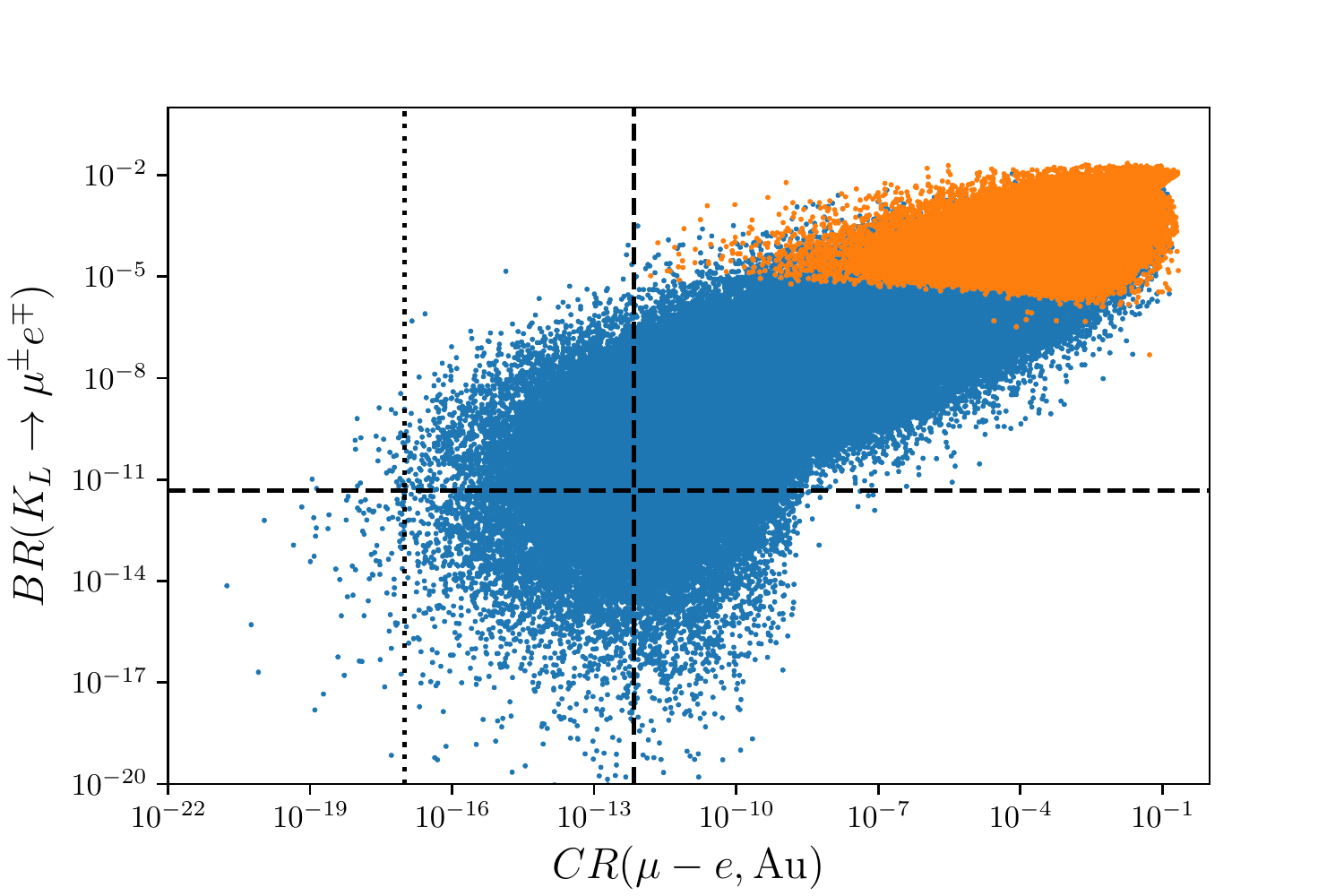}
\caption{Regions in the plane spanned by CR($\mu-e$,N) and
  BR($K_L \to \mu^\pm e^\mp$), accommodating $R_{K^{(*)}}$ (left
  panel) and $R_{D^{(*)}}$ (right panel), respectively for
  leptoquark masses in the intervals
  $m_V \in [15~\text {TeV}, 45~\text {TeV}]$ and
  $m_V \in [1~\text {TeV}, 6~\text {TeV}]$, in the framework of
  nonunitary leptoquark couplings induced by the presence of
  3 generations of {\it isosinglet heavy leptons}.
  Blue points satisfy  $R_{{K^{(*)}},\,{D^{(*)}}}$ at the 3 $\sigma$ level,
  yellow points are consistent with leptonic $Z$ decays, and
  green points are compatible with all imposed constraints,
  other than those depicted by the corresponding vertical and
  horizontal dashed lines (dotted ones denoting future
  sensitivities).
  In both panels, all mixing angles have been varied randomly between $- \pi$ and $\pi$.}
  \label{fig:uni}
\end{figure}

Finally, one should address the compatibility of the
considered SM leptoquark extension with the
constraints arising from EW precision tests; as discussed in the
previous subsection (and in Appendix~\ref{app:ew}), nonunitary
mixings can modify the couplings of the $Z$ boson. In particular,
for {\it isosinglet heavy leptons}, the entries of the matrix $A$ (see Eqs. \eqref{eq:Aalphaii} and \eqref{eq:Aalphaij}) are severely
constrained by the $Z$ width and by bounds on its cLFV decays
($Z \to \ell \ell^\prime$). This is shown on the lower row of
Fig.~\ref{fig:alphaii_ij:RDRKcLFV}, which illustrates the
tension between LFUV and $Z$ bounds - a tension which
ultimately leads to
disfavouring this class of extensions as a phenomenologically viable
NP model to explain both $R_{K^{(*)}}$ and
$R_{D^{(*)}}$ discrepancies.

If one foregoes a solution to the charged current anomalies (i.e.,
$R_{D^{(*)}}$), it is possible to accommodate $R_{K^{(*)}}$
in full agreement with constraints from flavour bounds, relying on
very mild deviations from unitarity, and thus evading constraints from
universality violation in $Z$ decays.
However, one would be led to regions with considerably heavier leptoquarks,
$m_V\gtrsim 15$~TeV. This is depicted on the left panel of
Fig.~\ref{fig:uni}, in which we display regimes complying with $R_{K^{(*)}}$
at the $3\sigma$ level) in the plane spanned by two particularly
constraining observables, BR($K_L \to \mu^\pm e^\mp$) and CR($\mu-e$,
N), for $15\text{ TeV}\lesssim m_V\lesssim 45\text{ TeV}$.
As can be verified, a small subset of points (consistent with $R_{K^{(*)}}$ and respecting universality in $Z$ decays) is compatible
with current bounds on the cLFV processes.
This is in agreement with the analyses of various UV-complete models,
such as~\cite{Balaji:2018zna,Fornal:2018dqn}.

For completeness, the right panel of Fig.~\ref{fig:uni} shows a similar study for
$R_{D^{(*)}}$. In order to accommodate $R_{D^{(*)}}$ data, lower
leptoquark masses are required (in this case we have taken
$1\text{ TeV}\lesssim m_V\lesssim 6\text{ TeV}$), and it is no longer
possible to evade $K_L \to \mu^\pm e^\mp$ and $\mu-e$ conversion
bounds while being consistent with leptonic $Z$-decay
universality. The data displayed in the panels of
Fig.~\ref{fig:alphaii_ij:RDRKcLFV} was obtained for vector leptoquark
masses $m_V\sim 1.5$~TeV; analogous conclusions can be inferred for
$m_V\sim\mathcal(1-3)$~TeV, albeit for different
$\alpha_{ij}$ ranges.

\begin{figure}[h]
  \includegraphics[width=0.52\textwidth, ]{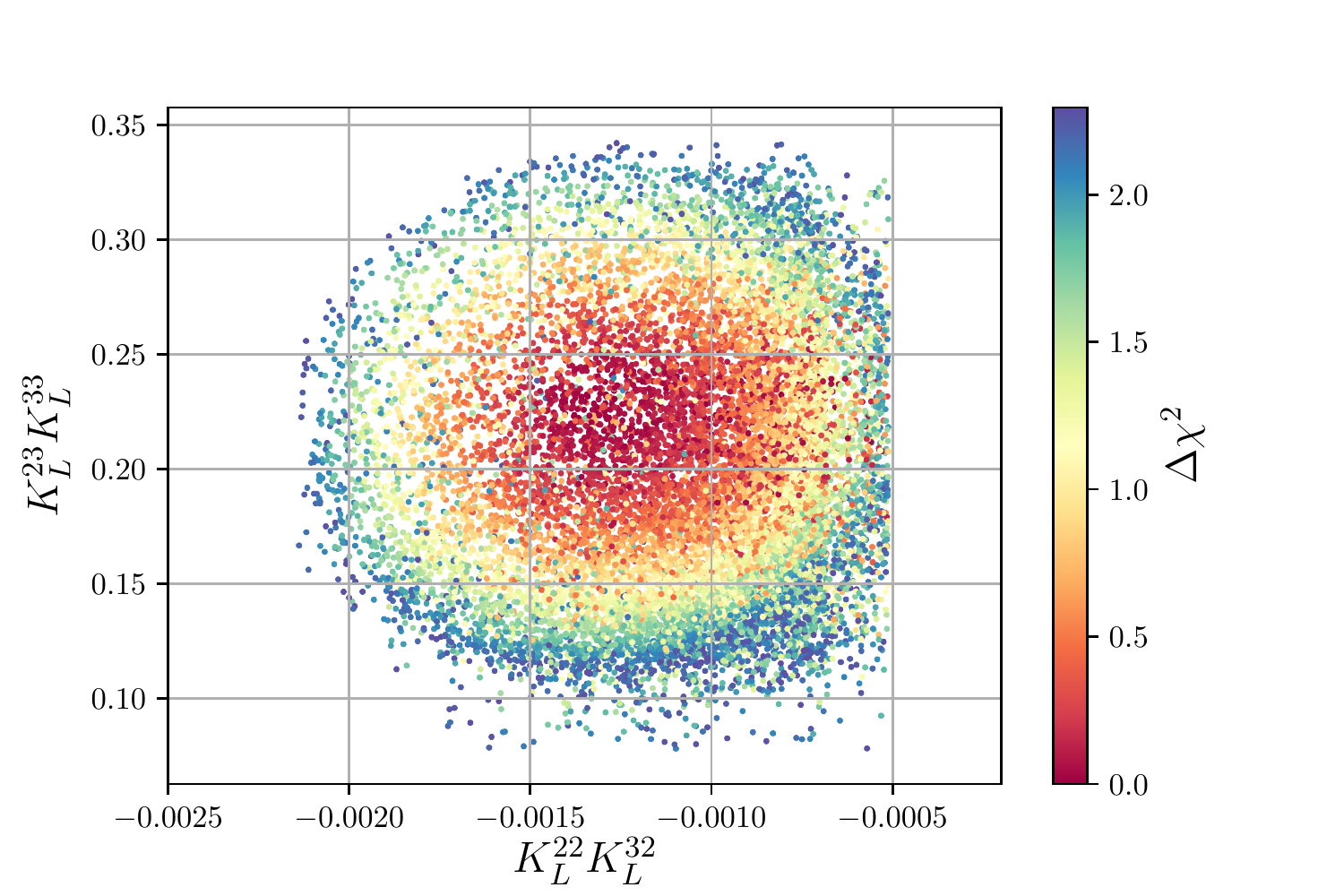}
  \includegraphics[width=0.52\textwidth, ]{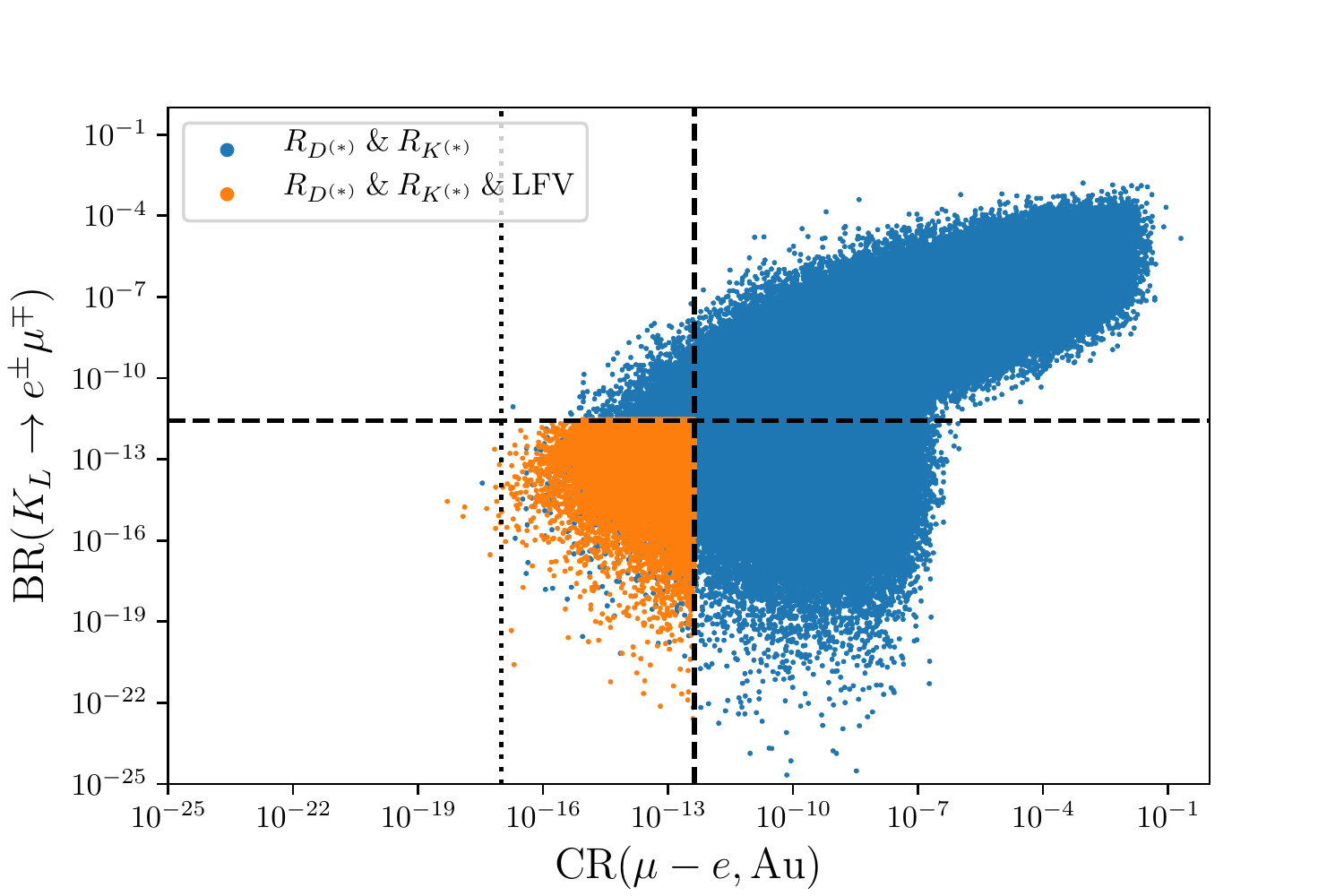}
\caption{On the left, $\Delta\chi^2$ distribution for the fit
  to $R_{K^{(*)}}$ and $R_{D^{(*)}}$ data ($1 \sigma$) in the plane of
  the $(K_L)_{ij}$ couplings.
  All points comply with the different (flavour)
  constraints. On the right,
  regions in the plane spanned by CR($\mu-e$,N) and
  BR($K_L \to \mu^\pm e^\mp$), accommodating both $R_{K^{(*)}}$ and
  $R_{D^{(*)}}$ (blue) and those in addition complying with LFV
  constraints (yellow).
  Both panels correspond to a heavy sector composed
  of {\it three isodoublet vector-like charged lepton states}, and to having set
  $m_V \sim 1.5$~TeV.
  The $\mathrm \Delta\chi^2$ corresponds to the $1\,\sigma$-region around the best fit point.}\label{fig:nuni}
\end{figure}

Since for the case of isosinglet leptons an explanation of $R_{D^{(*)}}$ is excluded by bounds on $Z$ decays, we now consider {\it isodoublet heavy charged leptons}. The nonunitarity in the couplings of the vector
leptoquark to the SM charged leptons can simultaneously
explain $R_{K^{(*)}}$ and $R_{D^{(*)}}$ data. Moreover, and by
construction, in the case of {\it isodoublet heavy charged lepton states}
$Z\ell\ell$ couplings remain universal
(in the absence of mixings between
right-handed SM charged leptons and vector-like doublets,
$\Delta g_R=0$, see Appendix~\ref{app:ew}); nevertheless,
flavour observables still play a crucial role, and are (as expected)
responsible for severe constraints on the NP degrees of freedom.

The left panel of Fig.~\ref{fig:nuni} offers a global view of this
case, showing the $\Delta\chi^2$ distribution for the fit
to $R_{K^{(*)}}$ and $R_{D^{(*)}}$ data, in the plane spanned by
$(K_L)_{ij}$ ``muon and tau couplings'' ($K_{22}K_{32}-K_{23}K_{33}$),
marginalising over the other couplings. The leptoquark mass is set to
$m_V \sim 1.5$~TeV.
We stress that leading to this plot all couplings
were determined by the underlying nonunitarity parametrisation
(with all mixing angles randomly sampled); in particular, we have not
set the leptoquark couplings to
the first generation of quark and leptons to zero. The displayed points
comply with {\it all} flavour bounds included in our study,
as described in Section~\ref{sec:constraints}.

The lowest $\Delta\chi^2$ region (dark red ellipsoid) suggests that the
best fit scenario corresponds to new physics dominantly coupling to
muons and taus. We stress that the patterns emerging from the $\Delta\chi^2$
distribution are not an artefact
of some particular assumption imposed on the couplings, but rather the result
of a very general scan over the full set of (mixing) parameters.

\medskip
The right panel of Fig.~\ref{fig:nuni} offers a projection of the
viable points (displayed on the left panel) in the plane of the most
constraining observables, CR($\mu-e$,N) and BR($K_L \to \mu^\pm e^\mp$).
It is interesting to notice that, to a very good approximation, most
of the currently phenomenologically viable
points lie within future reach of the upcoming muon-electron
conversion dedicated facilities (COMET and Mu2e).

In the near future, and should the $B$-meson decay anomalies be confirmed,
an explanation in terms of such a minimal leptoquark framework
could be probed via its impact for cLFV observables, in particular
$\mu-e$ conversion in nuclei. Although the cLFV bounds could be evaded
by increasing the mass of the vector leptoquark, this would however
prevent a viable explanation of the  $B$-meson decay anomalies,
especially of $R_{D^{(*)}}$.

\section{Concluding remarks}\label{sec:concs}
In this study we have considered a minimal SM extension via one
vector leptoquark $V_1$ and $n$ generations of heavy vector-like charged
leptons, as a candidate framework to explain the current $B$-meson
anomalies, $R_{K^{(*)}}$ and $R_{D^{(*)}}$.

Minimal extensions by a single $V_1$ leptoquark are in general
disfavoured due to the strong cLFV constraints on the (unitary) quark-lepton-$V_1$ couplings. Here we have suggested that
the pattern of mixings required to simultaneously address
$R_{K^{(*)}}$ and $R_{D^{(*)}}$ with a single $V_1$
could be interpreted within a
framework of nonunitary $V_1 \ell q$ couplings:
the mixings of the
SM charged leptons with the additional vector-like heavy
leptons can lead to effectively nonunitary $V_1 \ell q$ couplings,
offering the required amount of LFUV to  account for both anomalies.

As we have argued, the most
minimal nonunitary scenario (i.e. $n=1$) consistent with
both $R_{K^{(*)}}$ and
$R_{D^{(*)}}$, is ruled out as it leads to excessive contributions to cLFV
observables such as muon-electron conversion in nuclei.
We have thus considered
three families of vector-like heavy
leptons, and we have carried out a detailed analysis of the impact for
an extensive array of flavour violation and EW precision
observables.
Our findings revealed that the $SU(2)_L$ charges of the heavy charged
leptons are of paramount importance for the
model's viability:
for isosinglet heavy leptons, the mass of the leptoquark must be
sufficiently large to avoid excessive contributions to $Z$
decays, which then prevents an explanation of $R_{D^{(*)}}$.
This is expected to happen for heavy leptons of any $SU(2)_L$
representation except for isodoublets. In this particular case, the
$Z\ell \ell$ couplings remain universal, and we have shown that the
nonunitarity in the couplings allows to successfully explain both
sets of anomalies, while complying with all considered current bounds.

The nonunitary framework here considered in fact provides one of the
most economical options to simultaneously address $R_{K^{(\ast)}}$ and
$R_{D^{(\ast)}}$ data via a single vector leptoquark. Although our
results arise from a phenomenological analysis, our findings can be
easily related to well-motivated UV complete models, in which the
nonunitarity framework allows to obtain the required flavour
structure without further enlarging the gauge group.  Interestingly,
this is the case of the Pati-Salam
model~\cite{Calibbi:2017qbu,Bordone:2017bld,Blanke:2018sro,Bordone:2018nbg,Kumar:2018kmr,Angelescu:2018tyl,Balaji:2018zna,Fornal:2018dqn,Baker:2019sli,Cornella:2019hct},
in which the existence of a vector leptoquark with a mass around the
TeV scale (as required for a successful simultaneous explanation of
both the neutral and charged current anomalies) naturally motivates
the existence of such vector-like heavy leptons.

Finally, it is worth stressing that in view of the important progress
expected in the near future, cLFV searches can play a crucial role in
falsifying these minimal leptoquark frameworks, in view of
their strong impact for cLFV observables, in particular
$\mu-e$ conversion in nuclei.
As we have discussed, should the $B$-meson decay anomalies be verified, and should this be a valid framework to explain them, then one should expect the observation of
cLFV signals, in particular from $\mu-e$ conversion in Aluminium targets,
either at COMET or Mu2e.

\section*{Acknowledgements}
We acknowledge support within the framework of the
European Union’s Horizon 2020 research and innovation programme
under the Marie Sklodowska-Curie grant agreements No 690575 and No
674896.

\begin{appendix}

\section{Constraints from flavour violating rare meson decays and
  neutral meson mixing}\label{app:meson}

New physics frameworks including leptoquarks give rise to important
contributions to observables in the meson sector. These include flavour conserving and flavour violating
leptonic and semileptonic decays\footnote{We do not consider the
  $B_s\rightarrow \mu\mu$ decay explicitly, since it is already included
  in deriving the global ranges for the new physics contributions to
  $b\rightarrow s \ell\ell$ anomalies.} (including
final state neutrinos), as well as neutral meson mixings.
In this appendix we discuss the different leptoquark contributions to
leptonic and semi-leptonic meson decays (arising at tree-level) and to modes with final state neutrinos (at one-loop level). We recall that the SM predictions
and current experimental limits are summarised in
Table~\ref{tab:flavour}.

Note that the matching of the model's parameters with the Wilson coefficients is performed at the scale of the leptoquark mass, and then these are run down to the $b$-quark mass scale. In our analysis, the running from the scale of leptoquark mass to the $b$-quark mass scale, and to any other relevant process (observable) scale, is taken into account using the \texttt{wilson} package~\cite{Aebischer:2018bkb} in association with the \texttt{flavio} package~\cite{Straub:2018kue}.

\subsection{Exclusive $d_j \rightarrow d_i \ell^- \ell^{\prime +}$ decays}
The effective Hamiltonian for
$d_j \rightarrow d_i \ell^- \ell^{\prime +}$ transitions,
including the LFV operators, can be cast
as~\cite{Buchalla:1995vs,Bobeth:1999mk,Ali:2002jg,Hiller:2003js,Bobeth:2007dw,Bobeth:2010wg}
\begin{equation}\label{eqn:effHam}
\mathcal H_\text{eff} =
- \frac{4 G_F}{\sqrt{2}} V_{3j}V_{3i}^{\ast}\Big[ \sum_{
    \begin{array}{c}
      k=7,9,\\
      10,S,P
    \end{array}
  } \left( C_k (\mu) \mathcal{O}_k(\mu) + C_k^{'} (\mu)
  \mathcal{O}_k^{'} (\mu)\right) + C_T (\mu) \mathcal{O}_T(\mu) +
  C_{T_5} (\mu) \mathcal{O}_{T_5}(\mu)\Big].
\end{equation}
In the above, we recall that $V_{ij}$ denotes the CKM matrix;
the operators are defined as
\begin{align}\label{eqn:operators}
\mathcal O_7^{ij} &= \frac{e m_{d_j}}{(4\pi)^{2}}(\bar d_i
\sigma_{\mu\nu}P_R d_j)F^{\mu\nu}\:\text, \nonumber \\
\mathcal{O}_9^{ij;\ell \ell^{\prime}} &= \frac{e^{2}}{(4\pi)^2}(\bar
d_i \gamma^{\mu}P_L d_j)(\bar \ell \gamma_\mu \ell^\prime)\:\text,  \nonumber \\
\mathcal{O}_{10}^{ij;\ell \ell^{\prime}} &=
\frac{e^{2}}{(4\pi)^2}(\bar d_i \gamma^{\mu}P_L d_j)(\bar \ell
\gamma_\mu \gamma_5 \ell^\prime)\:\text, \nonumber \\
\mathcal{O}_S^{ij;\ell \ell^{\prime}} &= \frac{e^{2}}{(4\pi)^2}(\bar
d_i P_R d_j)(\bar\ell \ell^\prime)\:\text, \nonumber \\
\mathcal{O}_P^{ij;\ell \ell^{\prime}} &= \frac{e^{2}}{(4\pi)^2}(\bar
d_i P_R d_j)(\bar\ell\gamma_5 \ell^\prime)\:\text, \nonumber \\
\mathcal O_T^{ij;\ell \ell^{\prime}} &= \frac{e^{2}}{(4\pi)^{2}}(\bar
d_i \sigma_{\mu\nu} d_j)(\bar \ell \sigma^{\mu\nu}
\ell^\prime)\:\text, \nonumber \\
\mathcal O_{T5}^{ij;\ell \ell^{\prime}} &=
\frac{e^{2}}{(4\pi)^{2}}(\bar d_i \sigma_{\mu\nu} d_j)(\bar \ell
\sigma^{\mu\nu}\gamma_5 \ell^\prime)\:\text,
\end{align}
with the primed operators $\mathcal O_{7, 9, 10, S, P}^{\prime}$
following from the replacement $P_L \leftrightarrow P_R$.

Regarding lepton flavour violating processes, only the operators
$\mathcal O_{9,10,S,P}^{(\prime)}$ play a relevant role:
for vector leptoquarks, the complete set of associated
Wilson coefficients (not present in the SM) is given by~\cite{Dorsner:2016wpm}
\begin{align}
C^{ij;\ell \ell^{\prime}}_{9,10} &= \mp\frac{\pi}{\sqrt{2}G_F\,\alpha_\text{em}\,V_{3j}\,V_{3i}^{\ast} \,m_V^2}\left(K_L^{i
  \ell^\prime} \,K_L^{j\ell\ast} \right)\nonumber \\
C^{\,\prime\,ij;\ell \ell^{\prime}}_{9,10} &= -\frac{\pi}{\sqrt{2}G_F\,\alpha_\text{em} \,V_{3j}\,V_{3i}^{\ast} \,m_V^2}\left(K_R^{i
  \ell^\prime} \,K_R^{j\ell\ast} \right)\nonumber \\
C^{ij;\ell \ell^{\prime}}_{S,P} &= \pm\frac{\pi}{\sqrt{2}G_F\,\alpha_\text{em} \,V_{3j}\,V_{3i}^{\ast}\, m_V^2}\left(K_L^{i
  \ell^\prime} \,K_R^{j\ell\ast} \right)\nonumber \\
C^{\,\prime\, ij;\ell \ell^{\prime}}_{S,P} &= \frac{\pi}{\sqrt{2}G_F\,\alpha_\text{em} \,V_{3j}\,V_{3i}^{\ast}\, m_V^2}\left(K_R^{i
  \ell^\prime} \,K_L^{j\ell\ast} \right)\:\text,
\end{align}
where $K_L^{i\ell}$
($K_R^{i\ell}$) denoting the left-handed (right-handed) leptoquark couplings.
Here, and for completeness, we include the
$K_R$ mixing matrix, although we do not take it
into account in our analysis.

\subsection{$P \rightarrow \ell^- \ell^{\prime +}$ decays}
Leptonic decays of pseudoscalar mesons lead to important constraints
on the (vector) leptoquark couplings. Here, we summarise the
computation of the $P \rightarrow \ell^- \ell^{\prime +}$ rates,
following the derivation of~\cite{Becirevic:2016zri}.
With the standard decomposition of the hadronic matrix element
\begin{equation}
  \langle 0\,| \,\bar d_j \,\gamma_\mu\,\gamma_5 \,d_i|\,P(p)\rangle
  \,=\, i \,p_\mu \,f_{P}\,,
\end{equation}
in which $f_P$ denotes the $P$ meson decay constant,
the branching fraction\footnote{Notice that for the lepton flavour
  conserving decay $B_s \rightarrow \mu\mu$,
  the SM contribution and the renormalisation group running (as well as
  mixing of the SM operators) must be taken into account.} is given by
\begin{align}
\text{BR}(P &\rightarrow \ell^- \,\ell^{\prime +}) \,=\,
\frac{\tau_{P}}{64 \,\pi^3}\frac{\alpha^{2}
  \,G_F^{2}}{m_P^{3}}\,f_P^{2}\,|V_{3j}\,V_{3i}^{{\ast}}|^2\,
\lambda^{\frac{1}{2}}(m_P, m_{\ell}, m_{\ell^\prime})\nonumber \\
&\times\Bigg\{\left(m_P^{2} - \left(m_{\ell} + m_{\ell^\prime}
\right)^{2} \right)\Bigg|\left(C_9 - C_9^{\prime}\right)\left(m_{\ell}
- m_{\ell^\prime} \right) +\left(C_S - C_S^{\prime}
\right)\frac{m_P^{2}}{m_{d_j} + m_{d_i}}  \Bigg|^{2}\nonumber \\
&+\left(m_P^{2} - \left(m_{\ell} - m_{\ell^\prime} \right)^{2}
\right)\Bigg|\left(C_{10} - C_{10}^{\prime}\right)\left(m_{\ell} +
m_{\ell^\prime} \right) +\left(C_P -
C_P^{\prime}\right)\frac{m_P^{2}}{m_{d_j} + m_{d_i}}   \Bigg|^{2}
\Bigg\}\,,
\end{align}
where the K\"all\'en-function is defined as
\begin{equation}
\label{eq:kallenlambda}
\lambda(a,b,c) = \left(a^{2} - \left(b-c\right)^{2} \right)\left(a^{2} -
\left(b+c\right)^{2} \right)\,\text.
\end{equation}

\noindent
Since this FCNC transition is generated
at tree-level from $V_1$ interactions,
the modes leading to different final state lepton charge assignments
must both be included and treated separately.
Leptonic pseudoscalar meson decays thus provide extensive (and very tight) constraints
on the leptoquark couplings.

\subsection{$P \rightarrow P^{\prime} \ell^- \ell^{\prime +}$ decays}
Again working in the standard basis,
the hadronic matrix elements are parametrised as
\begin{align}
  &\langle \bar P^{\prime}(k)\,| \,\bar d_i \,\gamma_\mu \,
  d_j\,| \,\bar P (p)\rangle
= \left[(p+k)_\mu - \frac{m_P^{2} - m_{P^{\prime}}^{2}}{q^{2}}q_\mu
  \right]\,f_+(q^{2}) +\frac{m_P^{2} - m_{P^{\prime}}^{2}}{q^{2}}q_\mu
\,f_0(q^{2})\,,\\
&\langle \bar P^{\prime}(k)\,|\,\bar d_i  \,\sigma_{\mu\nu} \,d_j\,
|\,\bar P(p)\rangle = -i\,\left(p_\mu \,k_\nu - p_\nu \,k_\mu
\right)\frac{2}{m_{P}
  + m_{P^{\prime}}}\, f_T(q^{2},\mu)\,.
\end{align}
In the above, the hadronic form factors $f_{+,0,T}(q^{2})$
depend on the momentum transfer, which lies in the range
$(m_{\ell} + m_{\ell^\prime})^{2} \leq q^{2}\leq (m_P - m_{P^{\prime}})^{2}$.
In our evaluation of the form factors we closely follow the
results of~\cite{Khodjamirian:2010vf};
since we focus on decays with heavy-to-light meson transitions,
we further assume the scale to be $\mu = m_{d_j}$, i.e. $\mu = m_b$.
With the quantities
\begin{align}
\varphi_7(q^{2}) &= \frac{2\,m_{d_j}\,|f_T(q^{2})|^{2}}{(m_P +
  m_{P^{\prime}})^{2}}\,\lambda(m_P, m_{P^{\prime}},
\sqrt{q^{2}})\,\left[1 - \frac{(m_{\ell} - m_{\ell^\prime})^{2}}{q^{2}}
  - \frac{\lambda(\sqrt{q^{2}}, m_{\ell}, m_{\ell^\prime})}{3\,q^{4}}
  \right]\:\text, \nonumber\\
\varphi_{9(10)}(q^{2}) &= \frac{1}{2}\,|f_0(q^{2})|^{2}(m_{\ell} \mp
m_{\ell^\prime})^{2}\,\frac{(m_P^{2} -
  m_{P^{\prime}}^{2})^{2}}{q^{2}}\,\left[1 - \frac{(m_{\ell} \pm
    m_{\ell^\prime})^{2}}{q^{2}} \right]\nonumber\\
&+ \frac{1}{2}\,|f_+(q^{2})|^{2}\,\lambda(m_P, m_{P^{\prime}},
\sqrt{q^{2}})\,\left[1 - \frac{(m_{\ell} \mp
    m_{\ell^\prime})^{2}}{q^{2}} - \frac{\lambda(\sqrt{q^{2}},
    m_{\ell}, m_{\ell^\prime})}{3\,q^{4}} \right]\:\text,\nonumber\\
\varphi_{79}(q^{2}) &= \frac{2\,m_{d_j}\,f_+(q^{2})\,f_T(q^{2})}{m_P +
  m_{P^{\prime}}}\,\lambda(m_P, m_{P^{\prime}}, \sqrt{q^{2}})\,\left[1 -
  \frac{(m_{\ell} - m_{\ell^\prime})^{2}}{q^{2}} -
  \frac{\lambda(\sqrt{q^{2}}, m_{\ell}, m_{\ell^\prime})}{3\,q^{4}}
  \right]\:\text,\nonumber\\
\varphi_{S(P)}(q^{2}) &= \frac{q^{2}\,|f_0(q^{2})|^{2}}{2\,(m_{d_j} -
  m_{d_i})^{2}}\,\left(m_P^{2} - m_{P^{\prime}}^{2} \right)^{2}\,\left[1 -
  \frac{(m_{\ell} \pm m_{\ell^\prime})^{2}}{q^{2}} \right]\:\text,\nonumber\\
\varphi_{10P(9S)}(q^{2}) &= \frac{|f_0(q^{2})|^{2}}{m_{d_j} -
  m_{d_i}}\,(m_{\ell} \pm m_{\ell^\prime})(m_P^{2} -
m_{P^{\prime}}^{2})^{2}\,\left[1 - \frac{(m_{\ell} \mp
    m_{\ell^\prime})^{2}}{q^{2}} \right]\,,
\end{align}
and the normalisation factor
\begin{equation}
|\mathcal N_{P^{\prime}}(q^{2})|^{2} = \tau_{P}\,\frac{\alpha^{2}\,
  G_F^{2} \,|V_{3j} \,V_{3i}^{{\ast}}|^{2}}{512 \,\pi^5\,
  m_P^3}\,\frac{\lambda^{\frac{1}{2}}(\sqrt{q^{2}}, m_{\ell},
  m_{\ell^\prime})}{q^{2}}\,\lambda^{\frac{1}{2}}(\sqrt{q^{2}}, m_P,
m_{P^{\prime}})\,,
\end{equation}
the differential branching fraction can be cast as
\begin{align}
\frac{d\,\mathrm{BR} (P \rightarrow P^{\prime}
\ell^-\ell^{\prime +})}{d q^{2}} &= |\mathcal
N_{P^{\prime}}(q^{2})|^{2}\times\Big\{\varphi_7(q^{2})\,|C_7 +
C_7^{\prime}|^{2} + \varphi_9(q^{2})\,|C_9 + C_9^{\prime}|^{2}
+\varphi_{10}(q^{2})\,|C_{10} + C_{10}^{\prime}|^{2} \nonumber\\
&+  \varphi_S(q^{2})\,|C_S + C_S^{\prime}|^{2} + \varphi_P(q^{2})\,|C_P +
C_P^{\prime}|^{2}  + \varphi_{79}(q^{2})\,\mathrm{Re}\left[(C_7 +
  C_{7}^{\prime})\,(C_9 + C_{9}^{\prime})^{{\ast}} \right]\nonumber\\
&+ \varphi_{9S}(q^{2})\,\mathrm{Re}\left[(C_9 + C_{9}^{\prime})\,(C_S +
  C_{S}^{\prime})^{{\ast}} \right] +
\varphi_{10P}(q^{2})\,\mathrm{Re}\left[(C_{10} + C_{10}^{\prime})\,(C_P +
  C_{P}^{\prime})^{{\ast}} \right] \Big\}\:\text.
\end{align}

\subsection{Loop effects in neutrino modes}
Both $s\to d \nu\nu$ and $b\to s \nu\nu$ transitions are known
to provide some of the most important
constraints on NP scenarios aiming at addressing the anomalies in
$R_{K^{(\ast)}}$ and $R_{D^{(\ast)}}$ data.
Following the convention of~\cite{Buras:2014fpa}, at the quark level
the $|\Delta S|=1$ rare decays
$K^+\,(K_L)\to \pi^+\,(\pi^0)\,\nu_\ell \bar\nu_{\ell^\prime }$ and
$B\to  K^{(\ast)} \nu_\ell\bar \nu_{\ell^\prime}$
can be described by the following short-distance effective Hamiltonian
for $d_j\to d_i \nu_\ell\bar \nu_{\ell^\prime}$
transitions~\cite{Bobeth:2017ecx,Bordone:2017lsy}
\begin{eqnarray}\label{eq:eff-H-Ktopi}
-\mathcal{H}_\text{eff} =
&\frac{4 \,G_F}{\sqrt{2}} \,V_{3i}^\ast \,V_{3j}\,
\frac{\alpha_e}{2\,\pi}\,
\left[C_{L,ij}^{\ell\ell^\prime} \,\left(\bar d_i\,\gamma_\mu \,P_L\,
  d_j\right)\, \left(\bar \nu_\ell\,\gamma^\mu\,
  \,P_L\,\nu_{\ell^\prime}\right) \right. \nonumber\\
&+\left.
C_{R,ij}^{\ell\ell^\prime} \,\left(\bar d_i\,\gamma_\mu \,P_R\,
d_j\right)\, \left(\bar \nu_\ell \,\gamma^\mu
\,P_L\nu_{\ell^\prime}\right) \right] \,+\, \text{H.c.}\, ,
\end{eqnarray}
in which $i,j$ denote the down-type flavour content of the final and
initial state meson, respectively.
For vector leptoquarks, the contributions are generated at
one loop-level and are a priori divergent; consequently, the
calculation should be carried in a given gauge-embedding,
including the corresponding would-be Goldstone modes. Here we follow
the computation of~\cite{Crivellin:2018yvo}, and the
coefficient for $d_a \rightarrow d_f\bar\nu_i\nu_j$ is thus given by
\begin{align}
C_{L,fa}^{ij} = \sum_{k,l} -\frac{M_W^{2}}{2\,e^{2}\, V_{3a}\,V_{3f}^{\ast}
 \, m_V^2}\Bigg(&6\,
K_L^{fj}\,K_L^{ai\ast}\,{\ln}\left(\frac{M_W^{2}}{m_V^2} \right) +
V_{3f}^{\ast}\,V_{3k}\,K_L^{kj}\,V_{3a}\,V_{3l}^{\ast}\,
K_L^{li\ast}\,\frac{m_t^{2}}{M_W^{2}}
\nonumber\\
&+
3\left(V_{3a}\,V_{3k}^{*}\,K_L^{ki\ast}\,K_L^{fj} +
V_{3f}^{\ast}\,V_{3k}\,K_L^{kj}\,K_L^{ai\ast}
\right)\,\frac{m_t^{2}\,{\ln}\left(\frac{m_t^{2}}{M_W^{2}}
  \right)}{m_t^{2} - M_W^{2}}
\Bigg)\:\text,
\end{align}
where $M_W$ is the mass of the $W$ boson and $m_t$ the mass of the top quark.
The branching fractions for the neutral and charged kaon decay
modes are given by~\cite{Buras:2004qb,Buras:2015qea}
\begin{align}
&\text{BR}(K^ \pm \to \pi ^ \pm \nu \bar \nu)\, =\,
  \frac{1}{3}\left( 1 + \Delta _{EM} \right)\,\eta _\pm \times
  \sum_{f,i = 1}^3 \left\{ \left[
    \frac{\text{Im}\left(\lambda_t\,\tilde X_L^{fi} \right)}{
      \lambda^5}\right]^{2}+ \left[
    \frac{\text{Re} \left(\lambda_c \right)}{
      \lambda} \,P_c\, \delta _{fi}
    + \frac{\text{Re} (
      \lambda_t \,\tilde X_L^{fi})}{ \lambda ^5}
  \right]^{2}\right \},\nonumber\\
  &\text{BR}(K_L \to \pi \nu \bar \nu )\, = \frac{1}{3}{\eta_L}
  \sum_{f,i = 1}^3 \left[\frac{\text{Im}\left(\lambda_t\,
        \tilde X_L^{fi} \right)}{\lambda^5}\right]^{2}\,,
\end{align}
with
\begin{align}
\tilde X_L^{fi} &
  = X_{L}^{\text{SM},fi} - s_W^2\,C_{L,sd}^{fi}\,,\;\quad
P_c = 0.404 \pm 0.024 \nonumber\\
\eta_\pm  &=
\left( 5.173 \pm 0.025 \right)\times 10^{-11}
\left[\frac{\lambda}{0.225} \right]^8\,,\nonumber\\
  \eta_L &=\left( 2.231 \pm 0.013 \right)\times 10^{-10}
 \left[\frac{\lambda}{0.225} \right]^8\,,\nonumber\\
\Delta_{EM} &=  - 0.003\,,\; \quad
X_{L}^{\text{SM},fi} =
\left(1.481 \pm 0.005 \pm 0.008\right)\,\delta _{fi}\,.
\end{align}
In this convention, $\lambda$ is one of the Wolfenstein parameters related to the Cabibbo angle, and $\lambda_c = V_{cs}^\ast V_{cd}$ and $\lambda_t = V_{ts}^\ast V_{td}$. \\
The $B\to K^{(*)}\nu\bar{\nu}$ decay width has been derived
in~\cite{Buras:2014fpa}, leading to
$C_{L,sb}^{\text{SM},fi} \approx -1.47/s_W^2\delta_{fi}$.
Finally, it is convenient to express the
BRs normalised to the SM predictions,
\begin{equation}
R_{K^{(*)}}^{\nu\bar{\nu}} \,= \,
\frac{1}{3}\sum_{f,i=1}^3
\frac{ \big|C_{L,sb}^{fi} \big|^2}{
  \big|C_{L,sb}^{\text{SM},fi}\big|^2} \,\text.
\end{equation}

%%%%%%%%%%%%%%%%%%%%%%%%%%%%
\subsection{Loop effects in neutral meson mixing}

In this scenario a contribution to $\left|\Delta F\right|=2$ amplitudes is
generated at one-loop level. Contributions to
neutral meson mixings, $P-\bar P$  with $P= B^0_s, B^0_d ,K^0$,
arise both from SM box diagrams involving top quarks and $W$'s, and from
NP box diagrams involving leptons and
vector leptoquarks.
These contributions can be described
in terms of the following effective Hamiltonian for $|\Delta F| =2$
transitions
\begin{equation}\label{eq:PPmixing}
  \mathcal{H}^{P}_\text{eff} \,= \,
(C_P^\text{SM}+C_P^\text{NP})
\left(\bar d_i\,\gamma^\mu \,P_L \,d_j\right)
\left(\bar d_i \,\gamma_\mu \,P_L\, d_j\right)\,+\text{H.c.},
\end{equation}
with $\{i,j\}$ respectively denoting $\{b,s\}$, $\{b,d\}$ or $\{d,s\}$
for $P=B^0_s$, $B^0_d$ or $K^0$ mesons.
The $|\Delta F| =2$ transitions are sensitive to the
mass scale of the heavy vector-like fermions, and the widths
scale proportionally to $m_V^2$
(similarly to the SM contribution, itself proportional to $m_t^2$).
A complete evaluation of the contributions must further include the
effects of the (physical) Higgs fields; therefore
the computation of these observables requires
specifying a particular UV completion.
Nevertheless, it is possible to draw preliminary conclusions
on the mass scales of the vector leptoquarks and heavy leptonic states
(here denoted by $M$)
based on the new physics contribution to the diagrams
involving $V_1$.
For example, taking $P=B^0_s$,
one obtains~\cite{Buttazzo:2017ixm,Calibbi:2017qbu}
\begin{equation}
C_{B_s}^{\text{NP}} \,=\,
-\frac{K_L^{2\ell}\,K_L^{3\ell\ast}\,K_L^{2\ell^\prime}\,
  K_L^{3\ell^\prime\ast}}{16\,\pi^2}\,
\left(\frac{D_6}{4M^4} + D_2 - \frac{2\,D_4}{M^2} \right)\,.
\end{equation}
Here $\ell,\ell^\prime=1,...,6$
are the six fermions with the quantum numbers of charged leptons (6 physical eigenstates arising from the mixings of the light SM and heavy vector-like charged leptons).
The loop functions $D_x\equiv D_x\left(M,M,m_s,m_t \right)$
are given by
\begin{equation}
  D_x(m_1,m_2,m_3,m_4)\, =\,\frac{i}{16\,\pi ^2}
  \int{
    \frac{d^d k}{(2\,\pi)^d}
    \frac{(k^2)^{x/2}}{(k^2 - m_1^2)\,(k^2- m_2^2)\,(k^2 - m_3^2)
    \,(k^2 - m_4^2)}}\,\text.
\end{equation}
The $|\Delta F| = 2$ transitions thus lead to a (lower) bound on the heavy leptonic mass scales of around $500$~GeV, while the vector leptoquark mass should lie above the TeV
to keep new physics contributions to $\Delta M_{B_{s,d}}$ below
$\mathcal{O}(10\%)$, given the experimental constraints.

\section{Constraints from cLFV decays}{\label{app:cLFV}}
Due to the LFUV couplings of the leptoquarks, sizeable contributions
can be generated for cLFV
observables, including
radiative decays $\ell_i \rightarrow \ell_j \gamma$,
three-body decays $\ell_i \rightarrow 3\ell_j$, and neutrinoless
$\mu - e$ conversion in nuclei, which then
lead to important constraints.

As already mentioned in Section~\ref{sec:pheno:constraints}, although the radiative decays are
generated at loop level, anapole operators can induce additional
contributions to the
Wilson coefficients (comparable to the tree-level contributions to
neutrinoless $\mu - e$ conversion), and dipole operators can also be
responsible to significant contributions to radiative decays and
neutrinoless $\mu - e$ conversion.

Notice that the one-loop dipole and anapole contributions due to
the exchange of vector
bosons generically diverge, and a UV completion must be specified to
obtain a convergent result in a gauge independent manner. We have computed
the anapole and dipole contributions in the Feynman gauge, for which it is
necessary to include the relevant contributions from the Goldstone
modes. To consistently compute these contributions
for vector leptoquarks, we make the minimal working
assumption that the new state corresponds to a (non-abelian)
$SU(3)_c$ gauge extension, whose breaking gives rise to
a would-be Goldstone boson degree of freedom, subsequently absorbed by the
massive vector leptoquark. We thus include this Goldstone mode
(degenerate in mass with $V_1$) to obtain the gauge invariant (finite)
form factors for the dipole and anapole contributions.

\subsection{Radiative lepton decays $\ell_i \rightarrow \ell_j \gamma$}
The relevant terms of the effective Lagrangian for
radiative lepton decays $\ell_i \rightarrow \ell_j \gamma$
can be written as
\begin{equation}\label{eqn:radeff}
\mathcal{L}^{\ell_i \to \ell_j \gamma}_\text{eff} \,=\,
-\frac{4G_F}{\sqrt{2}}\,\bar\ell_j\,
\sigma^{\mu\nu}\,F_{\mu\nu}\,
\left(C_L^{\ell_i\ell_j} \,P_L\,  +\,
C_R^{\ell_i\ell_j}\,P_R\right)\,\ell_i \,+\, \text{H.c.}\,,
\end{equation}
in which $F_{\mu\nu}$ is the electromagnetic field strength tensor.
The corresponding Wilson coefficients are related to the form
factors $\sigma_{L(R)}^{\ell_i\ell_j}$  as
\begin{equation}
  C_{L(R)}^{\ell_i \ell_j} \,=\, \frac{e}{4\sqrt{2}\,G_F}\,
  \sigma_{L(R)}^{\ell_i\ell_j}\,,
\end{equation}
where the form factors $\sigma_{L(R)}^{\ell_i\ell_j}$
can be computed following the prescription of~\cite{Lavoura:2003xp}.
At the one-loop level, there are 10 distinct diagrams that must
be included. The gauge-invariant amplitude can be decomposed as
\begin{equation}
  i\mathcal{M}_{\text{dipole}}\,=\, ie \,\epsilon_\mu^{{\ast}}
  \,M_{\text{dipole}}^{\mu}\,,
\end{equation}
with
\begin{equation}
  M^{\mu}_{\text{dipole}} \,=\, \bar \ell_j\,\left[i\,\sigma^{\mu\nu}
    \,q_\nu\,(\sigma_L \,P_L + \sigma_R \,P_R)\right]\,\ell_i
\end{equation}
in which $\epsilon$ and $q$ denote the photon polarisation and its
momentum. The $\ell_i \rightarrow \ell_j \gamma$ decay width is then given by
\begin{equation}
\Gamma(\ell_i \rightarrow \ell_j\gamma) \,= \,\frac{\alpha\,(m_{\ell_i}^{2}
  - m_{\ell_j}^{2})^{3}}{4 \,m_{\ell_i}^{3}} \,
\left(|\sigma_L^{\ell_i\ell_j}|^{2} + |\sigma_R^{\ell_i\ell_j}|^2\right)\,.
\end{equation}
In the physical (mass) basis,
the relevant part of the Lagrangian leading to the computation of $\sigma_{L,R}$ can be written as
(cf.Eq.~(\ref{eq:lagrangian:Vql_phys3N}))
\begin{equation}
  \mathcal{L}_{V_1} \,=\,
  \sum_{i,j} \left[V_1^{\mu}\,\bar d_i \,\gamma_\mu\,
    \left(K_L^{ij} \,P_L + K_R^{ij} \,P_R \right)\,\ell_j
    + \text{H.c.} \right]\,,
\end{equation}
with $K_L$ ($K_R$) the left-handed (right-handed) coupling matrix.
The Goldstone ($\varphi$) interaction terms in the Lagrangian are then given by
\begin{equation}
  \mathcal{L}_\text{Goldstone}\, =\, \varphi\,\frac{i}{M}\sum_{i,j}
  \bar d_i \,\Big[(K_R^{ij}\,m_{\ell_j} - K_L^{ij}\, m_{d_i})\,P_L
    + (K_L^{ij} \,m_{\ell_j} - K_R^{ij}\,m_{d_i})\,P_R \Big]\,\ell_j +
  \text{H.c.}\,.
\end{equation}
The form factors $\sigma_{L,R}$ can be cast in the following (compact) form:
\begin{align}
\sigma_L^{\ell_i\ell_j} \,= \,-\frac{i\,
  N_c}{16\pi^{2}\,M^2}\sum_{k}\Bigg\{&\frac{2}{3}
\Big[\left(K_R^{kj{\ast}}\,K_R^{ki}\,m_{\ell_i}
  +K_L^{kj{\ast}}\,K_L^{ki}\,m_{\ell_j} \right) g(t_k) +
  K_R^{kj{\ast}}\,K_L^{ki}\,m_{d_k}\, j(t_k)\Big] \nonumber \\
-&\frac{1}{3}\Big[\left(K_R^{kj{\ast}}\,K_R^{ki}\,m_{\ell_i}
  +K_L^{kj{\ast}}\,K_L^{ki}\,m_{\ell_j} \right)\, f(t_k) +
  K_R^{kj{\ast}}\,K_L^{ki}\,m_{d_k}\, h(t_k)\Big] \Bigg\}\,,
\\
\sigma_R^{\ell_i\ell_j} \,= \,-\frac{i \,N_c}{16\pi^{2}\,M^2}
\sum_{k}\Bigg\{&\frac{2}{3}
\Big[\left(K_L^{kj{\ast}}\,K_L^{ki}\,m_{\ell_i} +
  K_R^{kj{\ast}}\,K_R^{ki}\,m_{\ell_j} \right) \,g(t_k) +
  K_L^{kj{\ast}}\,K_R^{ki}\,m_{d_k}\, j(t_k)\Big] \nonumber\\
-&\frac{1}{3}\Big[\left(K_L^{kj{\ast}}\,K_L^{ki}\,m_{\ell_i}
 +K_R^{kj{\ast}}\,K_R^{ki}\,m_{\ell_j} \right) \,f(t_k) +
 K_L^{kj{\ast}}\,K_R^{ki}\,m_{d_k} \,h(t_k)\Big] \Bigg\}\,,
\end{align}
with $t_k = \frac{m_{d_k}^{2}}{m_V^2}$ and $N_c$ the number of colours of the internal fermion.
The loop functions are given by
\begin{align}
f(t) &= \frac{-5\,t^{3} + 9\,t^{2} - 30\,t + 8}{12\,(t-1)^{3}} +
\frac{3\,t^{2}\,{\ln}(t)}{2\,(t-1)^{4}}\,, \nonumber\\
g(t) &= \frac{-4\,t^{3} + 45\,t^{2} - 33\,t + 10}{12\,(t-1)^{3}} -
\frac{3\,t^{3}\,{\ln}(t)}{2\,(t-1)^{4}}\,, \nonumber\\
h(t) &= \frac{t^{2} + t + 4}{2\,(t-1)^{2}} -
\frac{3t\,{\ln}(t)}{(t-1)^{3}}\,, \nonumber \\
j(t) &= \frac{t^{2} - 11\,t + 4}{2\,(t-1)^2} +
\frac{3\,t^{2}\,{\ln}(t)}{(t-1)^{3}}\,.
\end{align}

\subsection{Three body decays $\ell \to \ell' \ell' \ell'$}
At the loop level, three body decays can receive contributions from
photon penguins (dipole and off-shell ``anapole''), $Z$ penguins
and box diagrams, arising from flavour violating interactions
involving the vector leptoquark $V_1$ and quarks.
The relevant low-energy effective Lagrangian inducing
the four-fermion operators responsible for $\ell \to \ell' \ell'
\ell'$ decays can be written as~\cite{Okada:1999zk,Kuno:1999jp}
\begin{eqnarray}\label{eq:lto3l}
\mathcal{L}_{\ell \to \ell' \ell' \ell'} &=&
-\frac{4\,G_F}{\sqrt{2}} \left[
g_1 \,(\bar {\ell'}\,P_L\, \ell) (\bar {\ell'}
\,P_L\, \ell')\,+\,g_2 \,(\bar {\ell'} \,P_R\, \ell) (\bar {\ell'}\,
P_R \,\ell') \right. \,+\nonumber\\
&& \left.\,+\,g_3 \,(\bar {\ell'} \,\gamma^\mu \,P_R \,\ell) (\bar {\ell'}
\, \gamma_\mu \,P_R \,\ell')\,+\,
g_4\, (\bar {\ell'} \,\gamma^\mu \,P_L \,\ell) (\bar {\ell'}
\,\gamma_\mu \,P_L\, \ell') \,+\,\right. \nonumber\\
&& \left. \,+\, g_5\, (\bar {\ell'} \,\gamma^\mu \,P_R\, \ell) (\bar {\ell'}
\, \gamma_\mu \,P_L \,\ell')
\,+\, g_6 \,(\bar {\ell'} \,\gamma^\mu \,P_L \,\ell) (\bar {\ell'}
\,  \gamma_\mu \,P_R\, \ell') \right]\, +\,
\text{H.c.}\,,
\end{eqnarray}
to which the photonic dipole terms entering in
$\mathcal{L}^{\ell_i \to \ell_j \gamma}_\text{eff}$,
cf. Eq.~(\ref{eqn:radeff}), must be added; the corresponding
coefficients parametrised by
$C_{L(R)}^{\ell_i\ell_j}$ have already been
discussed in detail in the previous subsection.
Neglecting
Higgs-mediated exchanges, the off-shell anapole photon penguins, $Z$
penguins and box diagrams will give rise to non-vanishing
contributions to the above $g_3$, $g_4$, $g_5$ and $g_6$
coefficients. Note that in the large
$V_1$ mass limit, the off-shell anapole photon-penguin diagrams scale
proportionally to $ |K|^2 \ln(m_q^2/M^2)/M^2$, in contrast with the
contributions from the $Z$-penguins and box diagrams, which are (na\"ively)
proportional to $ |K|^2 m_q^2/M^4$ and  $|K|^4 m_q^2/M^4$
respectively~\cite{Gabrielli:2000te}.
Therefore we only include in our computation the
log-enhanced photonic anapole contributions, in addition to the dipole
ones. Neglecting right-handed couplings of
the leptoquark as before, the only non-vanishing coefficients (at
one-loop) are $g_4 = g_6$.
The relevant amplitude for the anapole contribution can
be written as
\begin{equation}
  i\,\mathcal M_{\text{anapole}} \,= \,i\,e \,\epsilon_\mu^{{\ast}}
  \,M_{\text{anapole}}^{\mu}\,,
\end{equation}
where $M^\mu_{\text{anapole}}$
can be parametrised in terms of a form factor $F^{\gamma \ell\ell^\prime}_L$ as
\begin{equation}
  \mathcal M_\text{anapole}^\mu\, = \,\frac{1}{(4\pi)^2} \,
  F^{\gamma \ell\ell^\prime}_L\bar\ell^\prime\left(\gamma^\mu \,q^2 - \slashed q \,
  q^\mu\right) P_L\ell\,,
\end{equation}
with $q$ the off-shell photon momentum.
In this convention the $F^{\gamma \ell\ell^\prime}_L$
form factor is independent of $q^2$.
After performing the calculation in the Feynman gauge,
we obtain (in the limit of vanishing external lepton masses)
\begin{equation}
  F^{\gamma \ell\ell^\prime}_L \,=\,
  \frac{N_c}{m^2_V}\sum_i K_L^{i\ell}\,K_L^{i\ell^\prime \ast} \,
  f_a(x_i)\,,
\end{equation}
in which $x_i = m_{d_i}^2/m_V^2$ and $N_c$ is the colour factor
(corresponding to the coloured fields entering in the loop).
Finally, the loop function $f_a(x)$ is given by
\begin{equation}
  f_a(x) \,= \,\frac{4 - 26 \,x + 15 \,x^2 + x^3}{12\,(1-x)^3}
  + \frac{4 - 16\,x - 15\,x^2 + 20\,x^3 - 2\,x^4}{18\,(1-x)^4}\ln(x)\,.
\end{equation}
The leptoquark-induced contributions to the 4-fermion operators are given by
\begin{equation}
  g_4 \,= \,g_6\, = \,
  -\frac{\sqrt{2}}{4 \,G_F}\,\frac{\alpha}{4\,\pi}
  \,Q_f \,F^{\gamma \ell\ell^\prime}_L\,.
\end{equation}
In the case of the $\ell \to 3\ell'$ decays, $Q_f=Q_\ell'$ denotes
the charge of the fermion pair at the end of the off-shell photon
(in units of $e$).
As an example, for the case of $\mu \to 3 e$ decays, one obtains the
following branching ratio~\cite{Okada:1999zk,Kuno:1999jp}
\begin{eqnarray}
\text{BR}(\mu \to e e e)&=&
2\,\left(|g_3|^2\,+\,|g_4|^2\right)
\,+\,|g_5|^2+|g_6|^2\,+\nonumber\\
&&+ 8\,e\, \text{Re}\left[C^{\mu e}_R\,
\left(2g_4^*\,+\,g_6^*\right)\,+\,C^{\mu e}_L \,
\left(2g_3^*\,+\,g_5^*\right)\right]\,+\nonumber\\
&&+ \frac{32\,e^2}{m_{\mu}^2}\,
\left\{\ln\frac{m_\mu^2}{m_e^2}\,-\,
\frac{11}{4}\right\}(\left|C_{R}^{\mu e}\right|^2\,+\,
\left|C_{L}^{\mu e}\right|^2)\,.
 \end{eqnarray}
Similar expressions can be easily inferred for the other cLFV 3-body
decay channels.

\subsection{Neutrinoless $\mu-e$ conversion}
In terms of the relevant effective Wilson coefficients,
the general contribution to the neutrinoless $\mu-e$ conversion
rate is given by~\cite{Dorsner:2016wpm}
\begin{align}
  \Gamma_{\mu -e, \text{N}} \,= \,2\, G_F^{2}
  \,\Big(\,\Big|&\frac{C_R^{\mu e
            \ast}}{m_\mu}\,D + \left(2 \,g_{LV}^{(u)} +
        g_{LV}^{(d)}\right)V^{(p)} + \left(g_{LV}^{(u)} +
        2\,g_{LV}^{(d)} \right)V^{(n)} \nonumber\\
	&+ (G_S^{(u,p)}\,g_{LS}^{(u)} +G_S^{(d,p)}\,g_{LS}^{(d)} +
        G_S^{(s,p)}\,g_{LS}^{(s)})\,S^{(p)} \nonumber \\
	&+ (G_S^{(u,n)}\,g_{LS}^{(u)} +G_S^{(d,n)}\,g_{LS}^{(d)} +
        G_S^{(s,n)}\,g_{LS}^{(s)})\,S^{(n)} \Big|^{2} + (L\leftrightarrow
        R)\Big)\,,
\end{align}
in which the photonic dipole Wilson coefficients
$C_{L(R)}^{\ell_i\ell_j}$ have been given in Eq.~(\ref{eqn:radeff});
the other non-vanishing coefficients,
induced by the tree-level leptoquark exchange or arising from the
photonic anapole contributions, are given by
\begin{align}
g_{LV}^{(d)} &= \frac{\sqrt{2}}{G_F}\left(\frac{1}{\,m_V^2}  \,K_L^{d e} \,
K_L^{d\mu\ast} + \frac{\alpha}{4 \,\pi}  \,Q_d  \,
F^{\gamma \mu e}_L \right) \nonumber\\
g_{LV}^{(u)} &= \frac{\sqrt{2}}{G_F}\left(\frac{\alpha}{4 \,\pi} \, Q_u \, F^{\gamma \mu e}_L
\right)\nonumber\\
g_{RV}^{(d)} &= \frac{\sqrt{2}}{G_F}\left(\frac{\alpha}{4 \,\pi} \, Q_d  \,F^{\gamma \mu e}_L
\right)
\nonumber\\
g_{RV}^{(u)} &= \frac{\sqrt{2}}{G_F}\left(\frac{\alpha}{4 \,\pi}  \,Q_u  \,
F^{\gamma \mu e}_L \right)\,,
\end{align}
with $Q_d = -\frac{1}{3}$ and $Q_u = \frac{2}{3}$.
The values for the overlap integrals ($D, V, S$) are given in
Table~\ref{TabTiAuData}~\cite{Kitano:2002mt},
and the scalar coefficients $G_S^{(d_i,N)}$ can be found
in~\cite{Kosmas:2001mv}.
We again emphasise here that the off-shell anapole contributions,
often neglected in the literature, can have a
contribution comparable to the tree-level leptoquark exchange,
and therefore should be included for a thorough estimation
of the rate of $\mu-e$ conversion in nuclei.
\begin{table}
{\small
  \begin{center}
  \begin{tabular}{lccc} \hline
    Nucleus & $D [m_\mu^{5/2}]$
    & $V^{(p)} [m_\mu^{5/2}]$ & $V^{(n)} [m_\mu^{5/2}] $\\ \hline
    $\text{\text{Ti}}^{48}_{22}$    & 0.0864 & 0.0396 & 0.0468  \\
    $\text{\text{Au}}^{197}_{79}$ &  0.189 & 0.0974 & 0.146 \\
   \hline
& $S^{(p)} [m_\mu^{5/2}]$ & $S^{(n)} [m_\mu^{5/2}] $ &
   $\Gamma_{\text{capture}}[10^6\:\mathrm{s}^{-1}]$ \\
  	\hline
 $\text{\text{Ti}}^{48}_{22}$ &   0.0368 & 0.0435 & 2.59\\
 $\text{\text{Au}}^{197}_{79}$& 0.0614 &  0.0918  & 13.07\\
\hline
\end{tabular}
\end{center}
}
\caption{Overlap integrals $D, V, S$ and $\Gamma_\text{capture}$
  for Gold and Titanium nuclei, as reported in~\cite{Kitano:2002mt}
  (tables I and VIII).}
\label{TabTiAuData}
\end{table}

\section{Electroweak precision observables}\label{app:ew}
As mentioned in the main body of the paper,
the hypothetical  heavy vector-like fermions
can modify the couplings of SM
fermions to gauge bosons\footnote{As before, parallels can be drawn
  with respect to  the case in which heavy neutral leptons, with non-negligible
  mixings to the SM neutrinos, are added to the SM field
  content.}
leading to deviations from the theoretical predictions of the SM, which are
mostly in remarkable agreement with EW precision data.

\subsection{Couplings of the $Z$ boson and photon}

If the heavy vector-like fermion states are $SU(2)_L$ singlets,
mixings with the light $SU(2)_L$ doublets can lead to
modified couplings of the latter to the $Z$ boson ($\bar f f Z$).
For the case of charged leptons,
the relevant couplings can be obtained from the kinetic terms,
\begin{eqnarray}
\mathcal{L}_\text{kin} \,\supset \,
\bar \ell^{0}_{La} \,i \slashed D_a \, \ell^{0}_{La}  \,+ \,
\bar \ell^{0}_{Ra} \,i \slashed D_a \, \ell^{0}_{Ra}   \, =  \,
\bar { \ell}_{L j}  \, (U^\ell_L)^\dagger_{j a} \, i \slashed D_a \,
(U_L^\ell)_{ ak} \,  \ell_{L k}  \,+ \,
\bar {\ell}_{R j} \, (U^\ell_R)^\dagger_{j a} \, i \slashed D_a \,
(U_R^\ell)_{a k} \,  \ell_{R k} \,,
\label{eq:kinE}
\end{eqnarray}
where $\{j, k\}$ denotes the physical fields and
$\{a,b\} = 1 ... 6$ the interaction
states (with $ a \in \{1,2,3\}$ corresponding to SM fermions, and
$a \in \{4,5,6\}$ to the new heavy vector-like states).
The covariant derivative associated with the charges of a given state
$a$ can thus be written
\begin{eqnarray}
D_{\mu,\, a} \,= \, \partial _\mu \,-  \,
i \frac{g}{\cos \theta_W} \, \left(T^3_a  \,- \, \sin ^2 \theta_W  \,
Q_a \right) \, Z_\mu  \,-  \,i  \,e  \,Q_a  \,A_\mu \,,
\end{eqnarray}
where $T^3$ and $Q$
respectively denote the weak isospin and the electric charge.
Since the electric charge is the same for all lepton states ($Q_a=-1$),
the couplings of the photon are not modified.

Let us now introduce the ``effective'' $Z$ boson couplings,
\begin{align}\label{eq:gZff:eff}
&(g^{Z\ell_j \ell_k}_L)_\text{eff}\,=\,
   \sum_{a=1}^6 \frac{g}{\cos \theta_W} \left(
   T^3_{L\,a}  \,- \, \sin ^2 \theta_W\, Q_a \right) \,
   (U^\ell_L)^\dagger_{j a} (U^\ell_L)_{a k}\,,
\nonumber \\
&(g^{Z\ell_j \ell_k}_R)_\text{eff}\,=\,
   \sum_{a=1}^6 \frac{g}{\cos \theta_W}\left(
   T^3_{R\,a}  \,- \, \sin ^2 \theta_W\, Q_a \right) \,
   (U^\ell_R)^\dagger_{j a} (U^\ell_R)_{a k}\,,
\end{align}
where $T^3_{L(R)}$ is the weak isospin of a left-handed
(right-handed) lepton.  Should the SM fermions and the heavy states
belong to the same $SU(2)_L$ representation, universality is trivially
restored (by unitarity) for both $g^{Z\ell_j \ell_k}_{L,R}$ effective
couplings, and one recovers the SM universal couplings. For heavy
isodoublet vector-like states, one has
\begin{equation}
(g^{Z\ell_j \ell_k}_L)_\text{eff}\,=\,\frac{g}{\cos \theta_W} \left(
-\frac{1}{2}\,+ \, \sin ^2 \theta_W\, \right) \, \delta_{jk}\,\text,
\end{equation}
However, if the new fields transform differently (have distinct charges)
under $SU(2)_L$,
the $g^{Z\ell_j \ell_k}_{L,R}$ couplings are modified. In
particular, in the presence of {\it isosinglet heavy states}, one finds
\begin{equation}\label{eq:gZff:eff:singlets}
(g^{Z\ell_j \ell_k}_L)_\text{eff}\,=\,
\,\frac{g}{\cos \theta_W} \left(
-\frac{1}{2}\,+ \, \sin ^2 \theta_W\, \right) \, \delta_{jk}
\,+\, \Delta g^{jk}_L\,,
\quad \quad
\text{with} \quad
\Delta g^{jk}_L\, = \,
\sum_{a=4}^6\,  \frac{1}{2}\,\frac{g}{\cos \theta_W}
   (U^\ell_L)^\dagger_{j a} (U^\ell_L)_{a k}\,.
\end{equation}
Likewise, vector-like doublets also lead to the modification of
the $g^{Z\ell_j \ell_k}_{R}$ couplings:
\begin{equation}\label{eq:gZff:eff:doublets}
(g^{Z\ell_j \ell_k}_R)_\text{eff}\,=\,
\,\frac{g}{\cos \theta_W} \sin ^2 \theta_W \, \delta_{jk}
\,+\, \Delta g^{jk}_R\,,
\quad \quad
\text{with} \quad
\Delta g^{jk}_R\, = \,
\sum_{a=4}^6  \,-\frac{1}{2}\,\frac{g}{\cos \theta_W}
   (U^\ell_R)^\dagger_{j a} (U^\ell_R)_{a k}\,.
\end{equation}

\subsection{Couplings of the $W$ boson}

The possible mixings with the heavy vector-like leptons can also
modify the couplings to the $W$ boson. The charged current interaction
terms can be written
\begin{align}
\mathcal{L}^\text{cc} \, &=\,
\frac{g}{\sqrt{2}} \,W_\mu  \, \bar {\nu}^0_{a} \,\gamma^\mu\,
\,\ell^0_{L a}\, + \,
\text{H.c.} \,,\nonumber \\
&=\,
\frac{g}{\sqrt{2}} \,W_\mu  \, \bar {\nu}_{j} \,\gamma^\mu\,
(U^{\nu\dagger}_L)_{ja} \,(U_L)_{ak}\,  \ell_{L k}  \,+\,
\text{H.c.} \,\text,
\end{align}
so that the corresponding charged current couplings are then given by

\begin{equation}
g^{W\nu_j \ell_k}_{L} \,= \,
\frac{g}{\sqrt{2}}\, (U^{\nu\dagger}_L)_{ja} \,(U^{\ell}_L)_{ak}
\,=\,
\frac{g}{\sqrt{2}}\,U^{\mathrm P\dagger}_{jk}\,,
\end{equation}
where
$U^{\mathrm P}$ denotes the (generalised) PMNS mixing matrix.
A priori, the branching ratios of $W\rightarrow \ell \nu$ can
constrain the mixings of the heavy leptons (see,
e.g.~\cite{Dermisek:2013gta,Poh:2017tfo}). However, these
strongly depend on the neutrino mass generation mechanism (as well
as on the structure of the Higgs sector), and in the present analysis we
will not take them into account;
we nevertheless mention that
for a given Higgs sector,
in the presence of additional isosinglet heavy neutrinos,
the modified charged current vertex can impact
several observables. Therefore, in addition to electroweak
precision measurements of
$\mathrm{BR}(W\rightarrow \ell\nu)$~\cite{Abada:2013aba},
other decays or collider processes with one or
two neutrinos in the final state, as for example
$\tau$ decays, leptonic and semileptonic meson
decays~\cite{Atre:2009rg,Abada:2017jjx,Abada:2019bac}, production and
decay of $W$ bosons to dilepton and two jets at
the LHC~\cite{Bray:2007ru,Dev:2019rxh}, can also lead to interesting
constraints depending on the mass scales of the additional isosinglet
neutral states.

\subsection{Constraining EWP observables}

Due to the tree-level modified $Z$-couplings (a consequence of the mixing
of SM fermions with the heavy vector-like fermions),
strong constraints from EWP observables are expected to arise
from the observed lepton universality in charged
leptonic $Z$-decays.

At tree level, the decay width of a massive vector boson to fermions is
given by \cite{Abada:2013aba}
\begin{align}\label{eqn:Vff}
&\Gamma(V\to f f^\prime) \,= \, \frac{\lambda^{1/2}(m_V, m_f,
  m_{f^\prime})}{48 \pi m_V^3} \nonumber\\
&\times \left[\left(| g_L^{f f^\prime}|^2 + | g_R^{f
      f^\prime}|^2 \right)\left(2m_V^2 - m_f^2 - m_{f^\prime}^2
    - \frac{\left(m_f^2 - m_{f^\prime}^2\right)^2}{m_V^2} \right) + 12
    m_f m_{f^\prime} \,\mathrm{Re} \left(g_L^{f f^\prime} g_R^{f
      f^\prime \ast}\right) \right]\,,
\end{align}
where $g_{L(R)}$ are the chiral couplings and the Käll\'en function is
defined in Eq. \eqref{eq:kallenlambda}.
In the case of the $Z$-boson, the relevant couplings have been introduced
in Eq.~(\ref{eq:gZff:eff}). From Eq.~(\ref{eqn:Vff}), and in view of
the very good agreement between the SM predictions and experiment (cf.
Table~\ref{tab:EWPO}), it is clear that any modification of the
tree-level couplings of the $Z$-boson will be subject to very
stringent constraints, which in turn translate into bounds on
the mixing parameters responsible for the $\Delta g_{L(R)}$ terms
(see Eqs.~(\ref{eq:gZff:eff:singlets}, \ref{eq:gZff:eff:doublets})).
Using Eq.~(\ref{eqn:Vff}), one can derive conservative
constraints on $\Delta g_{L(R)}$ from the requirement of compatibility
with the bounds of Table~\ref{tab:EWPO}. As an example, the current
experimental data on $\text{BR}(Z\to e^\pm\mu^\mp)$ and
$\Gamma(\ell^+\ell^-)$ leads to
\begin{align}\label{eq:DeltagLgR}
  |\Delta g^{e\mu}_L |^2 + |\Delta g^{e\mu}_R|^2
  &\lesssim 1.55\times 10^{-6}, \nonumber\\
|\Delta g^{\ell\ell}_L | & \lesssim 5.6 \times 10^{-4},\nonumber\\
|\Delta g^{\ell\ell}_R | & \lesssim 3.5 \times 10^{-4}\,.
\end{align}
The above constraints allow in turn to infer bounds on the elements of
the matrix $A$ (see Eqs.~(\ref{eq:UL:ARBS} - \ref{eq:Aalphaij})).
To illustrate this point, we consider the case of {\it isosinglet heavy
vector-like leptons} (for which $\Delta g_R=0$)\footnote{Note that in
  the presence of a nontrivial Higgs sector inducing mixings between
  right-handed SM charged leptons and vector-like doublets, one has
  non-vanishing contributions to $ \Delta g_R$, which can
  lead to constraints on the right-handed mixing matrix, parametrised as done for $K_L$, see Eq.~(\ref{eq:Ndescopm_C1}), for a given UV complete
  framework. However, a detailed analysis of such a scenario is beyond
  the scope of our current work. Here, we only include the
  conservative limits for left-handed mixing elements, which are of
  foremost importance to our analysis.}: the limits of
Eq.~(\ref{eq:DeltagLgR}) translate into
\begin{align}
1-|\alpha_{11}| &\lesssim 4 \times 10^{-4}\,, \nonumber\\
1-|\alpha_{22}| &\lesssim 3 \times 10^{-4}\,, \nonumber\\
|\alpha_{21}| &\lesssim 4.6 \times 10^{-4}\, .
\end{align}
\section{Details of the numerical analysis}
In what follows, we detail relevant aspects of the numerical analysis done in Sections \ref{sec:framework} and \ref{sec:pheno:constraints}, in particular concerning the global fits and the scans over the parameter space leading to the different plots.
The global fits displayed in Figures \ref{fig:RK_global} and \ref{fig:RK_RD_global} are obtained using the ``fastfit'' method of the \texttt{flavio} package\cite{Straub:2018kue}. This is based on the approximation which assumes the likelihood to be of the form $\mathcal{L}=e^{-\chi^{2}(\vec{\xi}) / 2}$ where
\begin{equation}
	\chi^{2}(\vec{\xi})=\vec{\Delta}^{T} C^{-1}(\vec{\xi}=\hat{\vec{\xi}}) \vec{\Delta}\, , \quad \Delta_{i}=\left(x^{\mathrm{exp}}-x_{i}^{\mathrm{th}}(\vec{\theta})\right)\,\text,
	\end{equation}
and $C(\vec{\xi})=C_{\exp }+C_{\operatorname{th}}(\vec{\xi})$ being the combined (theoretical and experimental) covariance matrix of the observables $x_i$ and the theoretical and experimental uncertainties are approximated as Gaussian. For a more detailed description of the statistical treatment, we refer the reader to Ref.\cite{Aebischer:2018iyb}. Likelihood contours are obtained by calculating the $\mathrm\Delta\chi^2$ deviation around the best-fit point.

\medskip To obtain the various scatter plots, we have varied all mixing parameters of the parametrisation as stated in the captions. The mass of the vector leptoquark is varied in the ranges described in the captions of the figures (or set to a fixed value).
The mixing angles were varied between $-\pi$ and $\pi$.
To overcome the difficulties associated with presenting a parameter space spanned by $12$ mixing angles in the nonunitary parametrisation, we perform a random scan using $\sim 10^{10}$ $n$-tuples of random numbers and compute the predictions of the model for a large number of flavour violating observables, including those shown in Tables \ref{tab:flavour} and \ref{tab:cLFV}.
If any of the predictions for a certain parameter $n$-tuple exceeds the experimental bounds in a given set of constraints, the point is filtered in the appropriate category.
Points complying with the imposed constraints are further processed to calculate their global likelihood regarding the $b\to s\ell\ell$ and $b\to c\ell\nu$ observables using \texttt{flavio}\cite{Straub:2018kue}. Relevant effects due to renormalisation group running, in all the relevant processes (which we have used as constraints or for fitting procedures in our analysis) were computed with the \texttt{wilson} package~\cite{Aebischer:2018bkb} in association with the \texttt{flavio} package~\cite{Straub:2018kue}.

\end{appendix}


\begin{thebibliography}{99}

%%%%%%%%%%% Introduction %%%%%%%%%%%%%
%EWPO

%\cite{Tanabashi:2018oca}
\bibitem{Tanabashi:2018oca}
  M.~Tanabashi {\it et al.} [Particle Data Group],
  %``Review of Particle Physics,''
  Phys.\ Rev.\ D {\bf 98} (2018) no.3,  030001.
%  doi:10.1103/PhysRevD.98.030001
  %%CITATION = doi:10.1103/PhysRevD.98.030001;%%
  %1806 citations counted in INSPIRE as of 13 Jun 2019
%\cite{ALEPH:2005ab}
\bibitem{ALEPH:2005ab}
  S.~Schael {\it et al.} [ALEPH and DELPHI and L3 and OPAL and SLD Collaborations and LEP Electroweak Working Group and SLD Electroweak Group and SLD Heavy Flavour Group],
  %``Precision electroweak measurements on the $Z$ resonance,''
  Phys.\ Rept.\  {\bf 427} (2006) 257
  % doi:10.1016/j.physrep.2005.12.006
  [hep-ex/0509008].
  %%CITATION = doi:10.1016/j.physrep.2005.12.006;%%
  %2011 citations counted in INSPIRE as of 26 Jun 2019


  %%%%%%%%.  RK-RD citations %%%%%%%%%%%%

%%%%%%%% EXP %%%%%%%%%%%%
%\cite{Lees:2012xj,Lees:2013uzd, Huschle:2015rga,Adachi:2009qg, Bozek:2010xy, Aaij:2015yra,Hirose:2016wfn,Abdesselam:2019dgh}

%\cite{Lees:2012xj}
\bibitem{Lees:2012xj}
  J.~P.~Lees {\it et al.} [BaBar Collaboration],
  %``Evidence for an excess of $\bar{B} \to D^{(*)} \tau^-\bar{\nu}_\tau$ decays,''
  Phys.\ Rev.\ Lett.\  {\bf 109} (2012) 101802
  % doi:10.1103/PhysRevLett.109.101802
  [arXiv:1205.5442 [hep-ex]].
  %%CITATION = doi:10.1103/PhysRevLett.109.101802;%%
  %683 citations counted in INSPIRE as of 26 Jun 2019

%\cite{Lees:2013uzd}
\bibitem{Lees:2013uzd}
  J.~P.~Lees {\it et al.} [BaBar Collaboration],
  %``Measurement of an Excess of $\bar{B} \to D^{(*)}\tau^- \bar{\nu}_\tau$ Decays and Implications for Charged Higgs Bosons,''
  Phys.\ Rev.\ D {\bf 88} (2013) no.7,  072012
  % doi:10.1103/PhysRevD.88.072012
  [arXiv:1303.0571 [hep-ex]].
  %%CITATION = doi:10.1103/PhysRevD.88.072012;%%
  %541 citations counted in INSPIRE as of 26 Jun 2019

%\cite{Huschle:2015rga}
\bibitem{Huschle:2015rga}
  M.~Huschle {\it et al.} [Belle Collaboration],
  %``Measurement of the branching ratio of $\bar{B} \to D^{(\ast)} \tau^- \bar{\nu}_\tau$ relative to $\bar{B} \to D^{(\ast)} \ell^- \bar{\nu}_\ell$ decays with hadronic tagging at Belle,''
  Phys.\ Rev.\ D {\bf 92} (2015) no.7,  072014
  % doi:10.1103/PhysRevD.92.072014
  [arXiv:1507.03233 [hep-ex]].
  %%CITATION = doi:10.1103/PhysRevD.92.072014;%%
  %496 citations counted in INSPIRE as of 26 Jun 2019

%\cite{Adachi:2009qg}
\bibitem{Adachi:2009qg}
  I.~Adachi {\it et al.} [Belle Collaboration],
  %``Measurement of B ---> D(*) tau nu using full reconstruction tags,''
  arXiv:0910.4301 [hep-ex].
  %%CITATION = ARXIV:0910.4301;%%
  %82 citations counted in INSPIRE as of 26 Jun 2019

%\cite{Bozek:2010xy}
\bibitem{Bozek:2010xy}
  A.~Bozek {\it et al.} [Belle Collaboration],
  %``Observation of B+ -> Dbar*0 tau+ nu_tau and Evidence for B+ -> Dbar^0 tau+ nu_tau at Belle,''
  Phys.\ Rev.\ D {\bf 82} (2010) 072005
  % doi:10.1103/PhysRevD.82.072005
  [arXiv:1005.2302 [hep-ex]].
  %%CITATION = doi:10.1103/PhysRevD.82.072005;%%
  %190 citations counted in INSPIRE as of 26 Jun 2019

%\cite{Aaij:2015yra}
\bibitem{Aaij:2015yra}
  R.~Aaij {\it et al.} [LHCb Collaboration],
  %``Measurement of the ratio of branching fractions $\mathcal{B}(\bar{B}^0 \to D^{*+}\tau^{-}\bar{\nu}_{\tau})/\mathcal{B}(\bar{B}^0 \to D^{*+}\mu^{-}\bar{\nu}_{\mu})$,''
  Phys.\ Rev.\ Lett.\  {\bf 115} (2015) no.11,  111803
   Erratum: [Phys.\ Rev.\ Lett.\  {\bf 115} (2015) no.15,  159901]
  % doi:10.1103/PhysRevLett.115.159901, 10.1103/PhysRevLett.115.111803
  [arXiv:1506.08614 [hep-ex]].
  %%CITATION = doi:10.1103/PhysRevLett.115.159901, 10.1103/PhysRevLett.115.111803;%%
  %602 citations counted in INSPIRE as of 26 Jun 2019

%\cite{Hirose:2016wfn}
\bibitem{Hirose:2016wfn}
  S.~Hirose {\it et al.} [Belle Collaboration],
  %``Measurement of the $\tau$ lepton polarization and $R(D^*)$ in the decay $\bar{B} \to D^* \tau^- \bar{\nu}_\tau$,''
  Phys.\ Rev.\ Lett.\  {\bf 118} (2017) no.21,  211801
  % doi:10.1103/PhysRevLett.118.211801
  [arXiv:1612.00529 [hep-ex]].
  %%CITATION = doi:10.1103/PhysRevLett.118.211801;%%
  %266 citations counted in INSPIRE as of 26 Jun 2019

%\cite{Abdesselam:2019dgh}
\bibitem{Abdesselam:2019dgh}
  A.~Abdesselam {\it et al.} [Belle Collaboration],
  %``Measurement of $\mathcal{R}(D)$ and $\mathcal{R}(D^{\ast})$ with a semileptonic tagging method,''
  arXiv:1904.08794 [hep-ex].
  %%CITATION = ARXIV:1904.08794;%%
  %21 citations counted in INSPIRE as of 26 Jun

%%%  R_K(*) experimental data refs %%%

%\cite{Aaij:2019wad, Aaij:2017vbb, Abdessekam:2019wac, Aaij:2015esa, Wehle:2016yoi}

%\cite{Aaij:2019wad}
\bibitem{Aaij:2019wad}
  R.~Aaij {\it et al.} [LHCb Collaboration],
  %``Search for lepton-universality violation in $B^+\to K^+\ell^+\ell^-$ decays,''
  Phys.\ Rev.\ Lett.\  {\bf 122} (2019) no.19,  191801
%  doi:10.1103/PhysRevLett.122.191801
  [arXiv:1903.09252 [hep-ex]].
  %%CITATION = doi:10.1103/PhysRevLett.122.191801;%%
  %31 citations counted in INSPIRE as of 13 Jun 2019

%\cite{Aaij:2017vbb}
\bibitem{Aaij:2017vbb}
  R.~Aaij {\it et al.} [LHCb Collaboration],
  %``Test of lepton universality with $B^{0} \rightarrow K^{*0}\ell^{+}\ell^{-}$ decays,''
  JHEP {\bf 1708} (2017) 055
%  doi:10.1007/JHEP08(2017)055
  [arXiv:1705.05802 [hep-ex]].
  %%CITATION = doi:10.1007/JHEP08(2017)055;%%
  %420 citations counted in INSPIRE as of 13 Jun 2019

%\cite{Abdesselam:2019wac}
\bibitem{Abdesselam:2019wac}
  A.~Abdesselam {\it et al.} [Belle Collaboration],
  %``Test of lepton flavor universality in ${B\to K^\ast\ell^+\ell^-}$ decays at Belle,''
  arXiv:1904.02440 [hep-ex].
  %%CITATION = ARXIV:1904.02440;%%
  %12 citations counted in INSPIRE as of 13 Jun 2019

%\cite{Aaij:2015esa}
\bibitem{Aaij:2015esa}
  R.~Aaij {\it et al.} [LHCb Collaboration],
  %``Angular analysis and differential branching fraction of the decay $B^0_s\to\phi\mu^+\mu^-$,''
  JHEP {\bf 1509} (2015) 179
%  doi:10.1007/JHEP09(2015)179
  [arXiv:1506.08777 [hep-ex]].
  %%CITATION = doi:10.1007/JHEP09(2015)179;%%
  %257 citations counted in INSPIRE as of 13 Jun 2019

%\cite{Wehle:2016yoi}
\bibitem{Wehle:2016yoi}
  S.~Wehle {\it et al.} [Belle Collaboration],
  %``Lepton-Flavor-Dependent Angular Analysis of $B\to K^\ast \ell^+\ell^-$,''
  Phys.\ Rev.\ Lett.\  {\bf 118} (2017) no.11,  111801
%  doi:10.1103/PhysRevLett.118.111801
  [arXiv:1612.05014 [hep-ex]].
  %%CITATION = doi:10.1103/PhysRevLett.118.111801;%%
  %202 citations counted in INSPIRE as of 13 Jun 2019

%%%%%%%%%% HFLAV with update %%%%%%%%%%
%\cite{Amhis:2016xyh}
\bibitem{Amhis:2016xyh}
  Y.~Amhis {\it et al.} [HFLAV Collaboration],
  %``Averages of $b$-hadron, $c$-hadron, and $\tau$-lepton properties as of summer 2016,''
  Eur.\ Phys.\ J.\ C {\bf 77} (2017) no.12,  895
  % doi:10.1140/epjc/s10052-017-5058-4
  [arXiv:1612.07233 [hep-ex]].
  Results Spring 2019: \href{https://hflav-eos.web.cern.ch/hflav-eos/semi/spring19/html/RDsDsstar/RDRDs.html}{{\texttt{https://hflav.web.cern.ch/}}}
  %%CITATION = doi:10.1140/epjc/s10052-017-5058-4;%%
  %636 citations counted in INSPIRE as of 26 Jun 2019


%%%%%%%%% SM predictions %%%%%%%%%%%%%%%
%\cite{Bigi:2016mdz,Bigi:2017jbd}
%\cite{Ligeti:2016npd,Crivellin:2016ejn}
%\cite{Bordone:2016gaq,Capdevila:2017bsm}

%\cite{Bigi:2016mdz}
\bibitem{Bigi:2016mdz}
  D.~Bigi and P.~Gambino,
  %``Revisiting $B\to D \ell \nu$,''
  Phys.\ Rev.\ D {\bf 94} (2016) no.9,  094008
  % doi:10.1103/PhysRevD.94.094008
  [arXiv:1606.08030 [hep-ph]].
  %%CITATION = doi:10.1103/PhysRevD.94.094008;%%
  %122 citations counted in INSPIRE as of 26 Jun 2019

  %\cite{Bigi:2017jbd}
  \bibitem{Bigi:2017jbd}
    D.~Bigi, P.~Gambino and S.~Schacht,
    %``$R(D^*)$, $|V_{cb}|$, and the Heavy Quark Symmetry relations between form factors,''
    JHEP {\bf 1711} (2017) 061
%    doi:10.1007/JHEP11(2017)061
    [arXiv:1707.09509 [hep-ph]].
    %%CITATION = doi:10.1007/JHEP11(2017)061;%%
    %113 citations counted in INSPIRE as of 13 Jun 2019

%\cite{Ligeti:2016npd}
\bibitem{Ligeti:2016npd}
  Z.~Ligeti, M.~Papucci and D.~J.~Robinson,
  %``New Physics in the Visible Final States of $B\to D^{(*)}\tau\nu$,''
  JHEP {\bf 1701} (2017) 083
%  doi:10.1007/JHEP01(2017)083
  [arXiv:1610.02045 [hep-ph]].
  %%CITATION = doi:10.1007/JHEP01(2017)083;%%
  %38 citations counted in INSPIRE as of 13 Jun 2019

%\cite{Crivellin:2016ejn}
\bibitem{Crivellin:2016ejn}
  A.~Crivellin, J.~Fuentes-Martin, A.~Greljo and G.~Isidori,
  %``Lepton Flavor Non-Universality in B decays from Dynamical Yukawas,''
  Phys.\ Lett.\ B {\bf 766} (2017) 77
  % doi:10.1016/j.physletb.2016.12.057
  [arXiv:1611.02703 [hep-ph]].
  %%CITATION = doi:10.1016/j.physletb.2016.12.057;%%
  %65 citations counted in INSPIRE as of 26 Jun 2019

%\cite{Bordone:2016gaq}
\bibitem{Bordone:2016gaq}
  M.~Bordone, G.~Isidori and A.~Pattori,
  %``On the Standard Model predictions for $R_K$ and $R_{K^*}$,''
  Eur.\ Phys.\ J.\ C {\bf 76} (2016) no.8,  440
  % doi:10.1140/epjc/s10052-016-4274-7
  [arXiv:1605.07633 [hep-ph]].
  %%CITATION = doi:10.1140/epjc/s10052-016-4274-7;%%
  %205 citations counted in INSPIRE as of 26 Jun 2019

%\cite{Capdevila:2017bsm}
\bibitem{Capdevila:2017bsm}
  B.~Capdevila, A.~Crivellin, S.~Descotes-Genon, J.~Matias and J.~Virto,
  %``Patterns of New Physics in $b\to s\ell^+\ell^-$ transitions in the light of recent data,''
  JHEP {\bf 1801} (2018) 093
  % doi:10.1007/JHEP01(2018)093
  [arXiv:1704.05340 [hep-ph]].
  %%CITATION = doi:10.1007/JHEP01(2018)093;%%
  %267 citations counted in INSPIRE as of 26 Jun 2019

%%%%%%%%%% Bs->phi mu mu %%%%%%%%%%%%%%
%~\cite{Aaij:2015esa,Altmannshofer:2014rta,Straub:2015ica}

%\cite{Altmannshofer:2014rta}
\bibitem{Altmannshofer:2014rta}
  W.~Altmannshofer and D.~M.~Straub,
  %``New physics in $b\rightarrow s$ transitions after LHC run 1,''
  Eur.\ Phys.\ J.\ C {\bf 75} (2015) no.8,  382
  % doi:10.1140/epjc/s10052-015-3602-7
  [arXiv:1411.3161 [hep-ph]].
  %%CITATION = doi:10.1140/epjc/s10052-015-3602-7;%%
  %310 citations counted in INSPIRE as of 26 Jun 2019

%\cite{Straub:2015ica}
\bibitem{Straub:2015ica}
  A.~Bharucha, D.~M.~Straub and R.~Zwicky,
  %``$B\to V\ell^+\ell^-$ in the Standard Model from light-cone sum rules,''
  JHEP {\bf 1608} (2016) 098
  % doi:10.1007/JHEP08(2016)098
  [arXiv:1503.05534 [hep-ph]].
  %%CITATION = doi:10.1007/JHEP08(2016)098;%%
  %234 citations counted in INSPIRE as of 26 Jun 2019



 %%%%% MODEL INDEPENDENT %%%%%%%%%%%%%%%%%%%
     %\cite{Alguero:2019ptt,Aebischer:2019mlg,Ciuchini:2019usw,Datta:2019zca,Arbey:2019duh,Shi:2019gxi,Bardhan:2019ljo,Ghosh:2014awa,Glashow:2014iga,Bhattacharya:2014wla,Freytsis:2015qca,Ligeti:2016npd,Ciuchini:2017mik,Bigi:2017jbd}


 \bibitem{Alguero:2019ptt}
   M.~Alguer\'{o}, B.~Capdevila, A.~Crivellin, S.~Descotes-Genon, P.~Masjuan, J.~Matias and J.~Virto,
   %``Emerging patterns of New Physics with and without Lepton Flavour Universal contributions,''
   arXiv:1903.09578 [hep-ph].
   %%CITATION = ARXIV:1903.09578;%%
   %13 citations counted in INSPIRE as of 13 Jun 2019

  %\cite{Aebischer:2019mlg}
  \bibitem{Aebischer:2019mlg}
    J.~Aebischer, W.~Altmannshofer, D.~Guadagnoli, M.~Reboud, P.~Stangl and D.~M.~Straub,
    %``$B$-decay discrepancies after Moriond 2019,''
    arXiv:1903.10434 [hep-ph].
    %%CITATION = ARXIV:1903.10434;%%
    %22 citations counted in INSPIRE as of 13 Jun 2019

  %\cite{Ciuchini:2019usw}
  \bibitem{Ciuchini:2019usw}
    M.~Ciuchini, A.~M.~Coutinho, M.~Fedele, E.~Franco, A.~Paul, L.~Silvestrini and M.~Valli,
    %``New Physics in $b \to s \ell^+ \ell^-$ confronts new data on Lepton Universality,''
    arXiv:1903.09632 [hep-ph].
    %%CITATION = ARXIV:1903.09632;%%
    %16 citations counted in INSPIRE as of 13 Jun 2019

%\cite{Datta:2019zca}
\bibitem{Datta:2019zca}
  A.~Datta, J.~Kumar and D.~London,
  %``The $B$ Anomalies and New Physics in $b \to s e^+ e^-$,''
  arXiv:1903.10086 [hep-ph].
  %%CITATION = ARXIV:1903.10086;%%
  %9 citations counted in INSPIRE as of 13 Jun 2019

%\cite{Arbey:2019duh}
\bibitem{Arbey:2019duh}
  A.~Arbey, T.~Hurth, F.~Mahmoudi, D.~M.~Santos and S.~Neshatpour,
  %``Update on the $b \rightarrow s$ anomalies,''
  arXiv:1904.08399 [hep-ph].
  %%CITATION = ARXIV:1904.08399;%%
  %6 citations counted in INSPIRE as of 13 Jun 2019

%\cite{Shi:2019gxi}
\bibitem{Shi:2019gxi}
  R.~X.~Shi, L.~S.~Geng, B.~Grinstein, S.~Jäger and J.~Martin Camalich,
  %``Revisiting the new-physics interpretation of the $b\to c\tau\nu$ data,''
  arXiv:1905.08498 [hep-ph].
  %%CITATION = ARXIV:1905.08498;%%

%\cite{Bardhan:2019ljo}
  \bibitem{Bardhan:2019ljo}
    D.~Bardhan and D.~Ghosh,
    %``B-meson charged current anomalies: the post-Moriond status,''
    arXiv:1904.10432 [hep-ph].
    %%CITATION = ARXIV:1904.10432;%%
    %4 citations counted in INSPIRE as of 13 Jun 2019

  %\cite{Alok:2019ufo}
  \bibitem{Alok:2019ufo}
    A.~K.~Alok, A.~Dighe, S.~Gangal and D.~Kumar,
    %``Continuing search for new physics in $b \to s \mu \mu$ decays: two operators at a time,''
    JHEP {\bf 1906} (2019) 089
  %  doi:10.1007/JHEP06(2019)089
    [arXiv:1903.09617 [hep-ph]].
    %%CITATION = doi:10.1007/JHEP06(2019)089;%%
    %17 citations counted in INSPIRE as of 23 Jul 2019

	%\cite{Alok:2017qsi}
	\bibitem{Alok:2017qsi}
	  A.~K.~Alok, D.~Kumar, J.~Kumar, S.~Kumbhakar and S.~U.~Sankar,
	  %``New physics solutions for $R_D$ and $R_{D^*}$,''
	  JHEP {\bf 1809} (2018) 152
	  % doi:10.1007/JHEP09(2018)152
	  [arXiv:1710.04127 [hep-ph]].
	  %%CITATION = doi:10.1007/JHEP09(2018)152;%%
	  %41 citations counted in INSPIRE as of 23 Jul 2019


\bibitem{Ghosh:2014awa}
  D.~Ghosh, M.~Nardecchia and S.~A.~Renner,
  %``Hint of Lepton Flavour Non-Universality in $B$ Meson Decays,''
  JHEP {\bf 1412}, 131 (2014)
  %doi:10.1007/JHEP12(2014)131
  [arXiv:1408.4097 [hep-ph]].
  %%CITATION = doi:10.1007/JHEP12(2014)131;%%
  %120 citations counted in INSPIRE as of 19 Feb 2018

  \bibitem{Glashow:2014iga}
  S.~L.~Glashow, D.~Guadagnoli and K.~Lane,
  %``Lepton Flavor Violation in $B$ Decays?,''
  Phys.\ Rev.\ Lett.\  {\bf 114}, 091801 (2015)
  %doi:10.1103/PhysRevLett.114.091801
  [arXiv:1411.0565 [hep-ph]].
  %%CITATION = doi:10.1103/PhysRevLett.114.091801;%%
  %132 citations counted in INSPIRE as of 19 Feb 2018

  \bibitem{Bhattacharya:2014wla}
  B.~Bhattacharya, A.~Datta, D.~London and S.~Shivashankara,
  %``Simultaneous Explanation of the $R_K$ and $R(D^{(*)})$ Puzzles,''
  Phys.\ Lett.\ B {\bf 742}, 370 (2015)
  %doi:10.1016/j.physletb.2015.02.011
  [arXiv:1412.7164 [hep-ph]].
  %%CITATION = doi:10.1016/j.physletb.2015.02.011;%%
  %118 citations counted in INSPIRE as of 19 Feb 2018

\bibitem{Freytsis:2015qca}
  M.~Freytsis, Z.~Ligeti and J.~T.~Ruderman,
  %``Flavor models for $\bar{B} \to D^{(*)} \tau \bar{\nu}$,''
  Phys.\ Rev.\ D {\bf 92}, no. 5, 054018 (2015)
  %doi:10.1103/PhysRevD.92.054018
  [arXiv:1506.08896 [hep-ph]].
  %%CITATION = doi:10.1103/PhysRevD.92.054018;%%
  %120 citations counted in INSPIRE as of 19 Feb 2018


  %\cite{Ciuchini:2017mik}
\bibitem{Ciuchini:2017mik}
  M.~Ciuchini, A.~M.~Coutinho, M.~Fedele, E.~Franco, A.~Paul, L.~Silvestrini and M.~Valli,
  %``On Flavourful Easter eggs for New Physics hunger and Lepton Flavour Universality violation,''
  Eur.\ Phys.\ J.\ C {\bf 77}, no. 10, 688 (2017)
%  doi:10.1140/epjc/s10052-017-5270-2
  [arXiv:1704.05447 [hep-ph]].
  %%CITATION = doi:10.1140/epjc/s10052-017-5270-2;%%
  %87 citations counted in INSPIRE as of 10 Jul 2018





 %%%%% ZPRIME %%%%%%%%%%%%%%%%%%%
 %\cite{Altmannshofer:2014cfa,Crivellin:2015mga,Crivellin:2015lwa,Sierra:2015fma,Crivellin:2015era,Celis:2015ara,Bhatia:2017tgo,Kamenik:2017tnu,Chen:2017usq,Camargo-Molina:2018cwu,Darme:2018hqg,Baek:2018aru,Biswas:2019twf,Allanach:2019iiy}

\bibitem{Altmannshofer:2014cfa}
  W.~Altmannshofer, S.~Gori, M.~Pospelov and I.~Yavin,
  %``Quark flavor transitions in $L_\mu-L_\tau$ models,''
  Phys.\ Rev.\ D {\bf 89}, 095033 (2014)
  %doi:10.1103/PhysRevD.89.095033
  [arXiv:1403.1269 [hep-ph]].
  %%CITATION = doi:10.1103/PhysRevD.89.095033;%%
  %212 citations counted in INSPIRE as of 19 Feb 2018

  \bibitem{Crivellin:2015mga}
  A.~Crivellin, G.~D'Ambrosio and J.~Heeck,
  %``Explaining $h\to\mu^\pm\tau^\mp$, $B\to K^* \mu^+\mu^-$ and $B\to K \mu^+\mu^-/B\to K e^+e^-$ in a two-Higgs-doublet model with gauged $L_\mu-L_\tau$,''
  Phys.\ Rev.\ Lett.\  {\bf 114}, 151801 (2015)
  %doi:10.1103/PhysRevLett.114.151801
  [arXiv:1501.00993 [hep-ph]].
  %%CITATION = doi:10.1103/PhysRevLett.114.151801;%%
  %232 citations counted in INSPIRE as of 19 Feb 2018

  \bibitem{Crivellin:2015lwa}
  A.~Crivellin, G.~D'Ambrosio and J.~Heeck,
  %``Addressing the LHC flavor anomalies with horizontal gauge symmetries,''
  Phys.\ Rev.\ D {\bf 91}, no. 7, 075006 (2015)
  %doi:10.1103/PhysRevD.91.075006
  [arXiv:1503.03477 [hep-ph]].
  %%CITATION = doi:10.1103/PhysRevD.91.075006;%%
  %178 citations counted in INSPIRE as of 19 Feb 2018

  \bibitem{Sierra:2015fma}
  D.~Aristizabal Sierra, F.~Staub and A.~Vicente,
  %``Shedding light on the $b\to s$ anomalies with a dark sector,''
  Phys.\ Rev.\ D {\bf 92}, no. 1, 015001 (2015)
  %doi:10.1103/PhysRevD.92.015001
  [arXiv:1503.06077 [hep-ph]].
  %%CITATION = doi:10.1103/PhysRevD.92.015001;%%
  %94 citations counted in INSPIRE as of 19 Feb 2018

  \bibitem{Crivellin:2015era}
  A.~Crivellin, L.~Hofer, J.~Matias, U.~Nierste, S.~Pokorski and J.~Rosiek,
  %``Lepton-flavour violating $B$ decays in generic $Z'$ models,''
  Phys.\ Rev.\ D {\bf 92}, no. 5, 054013 (2015)
  %doi:10.1103/PhysRevD.92.054013
  [arXiv:1504.07928 [hep-ph]].
  %%CITATION = doi:10.1103/PhysRevD.92.054013;%%
  %92 citations counted in INSPIRE as of 19 Feb 2018

\bibitem{Celis:2015ara}
  A.~Celis, J.~Fuentes-Martin, M.~Jung and H.~Serodio,
  %``Family nonuniversal Zâ² models with protected flavor-changing interactions,''
  Phys.\ Rev.\ D {\bf 92}, no. 1, 015007 (2015)
  %doi:10.1103/PhysRevD.92.015007
  [arXiv:1505.03079 [hep-ph]].
  %%CITATION = doi:10.1103/PhysRevD.92.015007;%%
  %83 citations counted in INSPIRE as of 19 Feb 2018


    %\cite{}
\bibitem{Bhatia:2017tgo}
  D.~Bhatia, S.~Chakraborty and A.~Dighe,
  %``Neutrino mixing and $R_K$ anomaly in U(1)$_X$ models: a bottom-up approach,''
  JHEP {\bf 1703}, 117 (2017)
  %doi:10.1007/JHEP03(2017)117
  [arXiv:1701.05825 [hep-ph]].
  %%CITATION = doi:10.1007/JHEP03(2017)117;%%
  %18 citations counted in INSPIRE as of 06 Jun 2018

  \bibitem{Kamenik:2017tnu}
  J.~F.~Kamenik, Y.~Soreq and J.~Zupan,
  %``Lepton flavor universality violation without new sources of quark flavor violation,''
  Phys.\ Rev.\ D {\bf 97}, no. 3, 035002 (2018)
  %doi:10.1103/PhysRevD.97.035002
  [arXiv:1704.06005 [hep-ph]].
  %%CITATION = doi:10.1103/PhysRevD.97.035002;%%
  %33 citations counted in INSPIRE as of 19 Feb 2018

%\cite{Chen:2017usq}
\bibitem{Chen:2017usq}
  C.~H.~Chen and T.~Nomura,
  %``Penguin $b \to s\ell'^+ \ell'^-$ and $B$-meson anomalies in a gauged ${L_\mu -L_\tau}$,''
  Phys.\ Lett.\ B {\bf 777} (2018) 420
%  doi:10.1016/j.physletb.2017.12.062
  [arXiv:1707.03249 [hep-ph]].
  %%CITATION = doi:10.1016/j.physletb.2017.12.062;%%
  %30 citations counted in INSPIRE as of 13 Jun 2019

  %\cite{Camargo-Molina:2018cwu}
\bibitem{Camargo-Molina:2018cwu}
  J.~E.~Camargo-Molina, A.~Celis and D.~A.~Faroughy,
  %``Anomalies in Bottom from new physics in Top,''
  arXiv:1805.04917 [hep-ph].
  %%CITATION = ARXIV:1805.04917;%%
  %2 citations counted in INSPIRE as of 10 Jul 2018

%\cite{Darme:2018hqg}
\bibitem{Darme:2018hqg}
  L.~Darm\'{e}, K.~Kowalska, L.~Roszkowski and E.~M.~Sessolo,
  %``Flavor anomalies and dark matter in SUSY with an extra U(1),''
  JHEP {\bf 1810} (2018) 052
%  doi:10.1007/JHEP10(2018)052
  [arXiv:1806.06036 [hep-ph]].
  %%CITATION = doi:10.1007/JHEP10(2018)052;%%
  %8 citations counted in INSPIRE as of 13 Jun 2019

%\cite{Baek:2018aru}
\bibitem{Baek:2018aru}
  S.~Baek and C.~Yu,
  %``Dark matter for $b\to s \mu^+ \mu^-$ anomaly in a gauged $U(1)_X$ model,''
  JHEP {\bf 1811} (2018) 054
%  doi:10.1007/JHEP11(2018)054
  [arXiv:1806.05967 [hep-ph]].
  %%CITATION = doi:10.1007/JHEP11(2018)054;%%
  %8 citations counted in INSPIRE as of 13 Jun 2019

%\cite{Biswas:2019twf}
\bibitem{Biswas:2019twf}
  A.~Biswas and A.~Shaw,
  %``Reconciling dark matter, $R_{K^{(*)}}$ anomalies and $(g-2)_{\mu}$ in an ${L_{\mu}-L_{\tau}}$ scenario,''
  JHEP {\bf 1905} (2019) 165
%  doi:10.1007/JHEP05(2019)165
  [arXiv:1903.08745 [hep-ph]].
  %%CITATION = doi:10.1007/JHEP05(2019)165;%%
  %3 citations counted in INSPIRE as of 13 Jun 2019

  %\cite{Allanach:2019iiy}
  \bibitem{Allanach:2019iiy}
    B.~C.~Allanach and J.~Davighi,
    %``Deforming the Third Family Hypercharge Model for Neutral Current $B-$Anomalies,''
    arXiv:1905.10327 [hep-ph].
    %%CITATION = ARXIV:1905.10327;%%


  %%%%%%%%. other Leptoquarks %%%%%%%%%%%%%
  %     \cite{Hiller:2014yaa,Gripaios:2014tna,Sahoo:2015wya,Varzielas:2015iva,Alonso:2015sja,Bauer:2015knc,Hati:2015awg,Fajfer:2015ycq,Das:2016vkr,Becirevic:2016yqi,Sahoo:2016pet,Cox:2016epl,Crivellin:2017zlb,Becirevic:2017jtw,Cai:2017wry,Dorsner:2017ufx,Greljo:2018tuh,Sahoo:2018ffv,Becirevic:2018afm,Hati:2018fzc,deMedeirosVarzielas:2018bcy,Aebischer:2018acj,deMedeirosVarzielas:2019okf,Yan:2019hpm,Bigaran:2019bqv}


\bibitem{Hiller:2014yaa}
  G.~Hiller and M.~Schmaltz,
  %``$R_K$ and future $b \to s \ell \ell$ physics beyond the standard model opportunities,''
  Phys.\ Rev.\ D {\bf 90}, 054014 (2014)
  %doi:10.1103/PhysRevD.90.054014
  [arXiv:1408.1627 [hep-ph]].
  %%CITATION = doi:10.1103/PhysRevD.90.054014;%%
  %203 citations counted in INSPIRE as of 19 Feb 2018

\bibitem{Gripaios:2014tna}
  B.~Gripaios, M.~Nardecchia and S.~A.~Renner,
  %``Composite leptoquarks and anomalies in $B$-meson decays,''
  JHEP {\bf 1505}, 006 (2015)
  %doi:10.1007/JHEP05(2015)006
  [arXiv:1412.1791 [hep-ph]].
  %%CITATION = doi:10.1007/JHEP05(2015)006;%%
  %132 citations counted in INSPIRE as of 19 Feb 2018

\bibitem{Sahoo:2015wya}
  S.~Sahoo and R.~Mohanta,
  %``Scalar leptoquarks and the rare $B$ meson decays,''
  Phys.\ Rev.\ D {\bf 91}, no. 9, 094019 (2015)
  %doi:10.1103/PhysRevD.91.094019
  [arXiv:1501.05193 [hep-ph]].
  %%CITATION = doi:10.1103/PhysRevD.91.094019;%%
  %68 citations counted in INSPIRE as of 06 Jun 2018

\bibitem{Varzielas:2015iva}
  I.~de Medeiros Varzielas and G.~Hiller,
  %``Clues for flavor from rare lepton and quark decays,''
  JHEP {\bf 1506}, 072 (2015)
  %doi:10.1007/JHEP06(2015)072
  [arXiv:1503.01084 [hep-ph]].
  %%CITATION = doi:10.1007/JHEP06(2015)072;%%
  %104 citations counted in INSPIRE as of 19 Feb 2018

\bibitem{Alonso:2015sja}
  R.~Alonso, B.~Grinstein and J.~Martin Camalich,
  %``Lepton universality violation and lepton flavor conservation in $B$-meson decays,''
  JHEP {\bf 1510}, 184 (2015)
 % doi:10.1007/JHEP10(2015)184
  [arXiv:1505.05164 [hep-ph]].
  %%CITATION = doi:10.1007/JHEP10(2015)184;%%
  %159 citations counted in INSPIRE as of 06 Jun 2018

\bibitem{Bauer:2015knc}
  M.~Bauer and M.~Neubert,
  %``Minimal Leptoquark Explanation for the R$_{D^{(*)}}$ , R$_K$ , and $(g-2)_g$ Anomalies,''
  Phys.\ Rev.\ Lett.\  {\bf 116}, no. 14, 141802 (2016)
  %doi:10.1103/PhysRevLett.116.141802
  [arXiv:1511.01900 [hep-ph]].
  %%CITATION = doi:10.1103/PhysRevLett.116.141802;%%
  %163 citations counted in INSPIRE as of 06 Jun 2018

\bibitem{Hati:2015awg}
  C.~Hati, G.~Kumar and N.~Mahajan,
  %``$\bar{B}\rightarrow D^{(\ast)}\tau \bar{\nu}$ excesses in ALRSM constrained from $B$, $D$ decays and $D^{0}-\bar{D}^{0}$ mixing,''
  JHEP {\bf 1601}, 117 (2016)
  %doi:10.1007/JHEP01(2016)117
  [arXiv:1511.03290 [hep-ph]].
  %%CITATION = doi:10.1007/JHEP01(2016)117;%%
  %34 citations counted in INSPIRE as of 19 Feb 2018


\bibitem{Fajfer:2015ycq}
  S.~Fajfer and N.~Kosnik,
  %``Vector leptoquark resolution of $R_K$ and $R_{D^{(*)}}$ puzzles,''
  Phys.\ Lett.\ B {\bf 755}, 270 (2016)
  %doi:10.1016/j.physletb.2016.02.018
  [arXiv:1511.06024 [hep-ph]].
  %%CITATION = doi:10.1016/j.physletb.2016.02.018;%%
  %116 citations counted in INSPIRE as of 19 Feb 2018


\bibitem{Das:2016vkr}
  D.~Das, C.~Hati, G.~Kumar and N.~Mahajan,
  %``Towards a unified explanation of $R_{D^{(\ast)}}$, $R_{K}$ and $(g-2)_{\mu}$ anomalies in a left-right model with leptoquarks,''
  Phys.\ Rev.\ D {\bf 94}, 055034 (2016)
  %doi:10.1103/PhysRevD.94.055034
  [arXiv:1605.06313 [hep-ph]].
  %%CITATION = doi:10.1103/PhysRevD.94.055034;%%
  %60 citations counted in INSPIRE as of 19 Feb 2018


\bibitem{Becirevic:2016yqi}
  D.~Be\v{c}irevi\'{c}, S.~Fajfer, N.~Kosnik and O.~Sumensari,
  %``Leptoquark model to explain the $B$-physics anomalies, $R_K$ and $R_D$,''
  Phys.\ Rev.\ D {\bf 94}, no. 11, 115021 (2016)
  %doi:10.1103/PhysRevD.94.115021
  [arXiv:1608.08501 [hep-ph]].
  %%CITATION = doi:10.1103/PhysRevD.94.115021;%%
  %90 citations counted in INSPIRE as of 19 Feb 2018


\bibitem{Sahoo:2016pet}
  S.~Sahoo, R.~Mohanta and A.~K.~Giri,
  %``Explaining the $R_{K}$ and $R_{D^{(*)}}$ anomalies with vector leptoquarks,''
  Phys.\ Rev.\ D {\bf 95}, no. 3, 035027 (2017)
 % doi:10.1103/PhysRevD.95.035027
  [arXiv:1609.04367 [hep-ph]].
  %%CITATION = doi:10.1103/PhysRevD.95.035027;%%
  %56 citations counted in INSPIRE as of 06 Jun 2018

\bibitem{Cox:2016epl}
  P.~Cox, A.~Kusenko, O.~Sumensari and T.~T.~Yanagida,
  %``SU(5) Unification with TeV-scale Leptoquarks,''
  JHEP {\bf 1703}, 035 (2017)
  %doi:10.1007/JHEP03(2017)035
  [arXiv:1612.03923 [hep-ph]].
  %%CITATION = doi:10.1007/JHEP03(2017)035;%%
  %16 citations counted in INSPIRE as of 19 Feb 2018

\bibitem{Crivellin:2017zlb}
  A.~Crivellin, D.~M\"{u}ller and T.~Ota,
  %``Simultaneous explanation of R(D$^{(â)}$) and bâsÎŒ$^{+}$ ÎŒ$^{â}$: the last scalar leptoquarks standing,''
  JHEP {\bf 1709}, 040 (2017)
  %doi:10.1007/JHEP09(2017)040
  [arXiv:1703.09226 [hep-ph]].
  %%CITATION = doi:10.1007/JHEP09(2017)040;%%
  %51 citations counted in INSPIRE as of 19 Feb 2018\

\bibitem{Becirevic:2017jtw}
  D.~Be\v{c}irevi\'{c} and O.~Sumensari,
  %``A leptoquark model to accommodate $R_K^\mathrm{exp} < R_K^\mathrm{SM}$ and $R_{K^\ast}^\mathrm{exp} < R_{K^\ast}^\mathrm{SM}$,''
  JHEP {\bf 1708}, 104 (2017)
  %doi:10.1007/JHEP08(2017)104
  [arXiv:1704.05835 [hep-ph]].
  %%CITATION = doi:10.1007/JHEP08(2017)104;%%
  %39 citations counted in INSPIRE as of 19 Feb 2018

  %\cite{Cai:2017wry}
\bibitem{Cai:2017wry}
  Y.~Cai, J.~Gargalionis, M.~A.~Schmidt and R.~R.~Volkas,
  %``Reconsidering the One Leptoquark solution: flavor anomalies and neutrino mass,''
  JHEP {\bf 1710}, 047 (2017)
%  doi:10.1007/JHEP10(2017)047
  [arXiv:1704.05849 [hep-ph]].
  %%CITATION = doi:10.1007/JHEP10(2017)047;%%
  %43 citations counted in INSPIRE as of 10 Jul 2018


  %\cite{}
\bibitem{Dorsner:2017ufx}
  I.~Dor\v{s}ner, S.~Fajfer, D.~A.~Faroughy and N.~Ko\v{s}nik,
  %``The role of the $S_3$ GUT leptoquark in flavor universality and collider searches,''
  JHEP {\bf 1710}, 188 (2017)
%  doi:10.1007/JHEP10(2017)188
  [arXiv:1706.07779 [hep-ph]].
  %%CITATION = doi:10.1007/JHEP10(2017)188;%%
  %29 citations counted in INSPIRE as of 06 Jun 2018


 %\cite{}
\bibitem{Greljo:2018tuh}
A.~Greljo and B.~A.~Stefanek,
  %``Third family quark?lepton unification at the TeV scale,''
  Phys.\ Lett.\ B {\bf 782}, 131 (2018)
%  doi:10.1016/j.physletb.2018.05.033
  [arXiv:1802.04274 [hep-ph]].
  %%CITATION = doi:10.1016/j.physletb.2018.05.033;%%
  %5 citations counted in INSPIRE as of 26 Jun 2018

%\cite{Sahoo:2018ffv}
\bibitem{Sahoo:2018ffv}
  S.~Sahoo and R.~Mohanta,
  %``Impact of vector leptoquark on $\bar B \to \bar K^* l^+ l^-$ anomalies,''
  J.\ Phys.\ G {\bf 45} (2018) no.8,  085003
%  doi:10.1088/1361-6471/aaca12
  [arXiv:1806.01048 [hep-ph]].
  %%CITATION = doi:10.1088/1361-6471/aaca12;%%
  %8 citations counted in INSPIRE as of 13 Jun 2019

%\cite{Becirevic:2018afm}
\bibitem{Becirevic:2018afm}
  D.~Be\v{c}irevi\'{c}, I.~Dor\v{s}ner, S.~Fajfer, N.~Ko\v{s}nik, D.~A.~Faroughy and O.~Sumensari,
  %``Scalar leptoquarks from grand unified theories to accommodate the $B$-physics anomalies,''
  Phys.\ Rev.\ D {\bf 98} (2018) no.5,  055003
%  doi:10.1103/PhysRevD.98.055003
  [arXiv:1806.05689 [hep-ph]].
  %%CITATION = doi:10.1103/PhysRevD.98.055003;%%
  %58 citations counted in INSPIRE as of 13 Jun 2019

%\cite{Hati:2018fzc,deMedeirosVarzielas:2018bcy,Angelescu:2018tyl,Aebischer:2018acj,deMedeirosVarzielas:2019okf,Yan:2019hpm,Bigaran:2019bqv}
\bibitem{Hati:2018fzc}
  C.~Hati, G.~Kumar, J.~Orloff and A.~M.~Teixeira,
  %``Reconciling $B$-meson decay anomalies with neutrino masses, dark matter and constraints from flavour violation,''
  JHEP {\bf 1811} (2018) 011
%  doi:10.1007/JHEP11(2018)011
  [arXiv:1806.10146 [hep-ph]].
  %%CITATION = doi:10.1007/JHEP11(2018)011;%%
  %17 citations counted in INSPIRE as of 13 Jun 2019

 %\cite{deMedeirosVarzielas:2018bcy}
 \bibitem{deMedeirosVarzielas:2018bcy}
   I.~de Medeiros Varzielas and S.~F.~King,
   %``$ {R}_{K^{\left(*\right)}} $ with leptoquarks and the origin of Yukawa couplings,''
   JHEP {\bf 1811} (2018) 100
%   doi:10.1007/JHEP11(2018)100
   [arXiv:1807.06023 [hep-ph]].
   %%CITATION = doi:10.1007/JHEP11(2018)100;%%
   %10 citations counted in INSPIRE as of 13 Jun 2019

%\cite{Aebischer:2018acj}
\bibitem{Aebischer:2018acj}
  J.~Aebischer, A.~Crivellin and C.~Greub,
  %``QCD improved matching for semileptonic B$$ decays with leptoquarks,''
  Phys.\ Rev.\ D {\bf 99} (2019) no.5,  055002
%  doi:10.1103/PhysRevD.99.055002
  [arXiv:1811.08907 [hep-ph]].
  %%CITATION = doi:10.1103/PhysRevD.99.055002;%%
  %7 citations counted in INSPIRE as of 13 Jun 2019

%\cite{deMedeirosVarzielas:2019okf}
\bibitem{deMedeirosVarzielas:2019okf}
  I.~De Medeiros Varzielas and S.~F.~King,
  %``Origin of Yukawa couplings for Higgs bosons and leptoquarks,''
  Phys.\ Rev.\ D {\bf 99} (2019) no.9,  095029
%  doi:10.1103/PhysRevD.99.095029
  [arXiv:1902.09266 [hep-ph]].
  %%CITATION = doi:10.1103/PhysRevD.99.095029;%%
  %2 citations counted in INSPIRE as of 13 Jun 2019

%\cite{Yan:2019hpm}
\bibitem{Yan:2019hpm}
  H.~Yan, Y.~D.~Yang and X.~B.~Yuan,
  %``Phenomenology of $b\to c\tau\bar\nu$ decays in a scalar leptoquark model,''
  arXiv:1905.01795 [hep-ph].
  %%CITATION = ARXIV:1905.01795;%%
  %1 citations counted in INSPIRE as of 13 Jun 2019

%\cite{Bigaran:2019bqv}
\bibitem{Bigaran:2019bqv}
  I.~Bigaran, J.~Gargalionis and R.~R.~Volkas,
  %``A near-minimal leptoquark model for reconciling flavour anomalies and generating radiative neutrino masses,''
  arXiv:1906.01870 [hep-ph].
  %%CITATION = ARXIV:1906.01870;%%

 %\cite{Popov:2019tyc}
 \bibitem{Popov:2019tyc}
   O.~Popov, M.~A.~Schmidt and G.~White,
   %``$R_2$ as a single leptoquark solution to $R_{D^{(*)}}$ and $R_{K^{(*)}}$,''
   arXiv:1905.06339 [hep-ph].
   %%CITATION = ARXIV:1905.06339;%%
   %2 citations counted in INSPIRE as of 23 Jul 2019

%%%%%%%%%%  RPV SUSY %%%%%%%%%%%%%%%%
%\cite{Deshpand:2016cpw,Altmannshofer:2017poe,Das:2017kfo,Earl:2018snx,Trifinopoulos:2018rna,Trifinopoulos:2019lyo}

%\cite{Deshpand:2016cpw}
\bibitem{Deshpand:2016cpw}
  N.~G.~Deshpande and X.~G.~He,
  %``Consequences of R-parity violating interactions for anomalies in $\bar B\to D^{(*)} \tau \bar \nu$ and $b\to s \mu^+\mu^-$,''
  Eur.\ Phys.\ J.\ C {\bf 77} (2017) no.2,  134
%  doi:10.1140/epjc/s10052-017-4707-y
  [arXiv:1608.04817 [hep-ph]].
  %%CITATION = doi:10.1140/epjc/s10052-017-4707-y;%%
  %51 citations counted in INSPIRE as of 13 Jun 2019

\bibitem{Altmannshofer:2017poe}
  W.~Altmannshofer, P.~S.~Bhupal Dev and A.~Soni,
  %``$R_{D^{(*)}}$ anomaly: A possible hint for natural supersymmetry with $R$-parity violation,''
  Phys.\ Rev.\ D {\bf 96}, no. 9, 095010 (2017)
%  doi:10.1103/PhysRevD.96.095010
  [arXiv:1704.06659 [hep-ph]].
  %%CITATION = doi:10.1103/PhysRevD.96.095010;%%
  %35 citations counted in INSPIRE as of 10 Jul 2018

\bibitem{Das:2017kfo}
  D.~Das, C.~Hati, G.~Kumar and N.~Mahajan,
  %``Scrutinizing $R$-parity violating interactions in light of $R_{K^{(\ast)}}$ data,''
  Phys.\ Rev.\ D {\bf 96}, no. 9, 095033 (2017)
  %doi:10.1103/PhysRevD.96.095033
  [arXiv:1705.09188 [hep-ph]].
  %%CITATION = doi:10.1103/PhysRevD.96.095033;%%
  %14 citations counted in INSPIRE as of 19 Feb 2018

\bibitem{Earl:2018snx}
  K.~Earl and T.~Gregoire,
  %``Contributions to ${b \rightarrow s \ell \ell}$ Anomalies from ${R}$-Parity Violating Interactions,''
  arXiv:1806.01343 [hep-ph].
  %%CITATION = ARXIV:1806.01343;%%

%\cite{Trifinopoulos:2018rna}
\bibitem{Trifinopoulos:2018rna}
  S.~Trifinopoulos,
  %``Revisiting R-parity violating interactions as an explanation of the B-physics anomalies,''
  Eur.\ Phys.\ J.\ C {\bf 78} (2018) no.10,  803
%  doi:10.1140/epjc/s10052-018-6280-4
  [arXiv:1807.01638 [hep-ph]].
  %%CITATION = doi:10.1140/epjc/s10052-018-6280-4;%%
  %8 citations counted in INSPIRE as of 13 Jun 2019

%\cite{Trifinopoulos:2019lyo}
\bibitem{Trifinopoulos:2019lyo}
  S.~Trifinopoulos,
  %``B-physics anomalies: The bridge between R-parity violating Supersymmetry and flavoured Dark Matter,''
  arXiv:1904.12940 [hep-ph].
  %%CITATION = ARXIV:1904.12940;%%

  %%%%%%% Others %%%%%%%%%%%%%%%%%%
  %\cite{Greljo:2015mma,Arnan:2017lxi,Geng:2017svp,Choudhury:2017qyt,Choudhury:2017ijp,Grinstein:2018fgb,Cerdeno:2019vpd,Bhattacharya:2019eji,Crivellin:2019dun,Arnan:2019uhr}
  \bibitem{Greljo:2015mma}
  A.~Greljo, G.~Isidori and D.~Marzocca,
  %``On the breaking of Lepton Flavor Universality in B decays,''
  JHEP {\bf 1507}, 142 (2015)
  %doi:10.1007/JHEP07(2015)142
  [arXiv:1506.01705 [hep-ph]].
  %%CITATION = doi:10.1007/JHEP07(2015)142;%%
  %126 citations counted in INSPIRE as of 19 Feb 2018


\bibitem{Arnan:2017lxi}
  P.~Arnan, D.~Be\v{c}irevi\'{c}, F.~Mescia and O.~Sumensari,
  %``Two Higgs doublet models and $b\rightarrow s$ exclusive decays,''
  Eur.\ Phys.\ J.\ C {\bf 77}, no. 11, 796 (2017)
  %doi:10.1140/epjc/s10052-017-5370-z
  [arXiv:1703.03426 [hep-ph]].
  %%CITATION = doi:10.1140/epjc/s10052-017-5370-z;%%
  %9 citations counted in INSPIRE as of 19 Feb 2018

\bibitem{Geng:2017svp}
  L.~S.~Geng, B.~Grinstein, S.~J\"{a}ger, J.~Martin Camalich, X.~L.~Ren and R.~X.~Shi,
  %``Towards the discovery of new physics with lepton-universality ratios of $b\to s\ell\ell$ decays,''
  Phys.\ Rev.\ D {\bf 96}, no. 9, 093006 (2017)
  %doi:10.1103/PhysRevD.96.093006
  [arXiv:1704.05446 [hep-ph]].
  %%CITATION = doi:10.1103/PhysRevD.96.093006;%%
  %94 citations counted in INSPIRE as of 06 Jun 2018

\bibitem{Choudhury:2017qyt}
  D.~Choudhury, A.~Kundu, R.~Mandal and R.~Sinha,
  %``Minimal unified resolution to $R_{K^{(*)}}$ and $R(D^{(*)})$ anomalies with lepton mixing,''
  Phys.\ Rev.\ Lett.\  {\bf 119}, no. 15, 151801 (2017)
  %doi:10.1103/PhysRevLett.119.151801
  [arXiv:1706.08437 [hep-ph]].
  %%CITATION = doi:10.1103/PhysRevLett.119.151801;%%
  %11 citations counted in INSPIRE as of 19 Feb 2018


\bibitem{Choudhury:2017ijp}
  D.~Choudhury, A.~Kundu, R.~Mandal and R.~Sinha,
  %``$R_{K^{(*)}}$ and $R(D^{(*)})$ anomalies resolved with lepton mixing,''
  arXiv:1712.01593 [hep-ph].
  %%CITATION = ARXIV:1712.01593;%%

  %\cite{Grinstein:2018fgb}
  \bibitem{Grinstein:2018fgb}
    B.~Grinstein, S.~Pokorski and G.~G.~Ross,
    %``Lepton non-universality in $B$ decays and fermion mass structure,''
    JHEP {\bf 1812} (2018) 079
%    doi:10.1007/JHEP12(2018)079
    [arXiv:1809.01766 [hep-ph]].
    %%CITATION = doi:10.1007/JHEP12(2018)079;%%
    %6 citations counted in INSPIRE as of 13 Jun 2019

%\cite{Cerdeno:2019vpd}
\bibitem{Cerdeno:2019vpd}
  D.~G.~Cerdeño, A.~Cheek, P.~Martín-Ramiro and J.~M.~Moreno,
  %``B anomalies and dark matter: a complex connection,''
  arXiv:1902.01789 [hep-ph].
  %%CITATION = ARXIV:1902.01789;%%
  %3 citations counted in INSPIRE as of 13 Jun 2019

%\cite{Bhattacharya:2019eji}
\bibitem{Bhattacharya:2019eji}
  S.~Bhattacharya, A.~Biswas, Z.~Calcuttawala and S.~K.~Patra,
  %``An in-depth analysis of $b\to c(s)$ semileptonic observables with possible $\mu - \tau$ mixing,''
  arXiv:1902.02796 [hep-ph].
  %%CITATION = ARXIV:1902.02796;%%
  %3 citations counted in INSPIRE as of 13 Jun 2019

%\cite{Crivellin:2019dun}
\bibitem{Crivellin:2019dun}
  A.~Crivellin, D.~M\"{u}ller and C.~Wiegand,
  %``$b\to s\ell^+\ell^-$ Transitions in Two-Higgs-Doublet Models,''
  arXiv:1903.10440 [hep-ph].
  %%CITATION = ARXIV:1903.10440;%%
  %2 citations counted in INSPIRE as of 13 Jun 2019

  %\cite{Arnan:2019uhr}
  \bibitem{Arnan:2019uhr}
    P.~Arnan, A.~Crivellin, M.~Fedele and F.~Mescia,
    %``Generic Loop Effects of New Scalars and Fermions in $b\to s\ell^+\ell^-$ and a Vector-like $4^{\rm th}$ Generation,''
    arXiv:1904.05890 [hep-ph].
    %%CITATION = ARXIV:1904.05890;%%
  %%%%%%%% vector leptoquark V_1 solutions%%%%%%%%%%%
%\cite{Assad:2017iib,Buttazzo:2017ixm,Calibbi:2017qbu,Bordone:2017bld,Blanke:2018sro,Bordone:2018nbg,Kumar:2018kmr,Angelescu:2018tyl,Balaji:2018zna,Fornal:2018dqn,Baker:2019sli,Cornella:2019hct,DaRold:2019fiw}

%\cite{Assad:2017iib}
\bibitem{Assad:2017iib}
  N.~Assad, B.~Fornal and B.~Grinstein,
  %``Baryon Number and Lepton Universality Violation in Leptoquark and Diquark Models,''
  Phys.\ Lett.\ B {\bf 777} (2018) 324
  % doi:10.1016/j.physletb.2017.12.042
  [arXiv:1708.06350 [hep-ph]].
  %%CITATION = doi:10.1016/j.physletb.2017.12.042;%%
  %89 citations counted in INSPIRE as of 23 Jul 2019

 %\cite{Buttazzo:2017ixm}
\bibitem{Buttazzo:2017ixm}
D.~Buttazzo, A.~Greljo, G.~Isidori and D.~Marzocca,
%``B-physics anomalies: a guide to combined explanations,''
JHEP {\bf 1711} (2017) 044
%doi:10.1007/JHEP11(2017)044
[arXiv:1706.07808 [hep-ph]].
%%CITATION = doi:10.1007/JHEP11(2017)044;%%
%156 citations counted in INSPIRE as of 13 Jun 2019



%\cite{Calibbi:2017qbu}
\bibitem{Calibbi:2017qbu}
L.~Calibbi, A.~Crivellin and T.~Li,
%``Model of vector leptoquarks in view of the $B$-physics anomalies,''
Phys.\ Rev.\ D {\bf 98} (2018) no.11,  115002
%doi:10.1103/PhysRevD.98.115002
[arXiv:1709.00692 [hep-ph]].
%%CITATION = doi:10.1103/PhysRevD.98.115002;%%
%95 citations counted in INSPIRE as of 13 Jun 2019

%\cite{Bordone:2017bld}
\bibitem{Bordone:2017bld}
M.~Bordone, C.~Cornella, J.~Fuentes-Martin and G.~Isidori,
%``A three-site gauge model for flavor hierarchies and flavor anomalies,''
Phys.\ Lett.\ B {\bf 779}, 317 (2018)
%doi:10.1016/j.physletb.2018.02.011
[arXiv:1712.01368 [hep-ph]].
%%CITATION = doi:10.1016/j.physletb.2018.02.011;%%
%76 citations counted in INSPIRE as of 13 Jun 2019

%\cite{Blanke:2018sro}
\bibitem{Blanke:2018sro}
M.~Blanke and A.~Crivellin,
%``$B$ Meson Anomalies in a Pati-Salam Model within the Randall-Sundrum Background,''
Phys.\ Rev.\ Lett.\  {\bf 121} (2018) no.1,  011801
%doi:10.1103/PhysRevLett.121.011801
[arXiv:1801.07256 [hep-ph]].
%%CITATION = doi:10.1103/PhysRevLett.121.011801;%%
%77 citations counted in INSPIRE as of 13 Jun 2019

%\cite{Bordone:2018nbg}
\bibitem{Bordone:2018nbg}
M.~Bordone, C.~Cornella, J.~Fuentes-Martín and G.~Isidori,
%``Low-energy signatures of the $\mathrm{PS}^3$ model: from $B$-physics anomalies to LFV,''
JHEP {\bf 1810} (2018) 148
 %    doi:10.1007/JHEP10(2018)148
[arXiv:1805.09328 [hep-ph]].
%%CITATION = doi:10.1007/JHEP10(2018)148;%%
%36 citations counted in INSPIRE as of 13 Jun 2019

%\cite{Kumar:2018kmr}
\bibitem{Kumar:2018kmr}
  J.~Kumar, D.~London and R.~Watanabe,
  %``Combined Explanations of the $b \to s \mu^+ \mu^-$ and $b \to c \tau^- {\bar\nu}$ Anomalies: a General Model Analysis,''
  Phys.\ Rev.\ D {\bf 99} (2019) no.1,  015007
%  doi:10.1103/PhysRevD.99.015007
  [arXiv:1806.07403 [hep-ph]].
  %%CITATION = doi:10.1103/PhysRevD.99.015007;%%
  %43 citations counted in INSPIRE as of 13 Jun 2019

%\cite{Angelescu:2018tyl}
\bibitem{Angelescu:2018tyl}
  A.~Angelescu, D.~Be\v{c}irevi\'{c}, D.~A.~Faroughy and O.~Sumensari,
  %``Closing the window on single leptoquark solutions to the $B$-physics anomalies,''
  JHEP {\bf 1810} (2018) 183
%  doi:10.1007/JHEP10(2018)183
  [arXiv:1808.08179 [hep-ph]].
  %%CITATION = doi:10.1007/JHEP10(2018)183;%%
  %49 citations counted in INSPIRE as of 13 Jun 2019

%\cite{Balaji:2018zna}
\bibitem{Balaji:2018zna}
  S.~Balaji, R.~Foot and M.~A.~Schmidt,
  %``Chiral SU(4) explanation of the $b\to s$ anomalies,''
  Phys.\ Rev.\ D {\bf 99} (2019) no.1,  015029
%  doi:10.1103/PhysRevD.99.015029
  [arXiv:1809.07562 [hep-ph]].
  %%CITATION = doi:10.1103/PhysRevD.99.015029;%%
  %4 citations counted in INSPIRE as of 13 Jun 2019

%\cite{Fornal:2018dqn}
\bibitem{Fornal:2018dqn}
  B.~Fornal, S.~A.~Gadam and B.~Grinstein,
  %``Left-Right SU(4) Vector Leptoquark Model for Flavor Anomalies,''
  Phys.\ Rev.\ D {\bf 99} (2019) no.5,  055025
%  doi:10.1103/PhysRevD.99.055025
  [arXiv:1812.01603 [hep-ph]].
  %%CITATION = doi:10.1103/PhysRevD.99.055025;%%
  %14 citations counted in INSPIRE as of 13 Jun 2019


%\cite{Baker:2019sli}
\bibitem{Baker:2019sli}
M.~J.~Baker, J.~Fuentes-Martín, G.~Isidori and M.~König,
%``High- $p_T$ signatures in vector–leptoquark models,''
Eur.\ Phys.\ J.\ C {\bf 79} (2019) no.4,  334
%doi:10.1140/epjc/s10052-019-6853-x
[arXiv:1901.10480 [hep-ph]].
%%CITATION = doi:10.1140/epjc/s10052-019-6853-x;%%
%6 citations counted in INSPIRE as of 13 Jun 2019

%\cite{Cornella:2019hct}
\bibitem{Cornella:2019hct}
  C.~Cornella, J.~Fuentes-Martin and G.~Isidori,
  %``Revisiting the vector leptoquark explanation of the B-physics anomalies,''
  arXiv:1903.11517 [hep-ph].
  %%CITATION = ARXIV:1903.11517;%%
  %11 citations counted in INSPIRE as of 13 Jun 2019

  %\cite{DaRold:2019fiw}
  \bibitem{DaRold:2019fiw}
    L.~Da Rold and F.~Lamagna,
    %``A vector leptoquark for the B-physics anomalies from a composite GUT,''
    arXiv:1906.11666 [hep-ph].
    %%CITATION = ARXIV:1906.11666;%%

%\cite{Barbieri:2016las}
\bibitem{Barbieri:2016las}
  R.~Barbieri, C.~W.~Murphy and F.~Senia,
  %``B-decay Anomalies in a Composite Leptoquark Model,''
  Eur.\ Phys.\ J.\ C {\bf 77} (2017) no.1,  8
  % doi:10.1140/epjc/s10052-016-4578-7
  [arXiv:1611.04930 [hep-ph]].
  %%CITATION = doi:10.1140/epjc/s10052-016-4578-7;%%
  %102 citations counted in INSPIRE as of 12 Sep 2019

% Refs for L_R symmetric PS VLQ coupling limits
%\cite{Smirnov:2007hv}
%\cite{Hung:1981pd,Valencia:1994cj,Smirnov:2007hv,Carpentier:2010ue,Kuznetsov:2012ai,Smirnov2018ske}

%\cite{Hung:1981pd}
\bibitem{Hung:1981pd}
  P.~Q.~Hung, A.~J.~Buras and J.~D.~Bjorken,
  %``Petite Unification of Quarks and Leptons,''
  Phys.\ Rev.\ D {\bf 25} (1982) 805.
  % doi:10.1103/PhysRevD.25.805
  %%CITATION = doi:10.1103/PhysRevD.25.805;%%
  %39 citations counted in INSPIRE as of 26 Jun 2019



%\cite{Valencia:1994cj}
\bibitem{Valencia:1994cj}
  G.~Valencia and S.~Willenbrock,
  %``Quark - lepton unification and rare meson decays,''
  Phys.\ Rev.\ D {\bf 50} (1994) 6843
%  doi:10.1103/PhysRevD.50.6843
  [hep-ph/9409201].
  %%CITATION = doi:10.1103/PhysRevD.50.6843;%%
  %78 citations counted in INSPIRE as of 10 Jun 2019

\bibitem{Smirnov:2007hv}
  A.~D.~Smirnov,
  %``Mass limits for scalar and gauge leptoquarks from K(L)0 ---> e-+ mu+-, B0 ---> e-+ tau+- decays,''
  Mod.\ Phys.\ Lett.\ A {\bf 22} (2007) 2353
%  doi:10.1142/S0217732307024401
  [arXiv:0705.0308 [hep-ph]].
  %%CITATION = doi:10.1142/S0217732307024401;%%
  %24 citations counted in INSPIRE as of 10 Jun 2019

  %\cite{Carpentier:2010ue}
  \bibitem{Carpentier:2010ue}
    M.~Carpentier and S.~Davidson,
    %``Constraints on two-lepton, two quark operators,''
    Eur.\ Phys.\ J.\ C {\bf 70} (2010) 1071
%    doi:10.1140/epjc/s10052-010-1482-4
    [arXiv:1008.0280 [hep-ph]].
    %%CITATION = doi:10.1140/epjc/s10052-010-1482-4;%%
    %84 citations counted in INSPIRE as of 10 Jun 2019

  %\cite{Kuznetsov:2012ai}
  \bibitem{Kuznetsov:2012ai}
    A.~V.~Kuznetsov, N.~V.~Mikheev and A.~V.~Serghienko,
    %``The third type of fermion mixing in the lepton and quark interactions with leptoquarks,''
    Int.\ J.\ Mod.\ Phys.\ A {\bf 27} (2012) 1250062
%   doi:10.1142/S0217751X12500625
    [arXiv:1203.0196 [hep-ph]].
    %%CITATION = doi:10.1142/S0217751X12500625;%%
    %13 citations counted in INSPIRE as of 10 Jun 2019

    %\cite{Smirnov:2018ske}
\bibitem{Smirnov:2018ske}
  A.~D.~Smirnov,
      %``Vector leptoquark mass limits and branching ratios of $ K_L^0, B^0, B_s \to l^+_i l^-_j $ decays with account of fermion mixing in leptoquark currents,''
  Mod.\ Phys.\ Lett.\ A {\bf 33} (2018) 1850019
  %    doi:10.1142/S0217732318500190
  [arXiv:1801.02895 [hep-ph]].
      %%CITATION = doi:10.1142/S0217732318500190;%%
      %12 citations counted in INSPIRE as of 10 Jun 2019

	%\cite{Feruglio:2017rjo}
	\bibitem{Feruglio:2017rjo}
	  F.~Feruglio, P.~Paradisi and A.~Pattori,
	  %``On the Importance of Electroweak Corrections for B Anomalies,''
	  JHEP {\bf 1709} (2017) 061
	  % doi:10.1007/JHEP09(2017)061
	  [arXiv:1705.00929 [hep-ph]].
	  %%CITATION = doi:10.1007/JHEP09(2017)061;%%
	  %93 citations counted in INSPIRE as of 16 Sep 2019


%%%%%  Nonunitarity in neutrinos refs %%%%%
%\cite{Xing:2007zj, Blennow:2016jkn, Fernandez-Martinez:2016lgt, Escrihuela:2015wra}
%\cite{Xing:2007zj}
\bibitem{Xing:2007zj}
  Z.~z.~Xing,
  %``Correlation between the Charged Current Interactions of Light and Heavy Majorana Neutrinos,''
  Phys.\ Lett.\ B {\bf 660} (2008) 515
%  doi:10.1016/j.physletb.2008.01.038
  [arXiv:0709.2220 [hep-ph]].
  %%CITATION = doi:10.1016/j.physletb.2008.01.038;%%
  %58 citations counted in INSPIRE as of 13 Jun 2019

%\cite{Blennow:2016jkn}
\bibitem{Blennow:2016jkn}
  M.~Blennow, P.~Coloma, E.~Fernandez-Martinez, J.~Hernandez-Garcia and J.~Lopez-Pavon,
  %``Non-Unitarity, sterile neutrinos, and Non-Standard neutrino Interactions,''
  JHEP {\bf 1704} (2017) 153
%  doi:10.1007/JHEP04(2017)153
  [arXiv:1609.08637 [hep-ph]].
  %%CITATION = doi:10.1007/JHEP04(2017)153;%%
  %67 citations counted in INSPIRE as of 13 Jun 2019

%\cite{Fernandez-Martinez:2016lgt}
\bibitem{Fernandez-Martinez:2016lgt}
  E.~Fernandez-Martinez, J.~Hernandez-Garcia and J.~Lopez-Pavon,
  %``Global constraints on heavy neutrino mixing,''
  JHEP {\bf 1608} (2016) 033
%  doi:10.1007/JHEP08(2016)033
  [arXiv:1605.08774 [hep-ph]].
  %%CITATION = doi:10.1007/JHEP08(2016)033;%%
  %107 citations counted in INSPIRE as of 13 Jun 2019

%\cite{Escrihuela:2015wra}
\bibitem{Escrihuela:2015wra}
  F.~J.~Escrihuela, D.~V.~Forero, O.~G.~Miranda, M.~Tortola and J.~W.~F.~Valle,
  %``On the description of nonunitary neutrino mixing,''
  Phys.\ Rev.\ D {\bf 92} (2015) no.5,  053009
   Erratum: [Phys.\ Rev.\ D {\bf 93} (2016) no.11,  119905]
%  doi:10.1103/PhysRevD.93.119905, 10.1103/PhysRevD.92.053009
  [arXiv:1503.08879 [hep-ph]].
  %%CITATION = doi:10.1103/PhysRevD.93.119905, 10.1103/PhysRevD.92.053009;%%
  %62 citations counted in INSPIRE as of 13 Jun 2019


%%%%%% Wilson %%%%%%%%%%%%
%\cite{Aebischer:2018bkb}
\bibitem{Aebischer:2018bkb}
  J.~Aebischer, J.~Kumar and D.~M.~Straub,
  %``Wilson: a Python package for the running and matching of Wilson coefficients above and below the electroweak scale,''
  Eur.\ Phys.\ J.\ C {\bf 78} (2018) no.12,  1026
  % doi:10.1140/epjc/s10052-018-6492-7
  [arXiv:1804.05033 [hep-ph]].
  %%CITATION = doi:10.1140/epjc/s10052-018-6492-7;%%
  %29 citations counted in INSPIRE as of 17 Sep 2019

%%%%%%% FLAVIO %%%%%%%%%%
%\cite{Straub:2018kue}
\bibitem{Straub:2018kue}
  D.~M.~Straub,
  %``flavio: a Python package for flavour and precision phenomenology in the Standard Model and beyond,''
  arXiv:1810.08132 [hep-ph].
  %%CITATION = ARXIV:1810.08132;%%
  %24 citations counted in INSPIRE as of 08 Jul 2019
%%%%%%%%%%%%%%%%%%%%%%%%%%%%%%%%%%%%%%%%%%%%%%%%%%%%%%%%%%%%%%%%%%%%%%%%%%%%


%%%% power/charm loop corrections refs  %%%

%\cite{Jager:2012uw}
\bibitem{Jager:2012uw}
  S.~J\"{a}ger and J.~Martin Camalich,
  %``On $B \to  V \ell \ell$ at small dilepton invariant mass, power corrections, and new physics,''
  JHEP {\bf 1305} (2013) 043
  %doi:10.1007/JHEP05(2013)043
  [arXiv:1212.2263 [hep-ph]].
  %%CITATION = doi:10.1007/JHEP05(2013)043;%%
  %205 citations counted in INSPIRE as of 11 Jun 2019

  %\cite{Jager:2014rwa}
  \bibitem{Jager:2014rwa}
    S.~J\"{a}ger and J.~Martin Camalich,
    %``Reassessing the discovery potential of the $B \to K^{*} \ell^+\ell^-$ decays in the large-recoil region: SM challenges and BSM opportunities,''
    Phys.\ Rev.\ D {\bf 93} (2016) no.1,  014028
   % doi:10.1103/PhysRevD.93.014028
    [arXiv:1412.3183 [hep-ph]].
    %%CITATION = doi:10.1103/PhysRevD.93.014028;%%
    %161 citations counted in INSPIRE as of 11 Jun 2019

  %\cite{Ciuchini:2015qxb}
  \bibitem{Ciuchini:2015qxb}
    M.~Ciuchini, M.~Fedele, E.~Franco, S.~Mishima, A.~Paul, L.~Silvestrini and M.~Valli,
    %``$B\to K^* \ell^+ \ell^-$ decays at large recoil in the Standard Model: a theoretical reappraisal,''
    JHEP {\bf 1606} (2016) 116
    %doi:10.1007/JHEP06(2016)116
    [arXiv:1512.07157 [hep-ph]].
    %%CITATION = doi:10.1007/JHEP06(2016)116;%%
    %153 citations counted in INSPIRE as of 11 Jun 2019

  %\cite{Ciuchini:2016weo}
  \bibitem{Ciuchini:2016weo}
    M.~Ciuchini, M.~Fedele, E.~Franco, S.~Mishima, A.~Paul, L.~Silvestrini and M.~Valli,
    %``$B\to K^*\ell^+\ell^-$ in the Standard Model: Elaborations and Interpretations,''
    PoS ICHEP {\bf 2016} (2016) 584
    %doi:10.22323/1.282.0584
    [arXiv:1611.04338 [hep-ph]].
    %%CITATION = doi:10.22323/1.282.0584;%%
    %20 citations counted in INSPIRE as of 11 Jun 2019


%%%%.  Fajfer review %%%%%%%
%\cite{Dorsner:2016wpm}
\bibitem{Dorsner:2016wpm}
  I.~Dor\v{s}ner, S.~Fajfer, A.~Greljo, J.~F.~Kamenik and N.~Ko\v{s}nik,
  %``Physics of leptoquarks in precision experiments and at particle colliders,''
  Phys.\ Rept.\  {\bf 641} (2016) 1
%  doi:10.1016/j.physrep.2016.06.001
  [arXiv:1603.04993 [hep-ph]].
  %%CITATION = doi:10.1016/j.physrep.2016.06.001;%%
  %199 citations counted in INSPIRE as of 13 Jun 2019

  %%% RD(*) Refs %%%

%\cite{Crivellin:2018yvo}
\bibitem{Crivellin:2018yvo}
  A.~Crivellin, C.~Greub, D.~M\"{u}ller and F.~Saturnino,
  %``Importance of Loop Effects in Explaining the Accumulated Evidence for New Physics in B Decays with a Vector Leptoquark,''
  Phys.\ Rev.\ Lett.\  {\bf 122} (2019) no.1,  011805
%  doi:10.1103/PhysRevLett.122.011805
  [arXiv:1807.02068 [hep-ph]].
  %%CITATION = doi:10.1103/PhysRevLett.122.011805;%%
  %34 citations counted in INSPIRE as of 13 Jun 2019

%%%%% Meson Table refs %%%%%%%

%\cite{Buras:2015qea}
\bibitem{Buras:2015qea}
  A.~J.~Buras, D.~Buttazzo, J.~Girrbach-Noe and R.~Knegjens,
  %``$ {K}^{+}\to {\pi}^{+}\nu \overline{\nu} $ and $ {K}_L\to {\pi}^0\nu \overline{\nu} $ in the Standard Model: status and perspectives,''
  JHEP {\bf 1511} (2015) 033
%  doi:10.1007/JHEP11(2015)033
  [arXiv:1503.02693 [hep-ph]].
  %%CITATION = doi:10.1007/JHEP11(2015)033;%%
  %194 citations counted in INSPIRE as of 13 Jun 2019

%\cite{Artamonov:2008qb}
\bibitem{Artamonov:2008qb}
  A.~V.~Artamonov {\it et al.} [E949 Collaboration],
  %``New measurement of the $K^{+} \to \pi^{+} \nu \bar{\nu}$ branching ratio,''
  Phys.\ Rev.\ Lett.\  {\bf 101} (2008) 191802
%  doi:10.1103/PhysRevLett.101.191802
  [arXiv:0808.2459 [hep-ex]].
  %%CITATION = doi:10.1103/PhysRevLett.101.191802;%%
  %252 citations counted in INSPIRE as of 13 Jun 2019

  \bibitem{Na62:2018}
NA62 Collaboration, Contribution to the ”53rd Rencontres de Moriond on Electroweak Interactions and Unified Theories (Moriond EW 2018)”, La Thuile, Italy, 10-17 March 2018.

%\cite{Ahn:2009gb}
\bibitem{Ahn:2009gb}
  J.~K.~Ahn {\it et al.} [E391a Collaboration],
  %``Experimental study of the decay K0(L) ---> pi0 nu nu-bar,''
  Phys.\ Rev.\ D {\bf 81} (2010) 072004
  %  doi:10.1103/PhysRevD.81.072004
  [arXiv:0911.4789 [hep-ex]].
  %%CITATION = doi:10.1103/PhysRevD.81.072004;%%
  %166 citations counted in INSPIRE as of 13 Jun 2019

%\cite{Grygier:2017tzo}
\bibitem{Grygier:2017tzo}
  J.~Grygier {\it et al.} [Belle Collaboration],
  %``Search for $\boldsymbol{B\to h\nu\bar{\nu}}$ decays with semileptonic tagging at Belle,''
  Phys.\ Rev.\ D {\bf 96} (2017) no.9,  091101
   Addendum: [Phys.\ Rev.\ D {\bf 97} (2018) no.9,  099902]
%  doi:10.1103/PhysRevD.97.099902, 10.1103/PhysRevD.96.091101
  [arXiv:1702.03224 [hep-ex]].
  %%CITATION = doi:10.1103/PhysRevD.97.099902, 10.1103/PhysRevD.96.091101;%%
  %61 citations counted in INSPIRE as of 13 Jun 2019

%\cite{Charles:2015gya}
\bibitem{Charles:2015gya}
  J.~Charles {\it et al.},
  %``Current status of the Standard Model CKM fit and constraints on $\Delta F=2$ New Physics,''
  Phys.\ Rev.\ D {\bf 91} (2015) no.7,  073007
%  doi:10.1103/PhysRevD.91.073007
  [arXiv:1501.05013 [hep-ph]].
  %%CITATION = doi:10.1103/PhysRevD.91.073007;%%
  %202 citations counted in INSPIRE as of 13 Jun 2019

%\cite{Brod:2011ty}
\bibitem{Brod:2011ty}
  J.~Brod and M.~Gorbahn,
  %``Next-to-Next-to-Leading-Order Charm-Quark Contribution to the CP Violation Parameter epsilon_K and Delta M_K,''
  Phys.\ Rev.\ Lett.\  {\bf 108} (2012) 121801
%  doi:10.1103/PhysRevLett.108.121801
  [arXiv:1108.2036 [hep-ph]].
  %%CITATION = doi:10.1103/PhysRevLett.108.121801;%%
  %142 citations counted in INSPIRE as of 13 Jun 2019

%\cite{Buras:2012ru}
\bibitem{Buras:2012ru}
  A.~J.~Buras, J.~Girrbach, D.~Guadagnoli and G.~Isidori,
  %``On the Standard Model prediction for BR(B{s,d} to mu+ mu-),''
  Eur.\ Phys.\ J.\ C {\bf 72} (2012) 2172
%  doi:10.1140/epjc/s10052-012-2172-1
  [arXiv:1208.0934 [hep-ph]].
  %%CITATION = doi:10.1140/epjc/s10052-012-2172-1;%%
  %226 citations counted in INSPIRE as of 13 Jun 2019

%%%%%%%% tau final state refs  %%%%%

%\cite{Capdevila:2017iqn}
\bibitem{Capdevila:2017iqn}
  B.~Capdevila, A.~Crivellin, S.~Descotes-Genon, L.~Hofer and J.~Matias,
  %``Searching for New Physics with $b\to s\tau^+\tau^-$ processes,''
  Phys.\ Rev.\ Lett.\  {\bf 120} (2018) no.18,  181802
%  doi:10.1103/PhysRevLett.120.181802
  [arXiv:1712.01919 [hep-ph]].
  %%CITATION = doi:10.1103/PhysRevLett.120.181802;%%
  %24 citations counted in INSPIRE as of 13 Jun 2019


  %%%%% LFV Table refs %%%%%%%%%%
%\cite{TheMEG:2016wtm}
\bibitem{TheMEG:2016wtm}
  A.~M.~Baldini {\it et al.} [MEG Collaboration],
  %``Search for the lepton flavour violating decay $\mu ^+ \rightarrow \mathrm {e}^+ \gamma $ with the full dataset of the MEG experiment,''
  Eur.\ Phys.\ J.\ C {\bf 76} (2016) no.8,  434
%  doi:10.1140/epjc/s10052-016-4271-x
  [arXiv:1605.05081 [hep-ex]].
  %%CITATION = doi:10.1140/epjc/s10052-016-4271-x;%%
  %315 citations counted in INSPIRE as of 13 Jun 2019

%\cite{Baldini:2018nnn}
\bibitem{Baldini:2018nnn}
  A.~M.~Baldini {\it et al.} [MEG II Collaboration],
  %``The design of the MEG II experiment,''
  Eur.\ Phys.\ J.\ C {\bf 78} (2018) no.5,  380
%  doi:10.1140/epjc/s10052-018-5845-6
  [arXiv:1801.04688 [physics.ins-det]].
  %%CITATION = doi:10.1140/epjc/s10052-018-5845-6;%%
  %42 citations counted in INSPIRE as of 13 Jun 2019

%\cite{Aubert:2009ag}
\bibitem{Aubert:2009ag}
  B.~Aubert {\it et al.} [BaBar Collaboration],
  %``Searches for Lepton Flavor Violation in the Decays tau+- ---> e+- gamma and tau+- ---> mu+- gamma,''
  Phys.\ Rev.\ Lett.\  {\bf 104} (2010) 021802
%  doi:10.1103/PhysRevLett.104.021802
  [arXiv:0908.2381 [hep-ex]].
  %%CITATION = doi:10.1103/PhysRevLett.104.021802;%%
  %428 citations counted in INSPIRE as of 13 Jun 2019

%\cite{Kou:2018nap}
\bibitem{Kou:2018nap}
  E.~Kou {\it et al.} [Belle-II Collaboration],
  %``The Belle II Physics Book,''
  arXiv:1808.10567 [hep-ex].
  %%CITATION = ARXIV:1808.10567;%%
  %158 citations counted in INSPIRE as of 13 Jun 2019

%\cite{Bellgardt:1987du}
\bibitem{Bellgardt:1987du}
  U.~Bellgardt {\it et al.} [SINDRUM Collaboration],
  %``Search for the Decay mu+ ---> e+ e+ e-,''
  Nucl.\ Phys.\ B {\bf 299} (1988) 1.
%  doi:10.1016/0550-3213(88)90462-2
  %%CITATION = doi:10.1016/0550-3213(88)90462-2;%%
  %646 citations counted in INSPIRE as of 13 Jun 2019

%\cite{Blondel:2013ia}
\bibitem{Blondel:2013ia}
  A.~Blondel {\it et al.},
  %``Research Proposal for an Experiment to Search for the Decay $\mu \to eee$,''
  arXiv:1301.6113 [physics.ins-det].
  %%CITATION = ARXIV:1301.6113;%%
  %216 citations counted in INSPIRE as of 13 Jun 2019

%\cite{Hayasaka:2010np}
\bibitem{Hayasaka:2010np}
  K.~Hayasaka {\it et al.},
  %``Search for Lepton Flavor Violating Tau Decays into Three Leptons with 719 Million Produced Tau+Tau- Pairs,''
  Phys.\ Lett.\ B {\bf 687} (2010) 139
%  doi:10.1016/j.physletb.2010.03.037
  [arXiv:1001.3221 [hep-ex]].
  %%CITATION = doi:10.1016/j.physletb.2010.03.037;%%
  %275 citations counted in INSPIRE as of 13 Jun 2019

%\cite{Bertl:2006up}
\bibitem{Bertl:2006up}
  W.~H.~Bertl {\it et al.} [SINDRUM II Collaboration],
  %``A Search for muon to electron conversion in muonic gold,''
  Eur.\ Phys.\ J.\ C {\bf 47}, 337 (2006).
%  doi:10.1140/epjc/s2006-02582-x
  %%CITATION = doi:10.1140/epjc/s2006-02582-x;%%
  %418 citations counted in INSPIRE as of 13 Jun 2019

%\cite{Nguyen:2015vkk}
\bibitem{Nguyen:2015vkk}
  T.~M.~Nguyen [DeeMe Collaboration],
  %``Search for µ − e conversion with DeeMe experiment at J-PARC MLF,''
  PoS FPCP {\bf 2015} (2015) 060.
%  doi:10.22323/1.248.0060
  %%CITATION = doi:10.22323/1.248.0060;%%
  %5 citations counted in INSPIRE as of 13 Jun 2019

%\cite{Krikler:2015msn}
\bibitem{Krikler:2015msn}
  B.~E.~Krikler [COMET Collaboration],
  %``An Overview of the COMET Experiment and its Recent Progress,''
  arXiv:1512.08564 [physics.ins-det].
  %%CITATION = ARXIV:1512.08564;%%
  %7 citations counted in INSPIRE as of 13 Jun 2019

  %\cite{KunoESPP19}
  \bibitem{KunoESPP19}
      Y.~Kuno,
 Presentation at the Flavour Session of the CERN Council Open Symposium on the Update of the European Strategy for Particle Physics. Granada, 13-16 May 2019.

%\cite{Bartoszek:2014mya}
\bibitem{Bartoszek:2014mya}
  L.~Bartoszek {\it et al.} [Mu2e Collaboration],
  %``Mu2e Technical Design Report,''
  arXiv:1501.05241 [physics.ins-det].
  %%CITATION = ARXIV:1501.05241;%%
  %168 citations counted in INSPIRE as of 13 Jun 2019

  %\cite{Abusalma:2018xem}
  \bibitem{Abusalma:2018xem}
    F.~Abusalma {\it et al.} [Mu2e Collaboration],
    %``Expression of Interest for Evolution of the Mu2e Experiment,''
    arXiv:1802.02599 [physics.ins-det].
    %%CITATION = ARXIV:1802.02599;%%
    %12 citations counted in INSPIRE as of 11 Jul 2019


  %%%%%%. LQ LHC refs %%%%%%%%
%\cite{Khachatryan:2014ura}
\bibitem{Khachatryan:2014ura}
  V.~Khachatryan {\it et al.} [CMS Collaboration],
  %``Search for pair production of third-generation scalar leptoquarks and top squarks in proton–proton collisions at $\sqrt{s}$=8 TeV,''
  Phys.\ Lett.\ B {\bf 739} (2014) 229
%  doi:10.1016/j.physletb.2014.10.063
  [arXiv:1408.0806 [hep-ex]].
  %%CITATION = doi:10.1016/j.physletb.2014.10.063;%%
  %88 citations counted in INSPIRE as of 13 Jun 2019

%\cite{Aad:2015caa}
\bibitem{Aad:2015caa}
  G.~Aad {\it et al.} [ATLAS Collaboration],
  %``Searches for scalar leptoquarks in pp collisions at $\sqrt{s}$ = 8 TeV with the ATLAS detector,''
  Eur.\ Phys.\ J.\ C {\bf 76} (2016) no.1,  5
%  doi:10.1140/epjc/s10052-015-3823-9
  [arXiv:1508.04735 [hep-ex]].
  %%CITATION = doi:10.1140/epjc/s10052-015-3823-9;%%
  %112 citations counted in INSPIRE as of 13 Jun 2019

%\cite{Sirunyan:2017yrk}
\bibitem{Sirunyan:2017yrk}
  A.~M.~Sirunyan {\it et al.} [CMS Collaboration],
  %``Search for third-generation scalar leptoquarks and heavy right-handed neutrinos in final states with two tau leptons and two jets in proton-proton collisions at $ \sqrt{s}=13 $ TeV,''
  JHEP {\bf 1707} (2017) 121
%  doi:10.1007/JHEP07(2017)121
  [arXiv:1703.03995 [hep-ex]].
  %%CITATION = doi:10.1007/JHEP07(2017)121;%%
  %77 citations counted in INSPIRE as of 13 Jun 2019

%\cite{Sirunyan:2018vhk}
\bibitem{Sirunyan:2018vhk}
  A.~M.~Sirunyan {\it et al.} [CMS Collaboration],
  %``Search for heavy neutrinos and third-generation leptoquarks in hadronic states of two $\tau$ leptons and two jets in proton-proton collisions at $\sqrt{s} =$ 13 TeV,''
  JHEP {\bf 1903} (2019) 170
%  doi:10.1007/JHEP03(2019)170
  [arXiv:1811.00806 [hep-ex]].
  %%CITATION = doi:10.1007/JHEP03(2019)170;%%
  %9 citations counted in INSPIRE as of 13 Jun 2019

%\cite{Sirunyan:2018kzh}
\bibitem{Sirunyan:2018kzh}
  A.~M.~Sirunyan {\it et al.} [CMS Collaboration],
  %``Constraints on models of scalar and vector leptoquarks decaying to a quark and a neutrino at $\sqrt{s}=$ 13 TeV,''
  Phys.\ Rev.\ D {\bf 98} (2018) no.3,  032005
%  doi:10.1103/PhysRevD.98.032005
  [arXiv:1805.10228 [hep-ex]].
  %%CITATION = doi:10.1103/PhysRevD.98.032005;%%
  %25 citations counted in INSPIRE as of 13 Jun 2019


%\cite{Cerri:2018ypt}
\bibitem{Cerri:2018ypt}
  A.~Cerri {\it et al.},
  %``Opportunities in Flavour Physics at the HL-LHC and HE-LHC,''
  arXiv:1812.07638 [hep-ph].
  %%CITATION = ARXIV:1812.07638;%%
  %28 citations counted in INSPIRE as of 13 Jun 2019

%\cite{CidVidal:2018eel}
\bibitem{CidVidal:2018eel}
  X.~Cid Vidal {\it et al.} [Working Group 3],
  %``Beyond the Standard Model Physics at the HL-LHC and HE-LHC,''
  arXiv:1812.07831 [hep-ph].
  %%CITATION = ARXIV:1812.07831;%%
  %26 citations counted in INSPIRE as of 13 Jun 2019


  %%%% Appendix refs start %%%%%%%%%%%
%\cite{Buchalla:1995vs}
\bibitem{Buchalla:1995vs}
  G.~Buchalla, A.~J.~Buras and M.~E.~Lautenbacher,
  %``Weak decays beyond leading logarithms,''
  Rev.\ Mod.\ Phys.\  {\bf 68} (1996) 1125
%  doi:10.1103/RevModPhys.68.1125
  [hep-ph/9512380].
  %%CITATION = doi:10.1103/RevModPhys.68.1125;%%
  %2397 citations counted in INSPIRE as of 13 Jun 2019

%\cite{Bobeth:1999mk}
\bibitem{Bobeth:1999mk}
  C.~Bobeth, M.~Misiak and J.~Urban,
  %``Photonic penguins at two loops and $m_t$ dependence of $BR[B \to  X_s l^+ l^-]$,''
  Nucl.\ Phys.\ B {\bf 574} (2000) 291
%  doi:10.1016/S0550-3213(00)00007-9
  [hep-ph/9910220].
  %%CITATION = doi:10.1016/S0550-3213(00)00007-9;%%
  %385 citations counted in INSPIRE as of 13 Jun 2019

%\cite{Ali:2002jg}
\bibitem{Ali:2002jg}
  A.~Ali, E.~Lunghi, C.~Greub and G.~Hiller,
  %``Improved model independent analysis of semileptonic and radiative rare $B$ decays,''
  Phys.\ Rev.\ D {\bf 66} (2002) 034002
%  doi:10.1103/PhysRevD.66.034002
  [hep-ph/0112300].
  %%CITATION = doi:10.1103/PhysRevD.66.034002;%%
  %381 citations counted in INSPIRE as of 13 Jun 2019

%\cite{Hiller:2003js}
\bibitem{Hiller:2003js}
  G.~Hiller and F.~Kruger,
  %``More model-independent analysis of $b \to s$ processes,''
  Phys.\ Rev.\ D {\bf 69} (2004) 074020
%  doi:10.1103/PhysRevD.69.074020
  [hep-ph/0310219].
  %%CITATION = doi:10.1103/PhysRevD.69.074020;%%
  %367 citations counted in INSPIRE as of 13 Jun 2019

%\cite{Bobeth:2007dw}
\bibitem{Bobeth:2007dw}
  C.~Bobeth, G.~Hiller and G.~Piranishvili,
  %``Angular distributions of $\bar{B} \to \bar{K} \ell^+\ell^-$ decays,''
  JHEP {\bf 0712} (2007) 040
%  doi:10.1088/1126-6708/2007/12/040
  [arXiv:0709.4174 [hep-ph]].
  %%CITATION = doi:10.1088/1126-6708/2007/12/040;%%
  %270 citations counted in INSPIRE as of 13 Jun 2019

%\cite{Bobeth:2010wg}
\bibitem{Bobeth:2010wg}
  C.~Bobeth, G.~Hiller and D.~van Dyk,
  %``The Benefits of $\bar{B} -> \bar{K}^* l^+ l^-$ Decays at Low Recoil,''
  JHEP {\bf 1007} (2010) 098
%  doi:10.1007/JHEP07(2010)098
  [arXiv:1006.5013 [hep-ph]].
  %%CITATION = doi:10.1007/JHEP07(2010)098;%%
  %184 citations counted in INSPIRE as of 13 Jun 2019

%\cite{Becirevic:2016zri}
\bibitem{Becirevic:2016zri}
  D.~Be\v{c}irevi\'{c}, O.~Sumensari and R.~Zukanovich Funchal,
  %``Lepton flavor violation in exclusive $b\rightarrow s$ decays,''
  Eur.\ Phys.\ J.\ C {\bf 76} (2016) no.3,  134
%  doi:10.1140/epjc/s10052-016-3985-0
  [arXiv:1602.00881 [hep-ph]].
  %%CITATION = doi:10.1140/epjc/s10052-016-3985-0;%%
  %71 citations counted in INSPIRE as of 13 Jun 2019

%\cite{Khodjamirian:2010vf}
\bibitem{Khodjamirian:2010vf}
  A.~Khodjamirian, T.~Mannel, A.~A.~Pivovarov and Y.-M.~Wang,
  %``Charm-loop effect in $B \to K^{(*)} \ell^{+} \ell^{-}$ and $B\to K^*\gamma$,''
  JHEP {\bf 1009} (2010) 089
%  doi:10.1007/JHEP09(2010)089
  [arXiv:1006.4945 [hep-ph]].
  %%CITATION = doi:10.1007/JHEP09(2010)089;%%
  %296 citations counted in INSPIRE as of 13 Jun 2019

%\cite{Buras:2014fpa}
\bibitem{Buras:2014fpa}
  A.~J.~Buras, J.~Girrbach-Noe, C.~Niehoff and D.~M.~Straub,
  %``$ B\to {K}^{\left(\ast \right)}\nu \overline{\nu} $ decays in the Standard Model and beyond,''
  JHEP {\bf 1502} (2015) 184
%  doi:10.1007/JHEP02(2015)184
  [arXiv:1409.4557 [hep-ph]].
  %%CITATION = doi:10.1007/JHEP02(2015)184;%%
  %157 citations counted in INSPIRE as of 13 Jun 2019

%\cite{Bobeth:2017ecx}
\bibitem{Bobeth:2017ecx}
  C.~Bobeth and A.~J.~Buras,
  %``Leptoquarks meet $\varepsilon'/\varepsilon$ and rare Kaon processes,''
  JHEP {\bf 1802} (2018) 101
%  doi:10.1007/JHEP02(2018)101
  [arXiv:1712.01295 [hep-ph]].
  %%CITATION = doi:10.1007/JHEP02(2018)101;%%
  %41 citations counted in INSPIRE as of 13 Jun 2019

%\cite{Bordone:2017lsy}
\bibitem{Bordone:2017lsy}
  M.~Bordone, D.~Buttazzo, G.~Isidori and J.~Monnard,
  %``Probing Lepton Flavour Universality with $K \to \pi \nu \bar\nu$ decays,''
  Eur.\ Phys.\ J.\ C {\bf 77} (2017) no.9,  618
%  doi:10.1140/epjc/s10052-017-5202-1
  [arXiv:1705.10729 [hep-ph]].
  %%CITATION = doi:10.1140/epjc/s10052-017-5202-1;%%
  %29 citations counted in INSPIRE as of 13 Jun 2019

%\cite{Buras:2004qb}
\bibitem{Buras:2004qb}
  A.~J.~Buras, T.~Ewerth, S.~Jager and J.~Rosiek,
  %``K+ ---> pi+ nu anti-nu and K(L) ---> pi0 nu anti-nu decays in the general MSSM,''
  Nucl.\ Phys.\ B {\bf 714} (2005) 103
%  doi:10.1016/j.nuclphysb.2005.02.014
  [hep-ph/0408142].
  %%CITATION = doi:10.1016/j.nuclphysb.2005.02.014;%%
  %104 citations counted in INSPIRE as of 13 Jun 2019

%\cite{Lavoura:2003xp}
\bibitem{Lavoura:2003xp}
  L.~Lavoura,
  %``General formulae for f(1) ---> f(2) gamma,''
  Eur.\ Phys.\ J.\ C {\bf 29} (2003) 191
%  doi:10.1140/epjc/s2003-01212-7
  [hep-ph/0302221].
  %%CITATION = doi:10.1140/epjc/s2003-01212-7;%%
  %106 citations counted in INSPIRE as of 13 Jun 2019

%\cite{Okada:1999zk}
\bibitem{Okada:1999zk}
  Y.~Okada, K.~i.~Okumura and Y.~Shimizu,
  %``Mu --> e gamma and mu --> 3 e processes with polarized muons and supersymmetric grand unified theories,''
  Phys.\ Rev.\ D {\bf 61} (2000) 094001
%  doi:10.1103/PhysRevD.61.094001
  [hep-ph/9906446].
  %%CITATION = doi:10.1103/PhysRevD.61.094001;%%
  %115 citations counted in INSPIRE as of 13 Jun 2019

%\cite{Kuno:1999jp}
\bibitem{Kuno:1999jp}
  Y.~Kuno and Y.~Okada,
  %``Muon decay and physics beyond the standard model,''
  Rev.\ Mod.\ Phys.\  {\bf 73} (2001) 151
%  doi:10.1103/RevModPhys.73.151
  [hep-ph/9909265].
  %%CITATION = doi:10.1103/RevModPhys.73.151;%%
  %523 citations counted in INSPIRE as of 13 Jun 2019

%\cite{Gabrielli:2000te}
\bibitem{Gabrielli:2000te}
  E.~Gabrielli,
  %``Model independent constraints on leptoquarks from rare muon and tau lepton processes,''
  Phys.\ Rev.\ D {\bf 62} (2000) 055009
%  doi:10.1103/PhysRevD.62.055009
  [hep-ph/9911539].
  %%CITATION = doi:10.1103/PhysRevD.62.055009;%%
  %35 citations counted in INSPIRE as of 13 Jun 2019

%\cite{Kitano:2002mt}
\bibitem{Kitano:2002mt}
  R.~Kitano, M.~Koike and Y.~Okada,
  %``Detailed calculation of lepton flavor violating muon electron conversion rate for various nuclei,''
  Phys.\ Rev.\ D {\bf 66} (2002) 096002
   Erratum: [Phys.\ Rev.\ D {\bf 76} (2007) 059902]
%  doi:10.1103/PhysRevD.76.059902, 10.1103/PhysRevD.66.096002
  [hep-ph/0203110].
  %%CITATION = doi:10.1103/PhysRevD.76.059902, 10.1103/PhysRevD.66.096002;%%
  %259 citations counted in INSPIRE as of 13 Jun 2019

%\cite{Kosmas:2001mv}
\bibitem{Kosmas:2001mv}
  T.~S.~Kosmas, S.~Kovalenko and I.~Schmidt,
  %``Nuclear muon- e- conversion in strange quark sea,''
  Phys.\ Lett.\ B {\bf 511} (2001) 203
%  doi:10.1016/S0370-2693(01)00657-8
  [hep-ph/0102101].
  %%CITATION = doi:10.1016/S0370-2693(01)00657-8;%%
  %45 citations counted in INSPIRE as of 13 Jun 2019

%\cite{Poh:2017tfo}
\bibitem{Poh:2017tfo}
  Z.~Poh and S.~Raby,
  %``Vectorlike leptons: Muon g-2 anomaly, lepton flavor violation, Higgs boson decays, and lepton nonuniversality,''
  Phys.\ Rev.\ D {\bf 96} (2017) no.1,  015032
%  doi:10.1103/PhysRevD.96.015032
  [arXiv:1705.07007 [hep-ph]].
  %%CITATION = doi:10.1103/PhysRevD.96.015032;%%
  %16 citations counted in INSPIRE as of 13 Jun 2019

%\cite{Dermisek:2013gta}
\bibitem{Dermisek:2013gta}
  R.~Dermisek and A.~Raval,
  %``Explanation of the Muon g-2 Anomaly with Vectorlike Leptons and its Implications for Higgs Decays,''
  Phys.\ Rev.\ D {\bf 88} (2013) 013017
%  doi:10.1103/PhysRevD.88.013017
  [arXiv:1305.3522 [hep-ph]].
  %%CITATION = doi:10.1103/PhysRevD.88.013017;%%
  %58 citations counted in INSPIRE as of 13 Jun 2019

%\cite{Abada:2013aba}
\bibitem{Abada:2013aba}
  A.~Abada, A.~M.~Teixeira, A.~Vicente and C.~Weiland,
  %``Sterile neutrinos in leptonic and semileptonic decays,''
  JHEP {\bf 1402} (2014) 091
%  doi:10.1007/JHEP02(2014)091
  [arXiv:1311.2830 [hep-ph]].
  %%CITATION = doi:10.1007/JHEP02(2014)091;%%
  %88 citations counted in INSPIRE as of 13 Jun 2019

%\cite{Atre:2009rg}
\bibitem{Atre:2009rg}
  A.~Atre, T.~Han, S.~Pascoli and B.~Zhang,
  %``The Search for Heavy Majorana Neutrinos,''
  JHEP {\bf 0905} (2009) 030
%  doi:10.1088/1126-6708/2009/05/030
  [arXiv:0901.3589 [hep-ph]].
  %%CITATION = doi:10.1088/1126-6708/2009/05/030;%%
  %505 citations counted in INSPIRE as of 26 Jun 2019

%\cite{Abada:2017jjx}
\bibitem{Abada:2017jjx}
  A.~Abada, V.~De Romeri, M.~Lucente, A.~M.~Teixeira and T.~Toma,
  %``Effective Majorana mass matrix from tau and pseudoscalar meson lepton number violating decays,''
  JHEP {\bf 1802} (2018) 169
%  doi:10.1007/JHEP02(2018)169
  [arXiv:1712.03984 [hep-ph]].
  %%CITATION = doi:10.1007/JHEP02(2018)169;%%
  %20 citations counted in INSPIRE as of 26 Jun 2019

  %\cite{Abada:2019bac}
  \bibitem{Abada:2019bac}
    A.~Abada, C.~Hati, X.~Marcano and A.~M.~Teixeira,
    %``Interference effects in LNV and LFV semileptonic decays: the Majorana hypothesis,''
    arXiv:1904.05367 [hep-ph].
    %%CITATION = ARXIV:1904.05367;%%

%\cite{Bray:2007ru}
\bibitem{Bray:2007ru}
  S.~Bray, J.~S.~Lee and A.~Pilaftsis,
  %``Resonant CP violation due to heavy neutrinos at the LHC,''
  Nucl.\ Phys.\ B {\bf 786} (2007) 95
%  doi:10.1016/j.nuclphysb.2007.07.002
  [hep-ph/0702294 [HEP-PH]].
  %%CITATION = doi:10.1016/j.nuclphysb.2007.07.002;%%
  %77 citations counted in INSPIRE as of 26 Jun 2019

%\cite{Dev:2019rxh}
\bibitem{Dev:2019rxh}
  P.~S.~Bhupal Dev, R.~N.~Mohapatra and Y.~Zhang,
  %``Probing heavy neutrino mixing and associated CP violation at future hadron colliders,''
  arXiv:1904.04787 [hep-ph].
  %%CITATION = ARXIV:1904.04787;%%
  %1 citations counted in INSPIRE as of 26 Jun 2019

%\cite{Aebischer:2018iyb}
\bibitem{Aebischer:2018iyb}
  J.~Aebischer, J.~Kumar, P.~Stangl and D.~M.~Straub,
  %``A Global Likelihood for Precision Constraints and Flavour Anomalies,''
  Eur.\ Phys.\ J.\ C {\bf 79} (2019) no.6,  509
  % doi:10.1140/epjc/s10052-019-6977-z
  [arXiv:1810.07698 [hep-ph]].
  %%CITATION = doi:10.1140/epjc/s10052-019-6977-z;%%
  %23 citations counted in INSPIRE as of 12 Sep 2019

\end{thebibliography}
\end{document}